\documentclass[12pt,letter,english,intoc,bibliography=totoc,index=totoc,BCOR10mm,captions=tableheading,nottitlepage,fleqn]{article}

\usepackage{setspace, geometry}
\usepackage{etex}
\usepackage{babel,newcent,textcomp}

\usepackage{scrextend}
\usepackage{booktabs, threeparttable}

\usepackage[autostyle]{csquotes}
\usepackage[style=bwl-FU, citestyle=authoryear, backend=biber, useprefix, maxbibnames=99]{biblatex}
\usepackage{lmodern}

\usepackage[T1]{fontenc}
\usepackage[latin9]{inputenc}
\usepackage{sgame}
\usepackage{fancyhdr}
\usepackage{amsmath,amssymb,bbm}
\usepackage[amsthm]{ntheorem}
\usepackage{dsfont}
\usepackage{tablefootnote}
\usepackage[labelfont=bf]{caption}
\usepackage{enumitem}
\usepackage{geometry}
\usepackage{lscape}
\usepackage[unicode=true,
 bookmarks=true,bookmarksnumbered=true,bookmarksopen=true,bookmarksopenlevel=1,
 breaklinks=false,pdfborder={0 0 0},backref=false,colorlinks=true,linkcolor=black,citecolor=black,anchorcolor=black,urlcolor=blue]
 {hyperref}
\hypersetup{pdftitle={Your title},
 pdfauthor={Your name},
 pdfpagelayout=OneColumn, pdfnewwindow=true, pdfstartview=XYZ, plainpages=false}

\usepackage{tikz}

\usetikzlibrary{shapes.geometric, arrows, positioning}
\tikzset{
    >=stealth',
    box/.style={
           rectangle,
           rounded corners,
           draw=black, very thick,
           text width=6.5em,
           minimum height=1.5em,
           text centered},
    boxsmall/.style={
           rectangle,
           rounded corners,
           draw=black, very thick,
           text width=6em,
           minimum height=1.5em,
           text centered},
     simple/.style={
           text centered},
    arrow/.style={
           ->,
           thick,
           shorten <=2pt,
           shorten >=2pt,}
}
\usepackage{graphicx}
\graphicspath{{""}}

\usepackage[multiple]{footmisc}
\usepackage{cleveref}
\usepackage[figure]{hypcap}
\usepackage{calc}

\let\mySection\section\renewcommand{\section}{\suppressfloats[t]\mySection}

\theoremstyle{break}
\newtheorem{theorem}{Theorem}

\newtheorem{hypothesis}{Hypothesis}
\newtheorem*{fact}[theorem]{Fact}

\makeatother

\begin{document}

\lhead[]{}

\rhead[]{}

\lfoot[\thepage]{}

\cfoot{}

\rfoot[]{\thepage}

\newpage

\begin{titlepage}
\title{The Fake News Effect: Experimentally Identifying \\ Motivated Reasoning Using Trust in News\thanks{\scriptsize{I would like to especially thank Alberto Alesina, Roland Benabou, Christine Exley, David Laibson, Matthew Rabin, and Leeat Yariv for invaluable advice and support. Thank you also to Hunt Allcott, Kyle Chauvin, Stefano DellaVigna, Liz Engle, Ben Enke, Harris Eppsteiner, Ed Glaeser, Michael Hiscox, Alejandro Lagomarsino, Bart Lipman, Alessandro Lizzeri, Muriel Niederle, Dave Rand, Gautam Rao, Marta Serra-Garcia, Jesse Shapiro, Nick Stagnaro, Richard Thaler, Mattie Toma, and numerous seminar participants for helpful comments. I am grateful for funding support from the Harvard Business School Research Fellowship and the Eric M. Mindich Research Fund for the Foundations of Human Behavior. The main experiment was registered on the AEA RCT Registry (AEARCTR-0004339). The replication was preregistered on the AEA RCT Registry (AEARCTR-0004401). Both were deemed exempt by the IRB at Harvard (IRB17-1725).}}}
\author{Michael Thaler\thanks{\scriptsize{Princeton University. Email: \href{michael.thaler@princeton.edu}{michael.thaler@princeton.edu}.}}}
\date{May 2022\vspace{-2mm}}

\maketitle

\abstract{Motivated reasoning posits that people distort how they process information in the direction of beliefs they find attractive. This paper creates a novel experimental design to identify motivated reasoning from Bayesian updating when people have preconceived beliefs. It analyzes how subjects assess the veracity of information sources that tell them the median of their belief distribution is too high or too low. Bayesians infer nothing about the source veracity, but motivated beliefs are evoked. Evidence supports politically-motivated reasoning about immigration, income mobility, crime, racial discrimination, gender, climate change, and gun laws. Motivated reasoning helps explain belief biases, polarization, and overconfidence.

\vspace{5mm}


}

\bigskip

\setcounter{page}{0}
\thispagestyle{empty}
\end{titlepage}
\pagebreak \newpage


\section{Introduction}




\noindent On many topics, people disagree about the answers to factual questions, and their beliefs are often inaccurate in predictable directions. People have differing beliefs about questions related to income mobility, crime rates, and racial discrimination in labor markets; tend to be biased in the direction that is more representative of their political party's stances; and often overestimate their own political knowledge (e.g. \cite{AST18}; \cite{FNR17}; \cite{OS15}). As shown by \textcite{GH09}, \textcite{MPSS-WP}, and Allcott et al. (2020), these partisan beliefs can affect consumer, financial, and public health behavior. Given the significance of these issues, why does such bias and belief polarization persist? 

After receiving a piece of news, people immediately form posterior beliefs in ways that incorporate both their prior beliefs and how they perceive the informativeness of the news. If we only observe beliefs at a snapshot in time, two people's disagreement can be consistent with many explanations: for instance, they may have started with different beliefs, they may receive news from different sources, or they may have differently-distorted inference processes. The first two channels are often relevant in politicized settings. First, Democrats and Republicans often have different priors, leading to differences in posteriors; this can be consistent with Bayes' rule. Second, Democrats and Republicans often consume different media, and may find arguments akin to those from MSNBC and from Fox News differentially informative.\footnote{There is ample evidence consistent with these channels (e.g. \cite{TL06}; \cite{KHBE12}; \cite{NR10}; \cite{NR13}; \cite{NRU13}).}

This paper introduces a new tool for detecting a third channel, \textit{motivated reasoning}, which posits that people distort their inference process in the direction of states they find more attractive. In many settings, we may think that people are motivated to hold beliefs that align with their political party's stances: Republicans are motivated to believe that increasing immigration is associated with higher crime rates; Democrats are motivated to believe that there is severe racial discrimination in labor markets; and both parties are motivated to believe that members of their party are more knowledgeable about these issues than are members of the opposing party. When people receive news, they are often reminded of what beliefs they currently hold, and of beliefs they find more attractive to hold (\textit{motives}). They then use these motives when making inferences from and about news sources. 

While there is an intuition in the literature that motivated reasoning plays an important role in inference, particularly in political settings, designing an experiment to identify this mechanism has faced two major challenges. The first challenge is that it is difficult to identify motivated reasoning from Bayesian updating with sufficient statistical power when subjects' \textit{unmotivated} biases outweigh motivated reasoning. As summarized by \textcite{B19}, designs in which people receive informative signals have often not been able to find statistically significant evidence for motivated reasoning: \textcite{MNNR22}; \textcite{ER11}; and \textcite{CD17} find that people update more from ``good news'' than ``bad news,'' while \textcite{E11}; \textcite{K14}; \textcite{BGvdW18}; \textcite{C18}; and \textcite{B20} do not find evidence supporting this hypothesis.\footnote{It is worth noting that there is more consistent evidence for motivated decision-making and memory. This includes information avoidance and moral wiggle room (\cite{OSD13}; \cite{DWK07}; \cite{GNW16}), risk- and ambiguity-driven distortions (\cite{E15}; \cite{HW10}), and recall of past information (\cite{Z-WP}; \cite{CHZ20}). However, these papers do not show that updating in response to \textit{new} information is a driving channel.} The typical design for these papers involves giving subjects partially-informative news and testing for asymmetric updating from ``Good,'' ``Bad,'' and ``Neutral'' signals, a design that has consistently led to substantial underupdating (\cite{B19}).\footnote{Another strand of literature has empirically and experimentally considered how varied incentives affect beliefs (e.g. \cite{BLIC95}; \cite{BL97}; \cite{STvdW-WP}). Using a different design, \textcite{EK-WP} show that subjects behave differently in moral domains when they receive two pieces of equivalent information in differently-complex ways. My paper focuses more on the inference process for new information.} 

This challenge has led some researchers to argue that the importance of motivated reasoning is dwarfed by unmotivated explanations like conservatism or ``lazy thinking'' (\cite{PR19}; \cite{B19}; \cite{TPR20a}; \cite{TPR20b}). While such arguments may be valid in settings where there is both Bayesian inference and motivated reasoning, I bypass this issue by constructing an experimental paradigm that gives subjects messages in which there is \textit{nothing} for a Bayesian to infer, but motivated reasoning still predicts directional distortions. My design eliminates the role that Bayes' rule, conservatism, and confirmatory biases can play in inference, allowing for a cleaner identification of motivated reasoning. It therefore provides a more precise treatment-effect estimate, which enables me to find novel evidence identifying motivated reasoning in numerous previously-hypothesized domains. 

The second challenge is that in many domains of interest, such as politics, people enter into experiments with preconceived beliefs. Eliciting motivated reasoning in such settings is difficult; researchers typically either restrict tests to questions that have binary answers (e.g. \cite{MNNR22}), ask numerous questions in order to elicit entire prior and posterior belief distributions (e.g. \cite{ER11}), or do not fully rule out Bayesian updating (e.g. \cite{SBLS-WP}). My design is able to identify motivated reasoning from Bayesian updating on questions that have numerical answers, while only requiring elicitation of one moment of subjects' belief distribution: their median. Researchers can use this design to test how people motivatedly reason about essentially any factual question with a numerical answer.

The design has two main steps. First, each subject is given a variety of factual questions with numerical answers. On each question, each subject selects a response that they think is equally likely to be above or below the correct answer; that is, the median of their belief distribution is elicited. Second, the subject is given one binary message that is chosen randomly from either a True News source or a Fake News source; the message tells her whether the answer was above or below her median. If the message is from True News, it is always accurate. If the message is from Fake News, it is always inaccurate. The subject is not told which source the message came from; instead, she is asked to make inferences about the source's veracity from the message she receives. 

Since messages relate the true answer to subjects' subjective median, a Bayesian (she) would believe that it is equally likely for each source to report either message. That is, she has stated that she believes the answer is equally likely to be greater than or less than her median; so, she believes the likelihood that a True News source would report that the answer is greater than her median is 1/2, and the likelihood that a Fake News source would report that the answer is less than her median is also equal to 1/2. Therefore, the Bayesian would find a ``greater than'' message to be completely uninformative about the veracity of the news source. Likewise, she would find a ``less than'' message to be uninformative.

On the other hand, a subject who engages in motivated reasoning (he) will trust the news more if it sends a message that supports what he is more motivated to believe. If he engages in \textit{politically-}motivated reasoning, he will assess messages that align with the beliefs of his political party (\textit{Pro-Party news}) to be more truthful, while assessing messages that misalign (\textit{Anti-Party news}) to be less truthful.

I test for motivated reasoning using a within-subject experiment on Amazon Mechanical Turk with approximately 1,000 people in the United States. As a testament to the portability of the design, I run the test on nine different economic, political, and social questions (\textit{politicized} topics), and on one question about own performance in the experiment. The list of topics and hypothesized motives is in \Cref{topic-motives}. 

\renewcommand\arraystretch{1.4}
\begin{table}
\begin{tabular}{l l l}  
\toprule
\textbf{Topic} & \textbf{Pro-Democrat Motives} & \textbf{Pro-Republican Motives} \\
\midrule
\hyperref[obama-crime-question]{US crime} & Got better under Obama & Got worse under Obama \\
\hyperref[mobility-question]{Upward mobility} & Low in US after tax cuts & High in US after tax cuts \\
\hyperref[race-question]{Racial discrimination} & Severe in labor market & Not severe in labor market \\
\hyperref[gender-question]{Gender} & Girls better at math & Boys better at math \\
\hyperref[refugees-question]{Refugees} & Decreased violent crime & Increased violent crime \\
\hyperref[global-warming-question]{Climate change} & Scientific consensus & No scientific consensus \\
\hyperref[guns-question]{Gun reform} & Decreased homicides & Didn't decrease homicides \\
\hyperref[media-question]{Media bias} & Media not mostly Dems & Media mostly Dems \\
\hyperref[party-performance-question]{Party performance} & Higher for Dems over Reps & Higher for Reps over Dems \\
\hyperref[own-performance-question]{Own performance} & Higher for self over others & Higher for self over others \\
\bottomrule
\end{tabular}
\begin{scriptsize}
\caption{The list of topics and hypothesized motives in the experiment.}
\label{topic-motives}
\end{scriptsize}
\end{table}

\renewcommand\arraystretch{1}

The main finding from the experiment is that Bayesian updating is strongly rejected in favor of politically-motivated reasoning on the politicized topics. While a Bayesian would believe that Pro-Party and Anti-Party news are equally likely to be True News on the politicized topics, subjects in the experiment believe that Pro-Party messages are 9 percentage points more likely than Anti-Party messages to come from the True News source. These effects are moderate but very precisely estimated (s.e. 0.6 percentage points; $p \approx 10^{-40}$). This test is precise enough to measure each topic individually; for eight of the nine politicized topics, the treatment effect is statistically significantly larger than zero ($p \leq 0.001$). To my knowledge, this experiment provides the first evidence for motivated reasoning on these questions that is not confounded by Bayesian updating or prior-confirming biases (like those in \cite{F57}; \cite{LRL79}; \cite{T83}; and \cite{RS99}).\footnote{Papers that find asymmetric responses to information on related topics include: \textcite{TL06} [gun laws]; \textcite{AST18} [upward mobility]; \textcite{CHT-WP} [responses to taxation]; \textcite{RH-WP2} [racial labor market discrimination]; \textcite{S-WP} and \textcite{KS00} [gender and performance]; \textcite{AMS-WP}, \textcite{RH-WP1}, and \textcite{DPS13} [immigration]; \textcite{NR13} and \textcite{NRU13} [perceptions of elected officials]; and \textcite{SBLS-WP} [climate change]. Many findings from these papers may be due to motivated reasoning.} Treatment effects are larger for subjects who are more partisan. I also find consistent evidence that subjects motivatedly reason about their performance. 

The new design could lead to potential confounds that are unique to this experiment, such as if subjects systematically misreport their median belief or misinterpret the experiment's definitions of True News and Fake News, but I find that results are unlikely to be explained by these alternative hypotheses.\footnote{As described in \Cref{robustness-checks} and appendices, predictions are identical if subjects mistakenly believe Fake News sends random messages instead of always-false messages, and results are not driven by subjects who have skewed belief distributions and misreport their median.} To account for unmotivated mistakes, I also compare behavior on political topics to three neutral topics, and find that news veracity assessments on neutral topics are lower than assessments on Pro-Party news and higher than assessments on Anti-Party news.

The second main finding is that people's systematically-biased beliefs about these topics is related to their motivated beliefs (as also discussed by \cite{ER11}). Since people who motivatedly reason about an issue will systematically distort their beliefs, we can partly infer what people's motives are by looking at their current beliefs. That is, the direction of one's \textit{error} predicts the direction of one's \textit{motive}. In the context of the experiment, this hypothesis predicts that people will give higher veracity assessments to news that (falsely) reinforces their error compared to news that (truthfully) brings them closer to the correct answer. Indeed, subjects in the experiment trust the error-reinforcing Fake News more than the error-correcting True News, and only do so on topics where motivated reasoning is expected to play a role. 

I also show how motivated reasoning can lead to apparent overprecision, arguably the most durable form of overconfidence (\cite{MH08}; \cite{MTH15}). Motivated reasoning may provide a link between overprecision and partisanship, a relationship documented in \textcite{OS15}, because partisans' belief distributions are more miscalibrated. Miscalibrations are more severe with stronger motives, and this leads 50-percent confidence intervals to contain the true answer less than 50 percent of the time, especially for partisans. 

Motivated reasoning not only affects how people trust or distrust news, but also impacts how people change their beliefs about the topics themselves. Despite being uninformative, the messages lead to further belief polarization: subjects are more likely to revise their beliefs away from the population mean than towards it. That is, informational content is not a necessary condition for belief polarization.\footnote{Linking this finding to the polarization in trust in news relates to a growing literature that discusses the relationship between trust in news and political partisanship (\cite{NCG15}; \cite{L13}; \cite{DLM18}).} Politically-motivated reasoning helps reconcile the notions that the ideological polarization of beliefs may be high, even if the ideological polarization of information acquisition is modest (\cite{GS11}).\footnote{\textcite{GS06} and \textcite{GWZ-WP} provide alternative theoretical explanations with Bayesian agents who have different priors, but these models do not predict updating from uninformative signals.} It can also explain why news providers may distort messages to appeal to their audience's motivated beliefs (\cite{T-WPc}).

There are no sizable demographic heterogeneities in politically-motivated reasoning, neither in direction nor magnitude, once party preference is controlled for. Differences in treatment effects are almost all statistically indistinguishable from zero, and estimates are precise enough to rule out modest effect sizes.\footnote{On the other hand, \textcite{T21} shows that performance-motivated reasoning has substantial heterogeneity by gender; only men systematically motivatedly reason to think their performance was higher, while women are Bayesian on average.} Motivated beliefs on these questions seem to be principally driven by politics.

Lastly, to complement the experiment, I discuss a model of motivated reasoning in which people distort their inference when they receive new information. This model adds to the decades-old theoretical literature (such as \cite{AD82}; \cite{CM00}; and \cite{BT02}) by emphasizing the non-strategic nature of the bias: agents in this model do not distort their beliefs for functional reasons.\footnote{This directly contrasts the model with optimal beliefs models such as \textcite{BP05} and \textcite{MNNR22}. It is more similar to the selective memory models of \textcite{BT02} and \textcite{BT11} and the Bayesian model of \textcite{M-WP}, but emphasizes distorting the processing of new information.} They make inferences using a modified Bayes' rule, weighting priors and likelihoods as a Bayesian would, but act as if they receive an extra signal that puts more weight on higher-motive states.\footnote{This is a similar theory as \textcite{K16a}, but differs from related economic models of motivated cognition like \textcite{BT11} and \textcite{MNNR22}, in that it assumes directional distortions but not underreaction to information.} This leads them to make inferences even when a Bayesian does not. 

To further ensure that experimental results were unlikely to be driven by noise, I retested the main hypotheses using a preregistered replication one year later on a new sample. The findings successfully replicated; for details, see Online Appendix D.

The rest of the paper proceeds as follows: \Cref{theory} develops the model of motivated reasoning. \Cref{design} presents the experimental design and main hypotheses. \Cref{results-assessments} analyzes experimental results, including the evidence for motivated reasoning, treatment effect heterogeneity, and overprecision. \Cref{conclusion} concludes and proposes directions for future work.

\section{Theory}\label{theory}
\subsection{Defining and Identifying Motivated Reasoning}\label{moreas-subsection}

This section outlines the model of motivated reasoning. Agents distort how they process new information by acting as if they receive an additional signal corresponding to their motivated beliefs. When signals are uninformative to a Bayesian, as will be the case in the experiment, this distortion can lead motivated reasoners to update in the direction that they find more attractive. 

There are a set of agents $i$ and a set of questions $q$. For each question, Nature draws a state $\theta_q$ from $\Theta_q \in \mathbb{R}$. For each agent and question, Nature draws a source from $\mathcal{S}$: the source is either True ($T_{iq}$) or Fake ($F_{iq}$). Sources are independently drawn with $P(T_{iq}) = p$, and agents receive data $x_{iq}$ about the source veracity. 

Agents engage in \textit{motivated reasoning}, which means that they form their posterior by incorporating prior, likelihood, and a motivated beliefs term: 
\begin{equation*}
\underbrace{\mathbb{P}(T_{iq}|x_{iq})}_{\text{posterior}} \propto \underbrace{\mathbb{P}(T_{iq})}_{\text{prior}} \cdot \underbrace{\mathbb{P}(x_{iq}|T_{iq})}_{\text{likelihood}} \cdot \underbrace{M_{iq}(T_{iq}, x_{iq}) ^ {\varphi}}_{\text{mot. reasoning}},
\end{equation*}
We take log odds ratios of both sides (\cite{G80}) to attain an additive form:\footnote{\textcite{M-WP} uses a similar formulation in which motivated beliefs affect priors instead of belief updating.}
\begin{equation}
\label{moreas-eq}
\text{logit }\mathbb{P}(T_{iq}|x_{iq}) = \text{logit }p + \log \left(\frac{\mathbb{P}(x_{iq}|T_{iq})}{\mathbb{P}(x_{iq}|F_{iq})}\right) + \varphi (m_{iq}(T_{iq},x_{iq}) - m_{iq}(F_{iq},x_{iq})).
\end{equation}
Motivated reasoners act as if they receive both the actual signal ($x_{iq}$) and a signal whose relative likelihood corresponds to how much they are motivated to believe the state is $T_{iq}$ given $x_{iq}$. $m_{iq}: \mathcal{S} \to \mathbb{R}$ is called the \textbf{motive} function. The weight put on this signal is $\varphi \geq 0$, called \textbf{susceptibility}. $\varphi = 0$ corresponds to Bayes' rule; $\varphi > 0$ corresponds to motivated reasoning. 

I now introduce an information structure in which the $x_{iq}$ are binary and uninformative about the news source veracity, but affect $m_{iq}$. The $x_{iq}$ are chosen such that $\mathbb{P}(x_{iq}|T_{iq}) = \mathbb{P}(x_{iq}|F_{iq}) = 1/2$. In such a setting, described below, a Bayesian will set $\mathbb{P}(T_{iq}|x_{iq}) = p$, but a motivated reasoner will set $\text{logit }\mathbb{P}(T_{iq}|x_{iq}) = \text{logit }p + \varphi (m_{iq}(T_{iq},x_{iq}) - m_{iq}(F_{iq},x_{iq})).$ That is, if the motivated reasoner finds $x_{iq}$ more attractive, then they will form a posterior that the message comes from a truthful source. Meanwhile, a Bayesian will not update from either message because the message is uninformative about the source. 

In many environments, beliefs about $\mathbb{P}(x_{iq}|T_{iq})$ relate to $m_{iq}(T_{iq},x_{iq})$ since agents who are motivated to believe $x_{iq}$ is true may currently believe that $x_{iq}$ is true. As such, it is necessary to construct a signal structure that tailors messages to agents' \textit{current} beliefs.

For each $q$, agents report a median belief $\mu_{iq}$ such that they believe $\theta_q$ is equally likely to be greater than $\mu_{iq}$ or less than $\mu_{iq}$.\footnote{We assume that there is no atom at $\mu_{iq}$ and that $\mathbb{P}(\mu_{iq} = \theta_q) = 0$. That is, the agent believes that the answer has probability zero of being exactly equal to $\mu$, and the probability is indeed zero.} Nature then chooses one of two news sources from $\mathcal{S}$. The news source sends a message to the agent that either says that $\theta_q > \mu_{iq}$ or says that $\theta_q < \mu_{iq}$: $x_{iq}(\mu_{iq}) \in \{G(\mu_{iq}), L(\mu_{iq}) \}$. The True source always sends the ``truthful'' message and the Fake source always sends the ``untruthful'' message:

\vspace{3mm}
\begin{center}
\begin{tabular}{| c | c | c |}
\hline
& $\theta_q > \mu_{iq}$ & $\theta_q < \mu_{iq}$ \\
\hline
True News sends & $G(\mu_{iq})$ & $L(\mu_{iq})$ \\
\hline
Fake News sends & $L(\mu_{iq})$ & $G(\mu_{iq})$ \\
\hline
\end{tabular}
\end{center}
\vspace{3mm}

After receiving $x_{iq}(\mu_{iq})$, agents report how likely they think it is that the message came from the True source: $a(x_{iq}; \mu_{iq})$. They also report a revised median belief $\mu'_{iq}(x_{iq}; \mu_{iq})$. Finally, the states $\theta_q$ and sources are revealed.

Agents may either be motivated to believe $\theta_q$ is higher than $\mu_{iq}$ or $\theta_q$ is lower than $\mu_{iq}$. An agent with a high-$\theta_q$ motive has $m_{iq}(T_{iq},G(\mu_{iq})) = m_{iq}(F_{iq},L(\mu_{iq})) > m_{iq}(T_{iq},L(\mu_{iq})) = m_{iq}(F_{iq},G(\mu_{iq}))$, and vice versa for an agent with a low-$\theta_q$ motive.

We now consider how a Bayesian and a motivated reasoner update their beliefs about the news source. Given message $G(\mu_{iq})$, the Bayesian updates according to \Cref{moreas-eq} with $\varphi = 0$: 
\begin{align*}
\text{logit } a_B(G(\mu_{iq})) = \text{logit } \mathbb{P}(T(\mu_{iq})|G(\mu_{iq})) &= \text{logit } \mathbb{P}(T(\mu_{iq})) + \log \left(\frac{\mathbb{P}(G(\mu_{iq}) | T(\mu_{iq}))}{\mathbb{P}(G(\mu_{iq}) | F(\mu_{iq}))}\right) \\
&= \text{logit } p + \log \left(\frac{\mathbb{P}(\theta_q > \mu_{iq})}{\mathbb{P}(\theta_q < \mu_{iq})}\right) \\
&= \text{logit } p. \\
\text{Therefore: } a_B(G(\mu_{iq})) &= p = a_B(L(\mu_{iq})).
\end{align*}
Since the Bayesian has stated her median belief $\mu_{iq}$, she thinks that both messages are equally likely ex ante and does not update in either direction.

Meanwhile, the motivated reasoner updates according to \Cref{moreas-eq} with $\varphi > 0$:
\begin{align*}
\text{logit } a(G(\mu_{iq})) &= \text{logit } \mathbb{P}(T(\mu_{iq})) + \log \left(\frac{\mathbb{P}(G(\mu_{iq}) | T(\mu_{iq}))}{\mathbb{P}(G(\mu_{iq}) | F(\mu_{iq}))}\right) \\
&\hspace{32mm}+ \varphi \Big[m_{iq}(T_{iq},G(\mu_{iq})) - m_{iq}(F_{iq},G(\mu_{iq}))\Big] \\
&= \text{logit } p + \varphi \Big[m_{iq}(T_{iq},G(\mu_{iq})) - m_{iq}(F_{iq},G(\mu_{iq}))\Big].
\end{align*}
That is, we have the following:
\begin{fact*}[Identifying motivated reasoning using news veracity assessments]
The procedure above identifies motivated reasoning from Bayesian updating:
\begin{itemize}
\item Bayesian updating ($\varphi=0$): $a_B(G(\mu_{iq})) = a_B(L(\mu_{iq}))$.
\item Motivated reasoning ($\varphi>0$): \[a(G(\mu_{iq})) > a(L(\mu_{iq})) \iff m_{iq}(T_{iq},G(\mu_{iq})) > m_{iq}(T_{iq},L(\mu_{iq})).\]
\end{itemize}
\end{fact*} 
This design identifies whether agents are more motivated to believe that the true state is above their median belief $\mu_{iq}$ or to believe that the true state is below $\mu_{iq}$. 

More precisely, this design provides a joint test of $\varphi > 0$ and the direction of motives. In the political context, we will consider agents who are either of a Republican type or of a Democratic type, assume that $m_{Rq}(S_{Rq},x(\mu_{Rq})) = -m_{Dq}(S_{Dq},x(\mu_{Dq}))$, and then test whether, given these motives, $\varphi > 0$.

\subsection{Discussion}

Since signals are uninformative, this design also identifies motivated reasoning from several other non-Bayesian, models of inference. Consider the following generalized class of updating rules: 
\begin{align*}
\text{logit } a(G(\mu_{iq})) &= \zeta \text{ logit } p + \kappa \log \left(\frac{\mathbb{P}(G(\mu_{iq}) | T(\mu_{iq}))}{\mathbb{P}(G(\mu_{iq}) | F(\mu_{iq}))}\right) \\
&= \zeta \text{ logit } p = \text{logit } a(L(\mu_{iq})).
\end{align*}
This class of updating rules includes a prior-confirming bias ($\zeta > 1$), conservatism ($\kappa < 1$), and base-rate neglect ($\zeta < 1$). In all these cases, agents form the same posterior after $G(\mu_{iq})$ and $L(\mu_{iq})$ (even if $\kappa$ differs for good and bad news).

In addition, by taking $\Theta_q$ as the domain for the motive function, the same logic as in \Cref{moreas-subsection} leads motivated reasoners to be more likely to revise their median beliefs in the higher-motive direction. A Bayesian will have $\mathbb{P}\big(\mu'_{iq}(x_{iq}) > \mu_{iq}\big) = \mathbb{P}\big(\mu'_{iq}(x_{iq}) < \mu_{iq}\big)$. However, a motivated reasoner who has $m_{iq}(\theta_q)$ increasing in $\theta_q$ will be more likely to update upwards: $\mathbb{P}\big(\mu'_{iq}(x_{iq}) > \mu_{iq}\big) > \mathbb{P}\big(\mu'_{iq}(x_{iq}) < \mu_{iq}\big)$, and vice versa if $m_{iq}(\theta_q)$ is decreasing in $\theta_q$.

Online Appendix C discusses extensions to the model that can generate further predictions. Agents who start out believing that $\mu_{iq}$ is larger, due to receiving signals prior to this game, will be motivated to believe that $\theta_q$ is even larger than $\mu_{iq}$, as current beliefs are reflective of motivated beliefs. With additional functional form assumptions, motivated reasoning also leads to overprecision and overconfidence.

\section{Experimental Design}\label{design}

This section details the experiment introduced above. \Cref{experiment-summary} outlines the experimental design, \Cref{experiment-pages} discusses specific pages that subjects see, and \Cref{experiment-hypotheses} generates the main hypotheses. \Cref{data} discusses the participant sample and further details of the data collected. Specific question wordings and screenshots for every page type, including instructions and scoring rules, are in the Online Appendices.

\subsection{Summary, Timeline, and Topics}
\label{experiment-summary}

To fix ideas, consider the following question, taken verbatim from the experiment: 

\vspace{5mm}
\begin{addmargin}[1cm]{1cm}
\textit{Some people believe that the Obama administration was too soft on crime and that violent crime increased during his presidency, while others believe that President Obama's pushes towards criminal justice reform and reducing incarceration did not increase violent crime.}

\textit{This question asks how murder and manslaughter rates changed during the Obama administration. In 2008 (before Obama became president), the murder and manslaughter rate was 54 per million Americans.}

\textit{In 2016 (at the end of Obama's presidency), what was the per-million murder and manslaughter rate?}
\end{addmargin}
\vspace{5mm}

\noindent This is a question for which we will hypothesize that Republicans are motivated to believe the answer is higher, and Democrats are motivated to believe the answer is lower. The test of motivated reasoning involves the following steps:

\begin{enumerate}
\item \textbf{Beliefs:} Subjects are asked to guess the answers to politicized questions like the one above. They are asked and incentivized to guess their median belief (such that they find it equally likely for the answer to be above or below their guess). They are also asked and incentivized for their interquartile range. Details on incentives are below.

\item \textbf{News:} On each question subjects receive one binary message from one of two randomly-chosen news sources: True News and Fake News. The message from True News is always correct, and the message from Fake News is always incorrect. This is the main (within-subject) treatment variation.

\hspace{5mm} The message says either ``The answer is \textbf{greater than} your previous guess of [previous guess].'' or ``The answer is \textbf{less than} your previous guess of [previous guess].'' Note that the exact messages are different for subjects who make different initial guesses. 

\hspace{5mm} For the question above, ``greater than'' corresponds to Pro-Republican News and ``less than'' to Pro-Democratic News. For subjects who support the Republican Party more than the Democratic Party, ``greater than'' is coded as Pro-Party news and ``less than'' is coded as Anti-Party news, and vice versa for subjects who support the Democratic Party more.

\item \textbf{Assessment:} After receiving the message, subjects assess the probability that the message came from True News using a scale from 0/10 to 10/10, and are incentivized to state their true belief. I test for motivated reasoning by looking at the treatment effect of variation in news direction on news veracity assessments. 
\end{enumerate}

Recall that since subjects receive messages that compare the answer to their median, a Bayesian would not change her assessment based on the message. If she had a prior that P(True News) = 1/2 before seeing the message, she would form a posterior that P(True $|$ ``greater than'') = P(True $|$ ``less than'') = 1/2. We attribute systematic treatment effects of the messages on assessments to motivated reasoning. For instance, if Republicans tend to state P(True $|$ ``greater than'') $>$ P(True $|$ ``less than'') and Democrats tend to state P(True $|$ ``greater than'') $<$ P(True $|$ ``less than'') on the question above, this would be coded as politically-motivated reasoning. 

Subjects see 14 questions in the experiment, including the ten topics in \Cref{topic-motives}, three neutral questions, and one comprehension check.\footnote{Neutral questions ask about the latitude and longitude of the center of the U.S. and about a random number drawn from 0-100. The comprehension check asks what the current year is.} The timing is as follows:
\vspace{5mm}

\begin{tikzpicture}[node distance=2cm, auto,]
\node[box] (demographics) {Demographics};
\node[boxsmall, above right = 1.7mm and 9mm of demographics] (q1) {Question 1}
edge[arrow, <-] (demographics);
\node[boxsmall, below = 10mm of q1] (n1) {News 1}
edge[arrow,<-] (q1);
\node[simple, above right = 3.6mm and 1.3mm of n1] (ellipses) {...}
edge[arrow,<-] (n1);
\node[boxsmall, right = 8.8mm of q1](q14) {Question 14}
edge[arrow,<-] (ellipses);
\node[boxsmall, below = 10mm of q14] (n14) {News 14}
edge[arrow,<-] (q14);
\node[boxsmall, above right = 1.7mm and 9mm of n14] (results) {Results}
edge[arrow, <-] (n14);
\end{tikzpicture}

\vspace{5mm}
The Demographics page includes questions about party ratings, party affiliation, ideology, gender, age, race and ethnicity, annual income, highest education level, state or territory of residence, and religion. Party ratings are used to categorize subjects' party preferences; subjects are asked to rate the Democratic and Republican parties using a 0-100 scale that is akin to the feeling thermometer used in the American National Election Studies. 

The Results page tells subjects what their overall performance was, what their score on each question and assessment was, and the correct answer to each question and assessment. Subjects are told that they will see this page at the beginning of the experiment, and they are forced to go through it before exiting the study and receiving payment. Forcing subjects to see the answers aims to limit subjects' scope for strategic self-deception.\footnote{Subjects spend 71 seconds on the results page on average, suggesting that they are reading it.}

The order of Questions 1-12 is randomized between subjects, but Questions 13 and 14 are fixed.\footnote{Main questions are equally likely to be selected in each round, but the comprehension check is restricted to be between Question 2-11. This restriction ensures that subjects still pay attention after the first question, and to make sure that a willingness-to-pay round, which occurs for Question 12, does not overlap with the comprehension check.} Question 13 asks subjects how they scored on Questions 1-12 relative to 100 pilot subjects. Question 14 asks subjects to compare Democratic pilot subjects' performance to Republican pilot subjects' performance on Questions 1-12.\footnote{Half of subjects are given the Democrats' score and asked to predict the Republicans'; half are given the Republicans' score and asked to predict the Democrats'.}

\subsection{Pages and Scoring Rules}
\label{experiment-pages}

\noindent \textbf{Overall Scoring Rule}

At the end of the experiment, subjects earn a show-up fee of \$3 and either receive an additional \$10 bonus or no additional bonus. As described below, in each round of the experiment subjects earn between 0-100 ``points'' based on their performance. These points correspond to the probability that the subject wins the bonus: a score of $x$ points corresponds to an $x/10$ percent chance of winning the bonus.\footnote{This earnings system is similar to the most broadly incentive-compatible one from \textcite{ACH18} in which subjects are paid randomly for one round. I use my procedure instead in order to allow for a clearer measure of ``performance'' that is used as a question in the experiment. I do not need to assume risk neutrality in order for the experiment to be incentive compatible, but I do need to assume linearity in probabilities.}

\bigskip
\noindent \textbf{Questions Page}

On question pages, subjects are given the round number, the topic, the text of the question, and are asked to input three numbers about their initial beliefs:
\begin{itemize}
\item \textit{My Guess}: This elicits the median of subjects' prior distribution.
\item \textit{My Lower Bound}: This elicits the 25th percentile of subjects' prior distribution.
\item \textit{My Upper Bound}: This elicits the 75th percentile of subjects' prior distribution.
\end{itemize}

The scoring rule for guesses is piecewise linear. Subjects earn $\max \{100-|c-g|, 0\}$ points for a guess of $g$ when the correct answer is $c$. Subjects are told that they will maximize expected points by stating the median of their belief distribution.

The scoring rule for bounds is piecewise linear with different slopes. For upper bound $ub$, subjects earn $\max \{100-3(c-ub), 0\}$ points if $c \geq ub$ and $\max \{100-(ub-c), 0\}$ points if $c \leq ub$. For lower bound $lb$, subjects earn $\max \{100-(c-lb), 0\}$ points if $c \geq lb$ and $\max \{100-3(lb-c), 0\}$ points if $c \leq lb$. Subjects maximize expected points by setting $ub$ to be the $75^\text{th}$ percentile and $lb$ to be the $25^\text{th}$ percentile of their belief distribution. Subjects are restricted to give answers for which My Lower Bound $\leq$ My Guess $\leq$ My Upper Bound; if they do not, they see an error message. 

\bigskip
\noindent \textbf{News Assessments Page}

After submitting \textit{My Guess}, subjects see a second page about the same question. At the top of the page is the text of the original question. Below the question, there is a message relating the correct answer to the number they submitted for \textit{My Guess}. This message says either: 

\vspace{2mm}
``The answer is \textbf{greater than} your previous guess of \textbf{[My Guess]}.'' or

\vspace{2mm}
``The answer is \textbf{less than} your previous guess of \textbf{[My Guess]}.''

\vspace{2mm}
Subjects are told that True News \textit{always} tells the truth and Fake News \textit{never} tells the truth, and that sources are independent across questions. Below the message, subjects are asked: ``Do you think this information is from True News or Fake News?'' and choose one of eleven radio buttons that say ``x/10 chance it's True News, (10-x)/10 chance it's Fake News'' from each x=0, 1, \dots, 10.

The scoring rule for assessments is quadratic. For assessment $a$, subjects earn $100(1 - (1-a)^2)$ points if the source is True News and $100(1 - a^2)$ points if it is Fake News. Subjects maximize expected points by stating the closest multiple of 0.1 to their belief. They are given a table with the points earned as a function of each assessment and news type.

Occasionally, a subject will correctly guess the answer. If this happens, she skips the news assessment page and moves on to the next question.\footnote{This is not the case for the comprehension check question, where the message says ``The answer is \textbf{equal} / \textbf{not equal} to your previous guess of \textbf{[My Guess]}.''}

Half of subjects also see a ``Second Guess'' part of the News Assessment page. For these subjects, below each news assessment question they are asked an additional question: ``After seeing this message and assessing its truthfulness, what is your guess of the answer to the original question?'' Subjects are given the same linear scoring rule as on the initial guess. They earn $\max \{100-|c-g|, 0\}$ points for a guess of $g$ when the correct answer is $c$. These second guesses will be used as robustness, as well as to test for belief polarization about how subjects change their guesses. 

The other half of subjects see an additional page between Question 12 and News 12, on which they are given instructions and asked to submit a willingness-to-pay (WTP) for a message. Due to space constraints, this group is not discussed in the main text. For instructions and results, see the Online Appendix.\footnote{I find that subjects have a positive valuation of these messages, but that they do not differentially value messages on politicized and neutral topics; results suggest that subjects are aware that they treat signals as informative, but not that they expect to distort their updating process.}

I randomly vary whether subjects are given a prior at the beginning of the experiment that P(True News) = 0.5 or not given any prior. One-third of subjects are given a prior. This randomization serves as a robustness check to ensure that results are not entirely driven by either updating over meta-priors over the probability of True News (as may be the case for the group who is not given a prior) or by anchoring to a 50-50 benchmark (as may be the case for the group who is given a prior). I find no statistically significant differences between the two groups on any main outcomes, so results in the main text of this paper pool all subjects. Results are robust to restricting the sample to individual groups, as shown in the Online Appendix.

\subsection{Hypotheses}
\label{experiment-hypotheses}

The main hypothesis is that a news veracity assessment will be larger when it leads to a higher motive. This is a joint test that (1) people motivatedly reason, giving higher assessments to news in the direction of higher motives than to news in the direction of lower motives, and (2) the predicted direction of motives is as in \Cref{topic-motives}. News on neutral topics is assumed to not affect motives. For politicized topics, the level of partisanship affects the steepness of the motive function.
\begin{hypothesis}[Motivated reasoning with political motives]\label{hypothesis-PMR}
\begin{itemize}
\item Assessments of Pro-Party news are greater than assessments of Anti-Party news.
\item Assessments of Neutral topic news lie in between those of Anti-Party news and those of Pro-Party news.
\item The gap in assessments of Pro-Party news and Anti-Party news increases in partisanship.
\end{itemize}
\end{hypothesis}
The hypothesis for Pro-Performance news and Anti-Performance news is similar. 

The next hypothesis is based on an observation that subjects may have different beliefs because these beliefs reflect past instances of motivated reasoning. When motives are unobservable, an experimenter can learn about subjects' motives by looking at their initial beliefs. Two subjects who motivatedly reason in different directions will hold different median beliefs: A motivated reasoner with an increasing motive function will be more likely to hold a median belief that is too high, and a motivated reasoner with a decreasing motive function will be more likely to hold a median belief that is too low. Therefore, a subject whose median is too high is more likely to have an increasing motive function as compared to a subject whose median is too low. 

If the two subjects make news assessments using the procedure above, they will trust news that \textit{reinforces} the error in their initial beliefs more than news that \textit{mitigates} the error. This behavior occurs even though signals are designed so that their interpretation is distinct from their beliefs. In the experiment, this behavior implies that subjects will give higher assessments to error-reinforcing news compared to error-mitigating news. Recalling that error-reinforcing news is Fake News and error-mitigating news is True News, we have the following:
\begin{hypothesis}[Motivated reasoning and trust in Fake News]\label{hypothesis-Fake-News}
\begin{itemize}
\item Assessments of Fake News are greater than assessments of True News on politicized topics, but not on neutral topics.
\item Assessments of Fake News are greater than assessments of True News, controlling for whether the news is Pro-Party or Anti-Party.
\end{itemize}
\end{hypothesis}

We can use the second-guess group to construct an alternative test of motivated reasoning. By comparing subjects' first and second guesses, we can retest the main politically-motivated reasoning prediction and also study a form of belief polarization. 

First, as in \Cref{hypothesis-PMR}, subjects are expected to more frequently adjust their guesses in the direction that favors their preferred party. Second, by a similar logic to \Cref{hypothesis-Fake-News}, motivated reasoning would lead subjects to be more likely to adjust their guesses away from the population mean than towards it. 
We define \textit{Follow Message} as the ternary variable that takes value: 
\begin{itemize}
\item 1 if the message says $G$ and the second guess is greater than $\mu$, or if the message says $L$ and the second guess is less than $\mu$;
\item 0 if the second guess equals $\mu$; and
\item -1 if the message says $G$ and the second guess is less than $\mu$, or if the message says $L$ and the second guess is greater than $\mu$.
\end{itemize}
Polarizing says $G$ if $\mu$ is greater than the population mean guess or $L$ if $\mu$ is less than the population mean guess. Anti-Polarizing news says the opposite. 
\begin{hypothesis}[Motivated reasoning and second guesses]\label{hypothesis-second-guesses}
\begin{itemize}
    \item Follow Message is larger for Pro-Party than for Anti-Party news.
\item Follow Message is larger for Polarizing than for Anti-Polarizing news.
\end{itemize}
\end{hypothesis}

\subsection{Data and Experiment Details}\label{data}

The experiment was conducted in June 2018 on Amazon's Mechanical Turk (MTurk) platform. MTurk is an online labor marketplace in which participants choose ``Human Intelligence Tasks'' to complete. MTurk has become a popular way to run economic experiments (e.g. \cite{HRZ11}; \cite{KNSS15}), and \textcite{LFD16} find that participants tend to have more diverse demographics than students in university laboratories with respect to politics. The experiment was coded using oTree, an open-source software based on the Django web application framework developed by \textcite{CSW16}. 

The study was offered to MTurk workers currently living in the United States. 1,387 subjects were recruited and answered at least one question, and 1,300 subjects completed the study. Of these subjects, 987 (76 percent) passed simple attention and comprehension checks, and the rest are dropped from the analyses.\footnote{The Online Appendix shows that main results are robust to inclusion of these subjects. In order to pass these checks, subjects needed to perfectly answer the attention check question in \Cref{comprehension-question} by giving a correct answer, correct bounds, and answering the news assessment with certainty. In addition, some questions had maximum and minimum possible answers (e.g. percentages, between 0 and 100). Subjects were dropped if any of their answers did not lie within these bounds.}

When asked for their party ratings, 627 subjects (64 percent) give a higher rating to the Democratic Party; 270 (27 percent) give a higher rating to the Republican Party; and 90 (9 percent) give the same rating to each party.\footnote{\textcite{LFD16} also find that subjects on MTurk are mostly Democratic.} These subjects are labeled as ``Pro-Dem,'' ``Pro-Rep,'' and ``Neutral,'' respectively, and for most analyses I drop the Neutral subjects.

2/3 of subjects were randomly assigned to not receive a prior about the veracity of the news source, and 1/3 of subjects were told that True News and Fake News were equally likely. Independently, 1/2 of subjects were randomly assigned to receive a WTP question and 1/2 were asked for their second guesses.

Each subject answered 13 questions; there are a total of 11,661 guesses to questions for the 897 non-neutral subjects. There are 11,443 news assessments. The discrepancy between these numbers is due to 143 subjects in the WTP group who did not receive a message in one round, and due to 75 (0.7 percent) correct guesses.\footnote{The low frequency of correct guesses indicates that subjects were generally not looking up answers. It also suggests that the model's assumption of an atomless belief distribution is reasonable.} I drop these observations for news assessments, leaving 7,902 observations on politicized topics, 891 on the performance topic, and 2,650 on neutral topics. 

The balance table for the Pro-Party / Anti-Party treatment is in \Cref{balance-table}. The treatment is well-balanced, and overall shares of Pro-Party and Anti-Party news are nearly identical, suggesting that there was no differential attrition by treatment.

In order to validate the main results, I ran a replication experiment one year later. As discussed in the Online Appendix, the replication confirms the results.\footnote{The replication was identical in structure, but added several new political questions and did not include neutral questions, so it is not easy to pool results. All main hypotheses on political questions were preregistered, as were results on polarization and overprecision.}

\section{Main Results}\label{results-assessments}

\subsection{Raw Data}

This subsection shows that the raw data support the main hypotheses, and the following subsection shows the relevant regressions. To validate that these questions are politicized, Appendix \Cref{appendix-prior} compares initial guesses and finds systematic differences in median beliefs by party consistent with the directions predicted in \Cref{topic-motives}. 

\begin{figure}[htb!]
\caption{CDF of Assessments of Pro-Party and Anti-Party News}
\begin{center}
\vspace{-5mm}
\includegraphics[width = .85\textwidth]{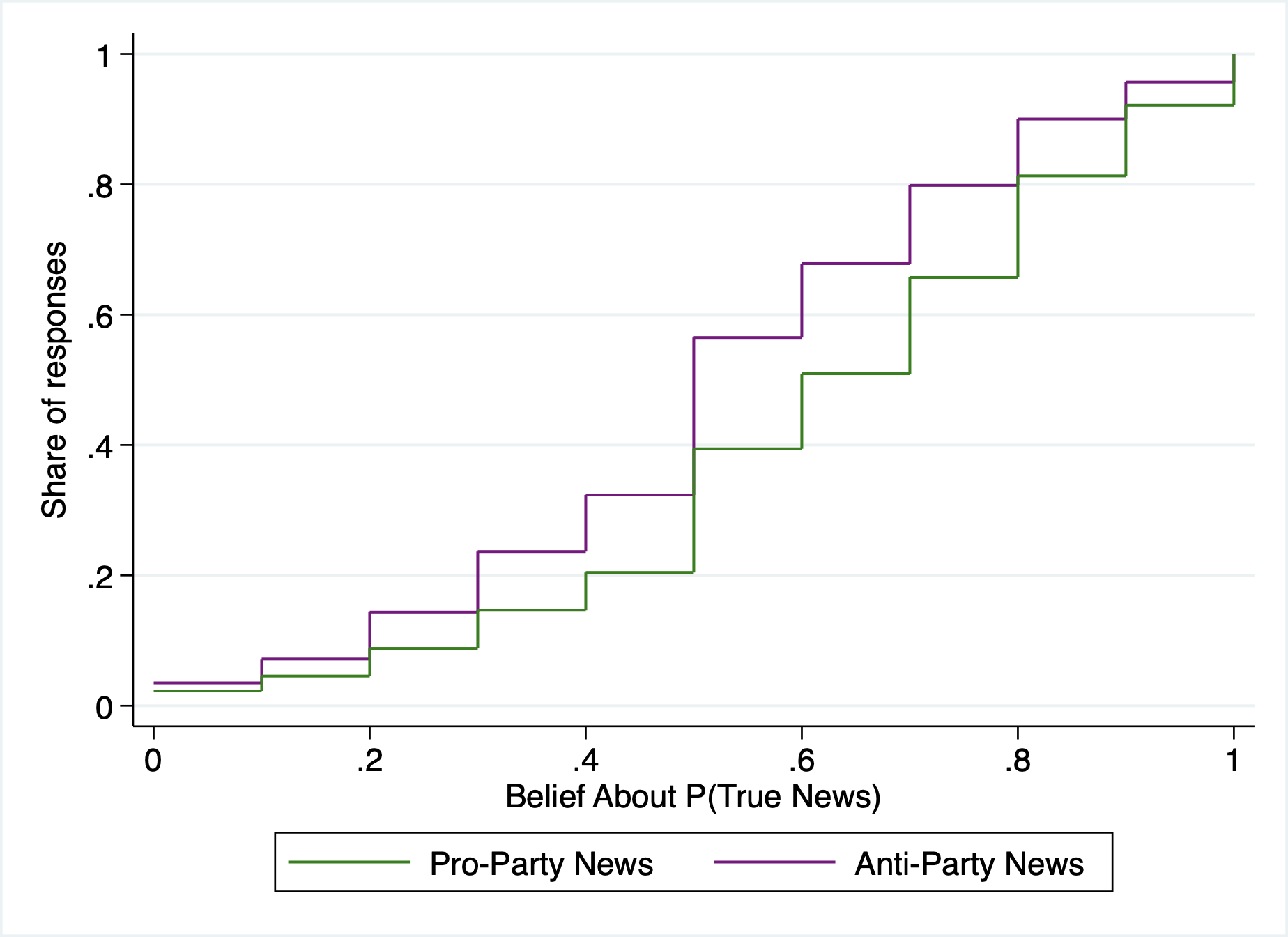}
\end{center}
\label{cdf-good-bad}
\begin{threeparttable}
\begin{tablenotes}
\begin{scriptsize}
\vspace{-10mm}
\item \textbf{Notes:} Pro-Party and Anti-Party news are defined in Table 1. This figure shows that subjects trust Pro-Party news more than Anti-Party news. The x-axis measures subjects' beliefs about P(True News | Pro-/Anti-Party News). The y-axis measures the share of respondents that give at most that high of an assessment. Bayesians would have the same trust in news for Pro-Party and Anti-Party news, and the residual is motivated reasoning. 
\end{scriptsize}
\end{tablenotes}
\end{threeparttable}
\end{figure}

In support of the first part of \Cref{hypothesis-PMR}, I show that subjects trust the Pro-Party news more than they trust the Anti-Party news. Subjects' average assessment that the likelihood of Pro-Party news is from the True News source is 9.1 percentage points higher than their average assessment that Anti-Party news is from the True News source (s.e. 0.6 pp; $p < 10^{-40}$). \Cref{cdf-good-bad} shows the CDF of assessments for Pro-Party and Anti-Party news; the Pro-Party distribution first-order stochastically dominates the Anti-Party distribution. Appendix \Cref{cdf-good-bad-performance} shows the same result for Pro-Performance and Anti-Performance news assessments.

\begin{figure}[htb!]
\caption{Politically-Motivated Reasoning: News Direction and Partisanship}
\begin{center}
\vspace{-5mm}
\includegraphics[width = .85\textwidth]{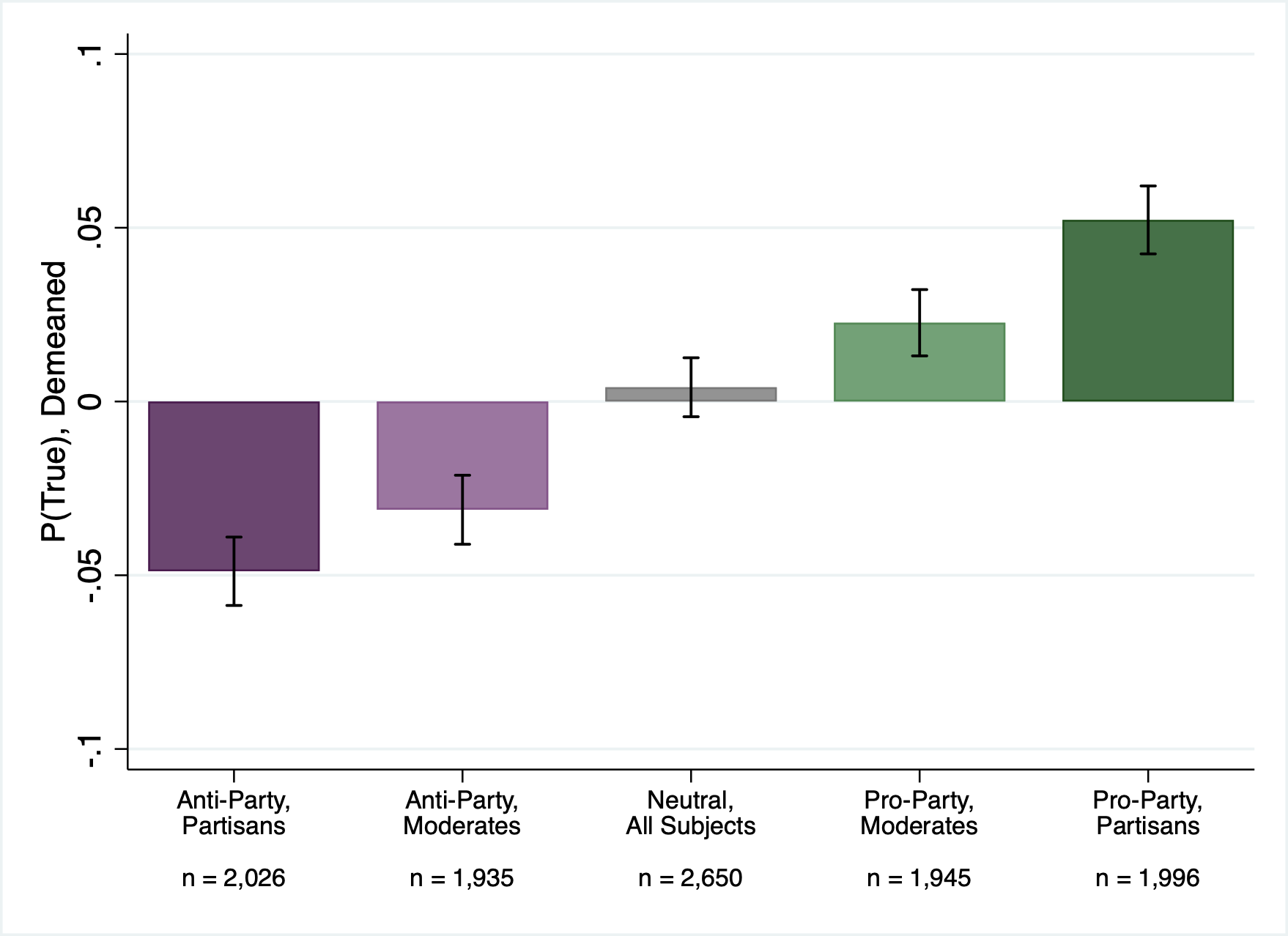}
\end{center}
\label{partisan-extremity}
\begin{threeparttable}
\begin{tablenotes}
\begin{scriptsize}
\vspace{-10mm}
\item \textbf{Notes:} The y-axis is motivated reasoning: stated P(True), demeaned at the subject level. News on partisan topics is classified as Pro-Party (Anti-Party) if it is more (less) representative of the subject's preferred political party, as defined in Table 1. A subject who is above the median value for abs(Republican Party rating - Democratic Party rating) is classified as Partisan; a subject who is not is classified as Moderate. Error bars correspond to 95 percent confidence intervals. 
\end{scriptsize}
\end{tablenotes}
\end{threeparttable}
\end{figure}

\Cref{partisan-extremity} shows the subject-demeaned assessments by news direction (Pro-Party; Anti-Party; news on neutral topics) and subject type (Partisan and Moderate, as defined by the absolute difference in party ratings). In support of all three parts of \Cref{hypothesis-PMR}, subjects believe that Pro-Party news is more likely than news on neutral topics to be true, believe that Anti-Party news is less likely than news on neutral topics to be true, and differences are larger for partisans. If neutral topics account for unmotivated mistakes in news assessments, then this figure shows that there is motivated reasoning in both positive and negative directions.

In support of \Cref{hypothesis-Fake-News}, subjects trust the Fake News more than they trust the True News. Subjects' average assessment that the likelihood of True News is actually from the True News source is 6.0 percentage points lower than their average assessment that Fake News is from the True News source (s.e. 0.6 pp; $p < 10^{-20}$). Appendix \Cref{cdf-true-fake} shows the CDF of assessments for True News and Fake News on the political questions; the Fake News distribution first-order stochastically dominates the True News distribution. 

\begin{figure}[htb!]
\caption{Motivated Reasoning and Assessments of Fake News}
\begin{center}
\vspace{-5mm}
\includegraphics[width = .85\textwidth]{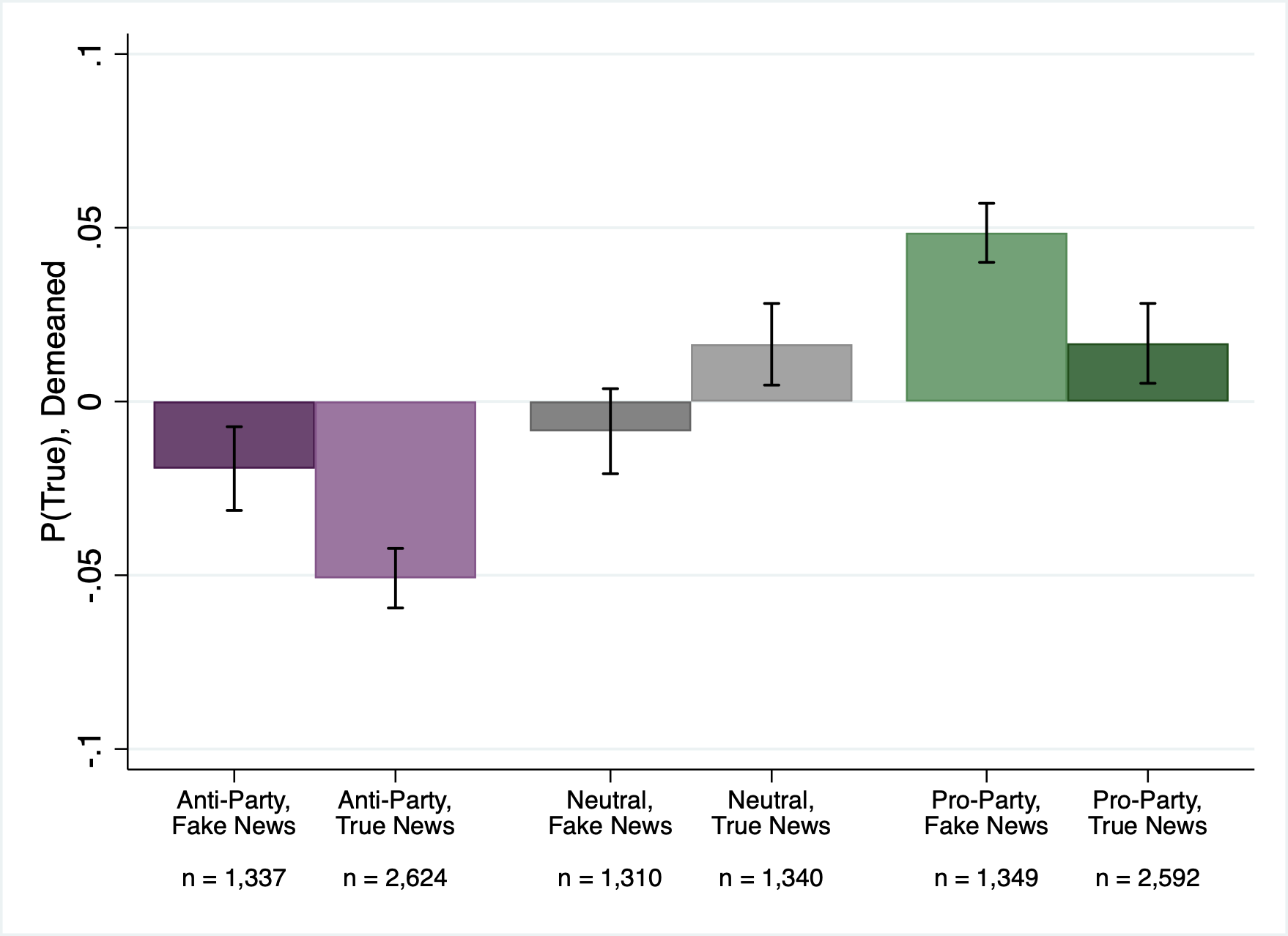}
\end{center}
\label{good-true-hist}
\begin{threeparttable}
\begin{tablenotes}
\begin{scriptsize}
\vspace{-10mm}
\item \textbf{Notes:} The y-axis is motivated reasoning: stated P(True), demeaned at the subject level. News on partisan topics is classified as Pro-Party (Anti-Party) if it is more (less) representative of the subject's preferred political party, as defined in Table 1. Fake News sends messages that reinforce the direction of subjects' error; True News sends messages that mitigate subjects' error. Error bars correspond to 95 percent confidence intervals. 
\end{scriptsize}
\end{tablenotes}
\end{threeparttable}
\end{figure}

This effect is not solely driven by whether the news is Pro-Party versus Anti-Party. \Cref{good-true-hist} shows the subject-demeaned assessments by news direction (Pro-Party; Anti-Party; Neutral) and news veracity (True News; Fake News). Subjects give higher assessments to Fake News than to True News on politicized topics, but they do not do so on neutral topics.\footnote{If anything, assessments are higher for True News than Fake News on neutral topics. Reflecting on the question may lead subjects to adjust \textit{towards} the truth in the absence of motivated beliefs.}

\subsection{Regression Specifications for News Assessments}
\label{main-specs}

The primary regression specification is within subject.\footnote{99.4 percent of non-neutral subjects receive at least one piece of Pro-Party and Anti-Party news. Three subjects randomly never receive Pro-Party news and two never receive Anti-Party news.} It regresses assessments $a$ on Pro-Party news for subject $i$, question topic $q$, and round $r$ with fixed effects for $i$, $q$, and $r$, restricting to news that is Pro-Party or Anti-Party:\footnote{The Online Appendix shows that results are qualitatively the same if we replace $a_{iqr}$ with logit($a_{iqr}$).}
\begin{equation*}
a_{iqr} = \alpha + \beta \cdot 1(\text{Pro-Party})_{iqr} + \gamma FE_i + \delta FE_q + \zeta FE_r + \epsilon_{iqr}
\end{equation*}
This specification is used in \Cref{regression-assessment}, column 2. A similar alternative specification, replacing individual-level fixed effects with demographic controls, is in column 1. 

\begin{center}
\begin{threeparttable}[htbp!]
\begin{footnotesize}
\caption{Motivated Reasoning and Perceived Truthfulness of News}
\begin{tabular}{l*{6}{c}}
\hline\hline
                    &\multicolumn{1}{c}{(1)}&\multicolumn{1}{c}{(2)}&\multicolumn{1}{c}{(3)}&\multicolumn{1}{c}{(4)}&\multicolumn{1}{c}{(5)}&\multicolumn{1}{c}{(6)}\\
\hline
Pro-Party News      &    0.092&    0.088&    0.041&    0.037&         &    0.077\\
                    &  (0.006)&  (0.006)&  (0.012)&  (0.006)&         &  (0.006)\\
Partisanship x Pro-Party \hspace{4mm}     &         &         &    0.099&         &         &         \\
           &         &         &  (0.022)&         &         &         \\
Anti-Party News     &         &         &         &   -0.048&         &         \\
                    &         &         &         &  (0.007)&         &         \\
True News           &         &         &         &         &   -0.059&   -0.034\\
                    &         &         &         &         &  (0.006)&  (0.006)\\
Question FE         &      Yes&      Yes&      Yes&       No&      Yes&      Yes\\
Round FE            &      Yes&      Yes&      Yes&      Yes&      Yes&      Yes\\
Subject FE          &       No&      Yes&      Yes&      Yes&      Yes&      Yes\\
Subject controls    &      Yes&       No&       No&       No&       No&       No\\
Neutral News        &       No&       No&       No&      Yes&       No&       No\\
\hline
Observations        &     7902&     7902&     7902&    10552&     7902&     7902\\
\(R^{2}\)           &     0.05&     0.25&     0.25&     0.21&     0.23&     0.25\\
Mean                &    0.574&    0.574&    0.574&    0.575&    0.574&    0.574\\
\hline\hline
\multicolumn{7}{l}{\footnotesize Standard errors in parentheses}\\
\end{tabular}

\label{regression-assessment}
\end{footnotesize}
\begin{tablenotes}
\begin{scriptsize}
\item \textbf{Notes:} OLS, errors clustered at subject level. Neutral News indicates that Pro-Party / Anti-Party news assessments are compared to assessments on Neutral topics. These classifications are in Table 1. Controls: race, gender, log(income), years of education, religion, and whether state voted for Trump or Clinton in 2016. Partisanship is abs(Republican Party rating - Democratic Party rating).
\end{scriptsize}
\vspace{5mm}
\end{tablenotes}
\end{threeparttable}
\end{center}

\Cref{hypothesis-PMR} claims that the Pro-Party / Anti-Party gap is increasing in partisanship, so column 3 interacts partisanship (the absolute difference in party ratings) with Pro-Party news. It also claims that motivated reasoning leads to both higher assessments for Pro-Party news and lower assessments for Anti-Party news; as such, column 4 includes indicators for both Pro-Party (vs. Neutral) news and Anti-Party (vs. Neutral) news. 
\Cref{hypothesis-Fake-News} posits that subjects will trust Fake News more than True News on politicized topics, so columns 5 and 6 regress assessments on a dummy for True News, controlling for and not controlling for Pro-Party news.

Hypotheses 1 and 2 are strongly supported. Assessments for Pro-Party news are higher than for Anti-Party news, and this effect increases in partisanship. There is evidence for motivated reasoning both towards Pro-Party and away from Anti-Party news, and Fake News assessments are higher than True News assessments.

\begin{figure}[htb!]
\caption{Motivated Reasoning by Topic}
\begin{center}
\vspace{-5mm}
\includegraphics[width = .85\textwidth]{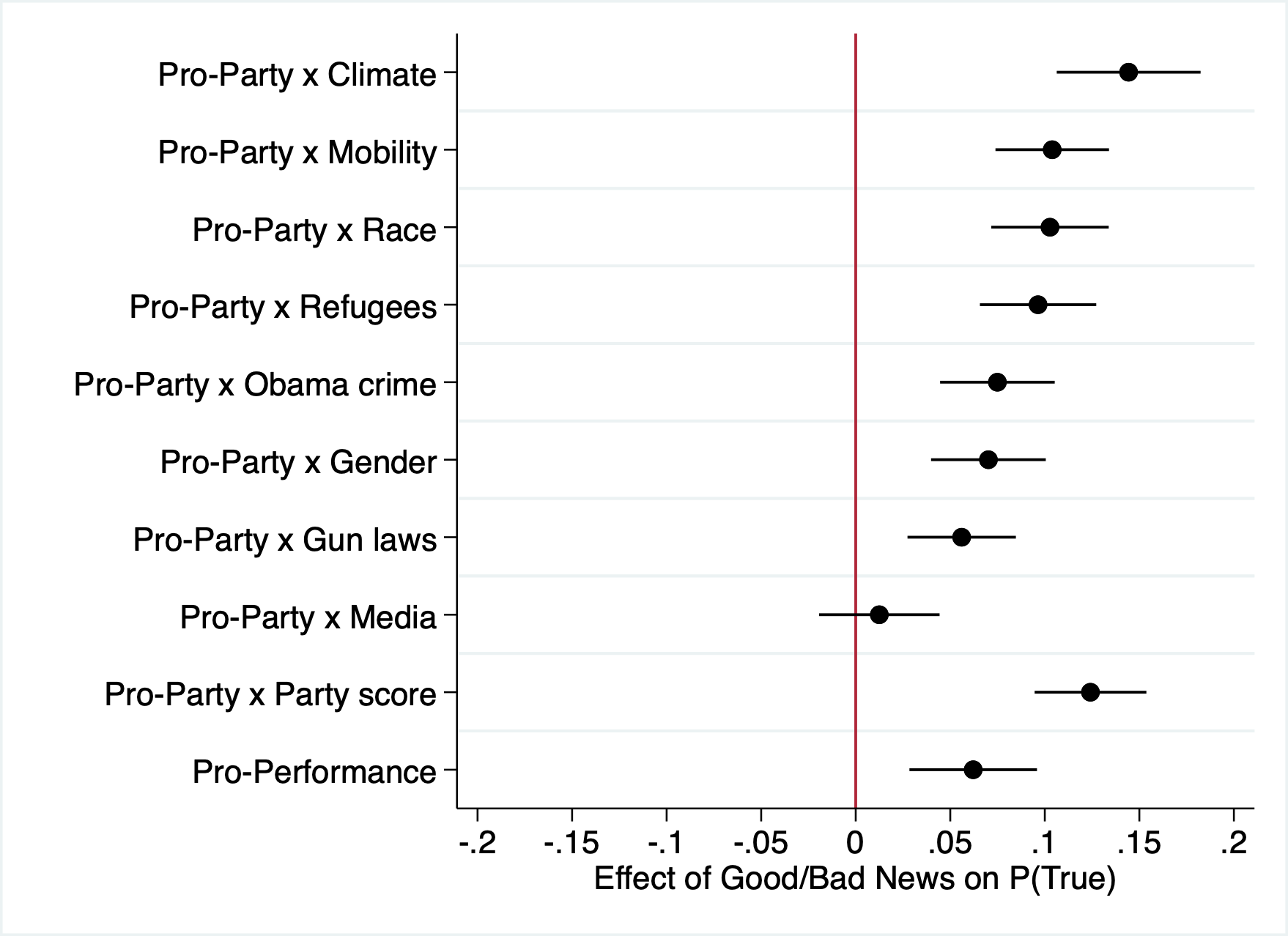}
\end{center}
\label{moreas-by-topic}
\begin{threeparttable}
\begin{tablenotes}
\begin{scriptsize}
\vspace{-10mm}
\item \textbf{Notes:} OLS regression coefficients, errors clustered at subject level. FE included for subject, round number, and topic. Pro-Party (vs. Anti-Party) news is defined in Table 1. Error bars correspond to 95 percent confidence intervals. 
\end{scriptsize}
\end{tablenotes}
\end{threeparttable}
\end{figure}

Next, we look at each topic individually by regressing on the interaction of treatment and topic dummies. \Cref{moreas-by-topic} shows evidence for politically-motivated reasoning on eight of the nine hypothesized topics ($p<0.001$ on each). It also shows that people motivatedly reason towards believing they outperformed others ($p<0.001$).

\subsection{Changing Guesses and Belief Polarization}
\label{changing-guesses}

To further validate the ``trust in news'' measure of motivated reasoning, I also show that subjects engage in asymmetric updating about the questions themselves. Half of subjects give a second guess to the initial question after receiving news, and I hypothesize that subjects are more likely to update in the Pro-Party direction than in the Anti-Party direction. This test is both useful as a robustness check as well as useful to better understand belief polarization on these questions. 

Column 1 of \Cref{change-guess-message} shows that subjects are more likely to update their median in the direction of Pro-Party messages than they are from Anti-Party messages. Column 2 shows that on politicized topics, subjects are also more likely to change their guesses in the direction of a Polarizing message (one that tells them their guess is farther away from the mean) than from an Anti-Polarizing message.

\begin{center}
\begin{threeparttable}[htb!]
\begin{footnotesize}
\caption{Motivated Reasoning and Following the Message Sent}
{
\def\sym#1{\ifmmode^{#1}\else\(^{#1}\)\fi}
\begin{tabular}{l*{6}{c}}
\hline\hline
                    &\multicolumn{1}{c}{(1)}         &\multicolumn{1}{c}{(2)}         &\multicolumn{1}{c}{(3)}         &\multicolumn{1}{c}{(4)}         &\multicolumn{1}{c}{(5)}         &\multicolumn{1}{c}{(6)}         \\
\hline
Pro-Party News      &    0.122 &                  &    0.114 &    0.018         &                  &    0.024         \\
                    &  (0.021)         &                  &  (0.021)         &  (0.018)         &                  &  (0.018)         \\
Polarizing News \hspace{9mm}    &                  &    0.061 &    0.032 &                  &   -0.017         &   -0.022         \\
                    &                  &  (0.019)         &  (0.019)         &                  &  (0.016)         &  (0.016)         \\
P(True)             &                  &                  &                  &    1.126 &    1.139 &    1.131 \\
                    &                  &                  &                  &  (0.062)         &  (0.061)         &  (0.063)         \\
Question FE         &      Yes         &      Yes         &      Yes         &      Yes         &      Yes         &      Yes         \\
Round FE            &      Yes         &      Yes         &      Yes         &      Yes         &      Yes         &      Yes         \\
Subject FE          &      Yes         &      Yes         &      Yes         &      Yes         &      Yes         &      Yes         \\
\hline
Observations        &     4085         &     4085         &     4085         &     4085         &     4085         &     4085         \\
\(R^{2}\)           &     0.28         &     0.28         &     0.28         &     0.45         &     0.45         &     0.45         \\
Mean                &    0.659         &    0.659         &    0.659         &    0.659         &    0.659         &    0.659         \\
\hline\hline
\multicolumn{7}{l}{\footnotesize Standard errors in parentheses}\\
\end{tabular}
}

\label{change-guess-message}
\end{footnotesize}
\begin{tablenotes}
\begin{scriptsize}
\item \textbf{Notes:} OLS, errors clustered at subject level. Only subjects from the Second-Guess group. Only Pro-Party / Anti-Party news observations, as defined in Table 1. Polarizing News is news that tells subjects that, compared to their initial guess, the answer is in the opposite direction from the population mean. Dependent variable is 1 if subjects change their guess upwards when the message says ``Greater Than'' or downwards when the message says ``Less Than,'' -1 if they change their guess in the opposite direction, and 0 if they do not change their guess.
\vspace{5mm}
\end{scriptsize}
\end{tablenotes}
\end{threeparttable}
\end{center}

Columns 4-6 of \Cref{change-guess-message} show that discrepancies in both motivated posteriors and belief polarization can be explained by differences in news assessments. After controlling for assessments, guess changes are not statistically significantly affected by Pro-Party / Anti-Party messages, nor are they statistically significantly affected by polarizing messages, and the point estimates are close to zero. Evidence suggests that subjects change their beliefs to follow Pro-Party news more exactly because they trust that news source more.

More broadly, these results give a stark prediction about how people change their beliefs. They show that, in environments where signals remind people of their motivated beliefs, informational content is not a necessary condition for belief polarization.

\subsection{Alternative Explanations and Robustness Checks}
\label{robustness-checks}

There are features in this design that may lead subjects to behave in ways that are consistent with motivated reasoning but are also consistent with unmotivated confounding hypotheses. This subsection discusses a number of potential confounds, and shows how it is possible to test these alternative hypotheses. I argue that it is unlikely that these confounds can fully explain the results above, strengthening the interpretation that motivated reasoning is what is being identified in the experiment. 

\subsubsection{Misunderstanding medians and skewed belief distributions}

It is reasonable to expect that subjects do not fully understand the concept of a median. For instance, they may answer with their mean belief instead. This would not directionally impact the news assessment results in a systematic direction, unless the prior distribution were notably skewed. We can use where the initial guess $\mu_q$ lies in subjects' confidence intervals as a proxy for skewness, and see that the main results hold for subjects whose distributions are unskewed. 

On the politicized questions, 32 percent of subjects' guesses are exactly halfway between their upper and lower bounds. \Cref{unskewed-priors} uses the same specification as the main regression but interacts Pro-Party news, Anti-Party news, and True News with a dummy for having ``unskewed'' priors. The treatment effects are qualitatively and quantitatively similar, indicating that skewness does not directionally affect results.

\subsubsection{The independence of news sources}

We have so far assumed that subjects treat news sources as being drawn from independent distributions. While subjects are explicitly told to do this in the instructions, it is useful to show that they are not using previous pieces of news to update about current pieces of news.

In \Cref{news-independence}, I modify the main regression table to account for the relative number of Pro- and Anti-Party news in previous rounds. The effect of previous rounds' Pro- and Anti-Party news have no effect on the current round's assessment, and the main treatment effects are unchanged, suggesting that subjects indeed treat news sources as independent.

\subsubsection{Misunderstanding ``Fake News''}

Using the terminology ``Fake News'' can lead to greater subject engagement, but may also evoke other commonly-used definitions (such as fake articles, as in \cite{AG17}). This subsection shows that results cannot be explained by two plausible misinterpretations of Fake News.

First, suppose that subjects believe that messages from Fake News are equally likely to send true and false messages, instead of \textit{always} sending false messages. In this experiment, no predictions about assessments would change. A Bayesian would still have an ex-ante prior that Pro-Party and Anti-Party messages are equally likely, and would not infer anything about P(True) given either message. A motivated reasoner who is motivated to believe that the answer is large would still infer that P(True $|$ Pro-Party) $>$ P(True $|$ Anti-Party). The same argument can be made if subjects believe Fake News randomly sends false messages with any interior probability.

A more complicated situation involves subjects who believe that messages from Fake News are actually from a news source that is biased against their party. That is, suppose that subjects believe that Fake News is more likely to report Anti-Party news given a Pro-Party truth than Pro-Party news given an Anti-Party truth.

To test this, we can again look at how subjects change their guesses. In particular, suppose that subjects were Bayesian but used this asymmetrically wrong definition of Fake News. Then, they would find Pro-Party ``Fake News'' messages to be more informative than Anti-Party ``Fake News'' messages, since ``Fake News'' is expected to usually send the Anti-Party message. (The quotes here indicate that these subjects are using the wrong definition.) Such subjects would then update more from Pro-Party than Anti-Party news, \textit{conditional on their assessment of P(True News)}. In \Cref{change-guess-message}, we see that subjects are similarly likely to update from Pro-Party and Anti-Party news after controlling for their assessments. While the data are too imprecise to rule out the existence of subjects who treat Fake News as biased, this story is insufficient for explaining the main effects.

\subsubsection{Incorrect initial guesses}
\label{robustness-priors}

While it can sometimes be in subjects' best interests to strategically misreport their median in order to earn more points on news assessment questions, I find no evidence that subjects do this.\footnote{For instance, if a subject believes that $\theta \sim U[0,20]$, then she will expect to earn more points on the news assessment question by guessing 20 than by guessing 10 since she is confident the correct answer is not greater than 20. She would expect to earn 90 points on the question and 100 points on news, as opposed to earning 95 points on the question and 75 points on news.} In Round 1 of the experiment, subjects do not yet know that they will be seeing a news assessment page. If subjects were strategically mis-guessing to earn more assessment points, they would perform worse in Round 1 than in subsequent rounds on assessments and better in Round 1 than in subsequent rounds on guesses. 

There are no statistically significant differences in assessment scores in Round 1. Subjects score 67.2 points (s.e. 0.9) in Round 1 and 66.4 points (s.e. 0.3) in Rounds 2-12; the difference is 0.8 points (s.e. 1.0; $p=0.383$).\footnote{I exclude scoring on Rounds 13-14 since the questions are not randomly assigned in those rounds; the result is similar if they are included. I also exclude scoring on comprehension check questions.} There are also no statistically significant differences in guess scores in Round 1. Subjects score 76.2 points (s.e. 1.0) in Round 1 and 75.9 points (s.e. 0.2) in Rounds 2-12; the difference is 0.3 points (s.e. 1.0; $p=0.758$).

Non-strategic forms of incorrect initial guesses are more complicated to rule out. If there is symmetric noise in guesses, such that the probability that a subject is equally likely to state her $Q$ quantile and her $1-Q$ quantile for $Q \neq 1/2$, then the main results do not change. Results are also not consistent with subjects biasing their initial guesses towards the population mean. While such behavior can explain why subjects trust error-reinforcing news more than error-mitigating news on politicized, and why they trust Pro-Party news more than Anti-Party news, it incorrectly predicts the same pattern on neutral topics. 

The one form of misreporting that can be consistent with both Bayesian updating and results from the experiment involves subjects systematically misreporting initial guesses in a way that is biased in the \textit{opposite} direction from their party. One potential reason for such a bias is that subjects do not sufficiently think about the question; and, given more time, they update towards their actual (more Pro-Party) belief. It is possible that seeing the second screen causes subjects to think harder about the original question, and thinking harder leads to more Pro-Party beliefs.\footnote{The psychology behind this explanation overlaps with this theory of motivated reasoning, as the second page evokes the motive, and further work could better elucidate what qualifies as a signal.} However, thinking harder does not seem to be asymmetric, as I find no evidence that subjects prefer to spend more time thinking about good news.\footnote{The mean time spent on the assessment page with Pro-Party news is 14.6 seconds (s.e. 0.3 seconds), and the mean time spent on the assessment page with Anti-Party news is 14.8 seconds (s.e. 0.3 seconds).}

\subsubsection{Expressive preferences}
\label{expressive-robustness}

\textcite{BCFGHY-WP} provides recent evidence showing that people in experiments may forgo payment in order to make political statements. In this experiment, if subjects have a preference for stating Pro-Party signals, then both their initial guesses and their news assessments will be biased in the Pro-Party direction, consistent with the data. However, if they are Bayesian, how they \textit{change} their guesses will not be directional, since they have already stated their preferred belief. 

In \Cref{change-guess-message}, subjects are more likely to update their guesses in the Pro-Party direction than in the Anti-Party direction, even though they are equally likely to receive Pro-Party and Anti-Party news. This is consistent with subjects sincerely trusting the Pro-Party news more; it is not consistent with expressive Bayesian updating.

\subsubsection{Other robustness tests}

As discussed in \Cref{data}, subjects either were told that P(True News) = 1/2 in the instructions or were not told this. In the Online Appendix, I restrict the regressions from \Cref{regression-assessment} to subjects in each randomization arm, and find similar results in the two arms. Likewise, seeing the WTP page or having a second guess has little effect.

It is possible that subjects learn over the course of the experiment that they motivatedly reason and debias themselves. I interact the main effect with dummies for each round number, and do not find evidence for this. In each round, subjects give larger assessments to Pro-Party than to Anti-Party news.

\subsection{Heterogeneity in Motivated Reasoning}
\label{heterogeneity-demographics}

There are two types of heterogeneity to consider: heterogeneity in the direction of motivated reasoning, and heterogeneity in its magnitude.

First, we consider the direction of heterogeneity. To do this, \Cref{heterogeneity-direction} shows the coefficients from the regression of news assessments on the interaction of the political direction of the news (Pro-Rep vs. Pro-Dem) and party preferences, as well as on binarized observable demographics. Non-political demographics are race, gender, income, age, education, whether the subject's state voted for Trump or Clinton in 2016, and religious affiliation. 

\begin{figure}[htb!]
\caption{Heterogeneity in the Partisan Direction of Motivated Reasoning}
\begin{center}
\vspace{-5mm}
\includegraphics[width = .85\textwidth]{fne_heterogeneity_direction.png}
\end{center}
\label{heterogeneity-direction}
\begin{threeparttable}
\begin{tablenotes}
\begin{scriptsize}
\vspace{-10mm}
\item \textbf{Notes:} This figure plots the relative treatment effect of seeing Pro-Rep versus Pro-Dem news on subjects' news assessments by binary demographics. These are OLS regression coefficients, errors clustered at subject level. FE included for subject, round number, and topic. Only Pro-Party / Anti-Party news observations, as defined in Table 1. Pro-Rep: higher rating for Republican than Democratic Party. Older: above the median age in the experiment. High income: above median income in the experiment. Red State: state voted for Trump in 2016. Religious: subject affiliates with any religion. 
\end{scriptsize}
\end{tablenotes}
\end{threeparttable}
\end{figure}

After controlling for party preference, none of the other demographics have a statistically significant effect on the direction of motivated reasoning. All coefficients are between plus and minus 0.03, a magnitude less than one-third than for politics.

In the Online Appendix, we consider the magnitude of motivated reasoning, acknowledging that this design is unable to disentangle magnitude of bias and strength of motive. Partisans have a larger treatment effect than moderates, but demographics besides partisanship do not notably affect the magnitude of the bias; all effects are between +/- 0.03.\footnote{There is suggestive evidence that Pro-Dem subjects motivatedly reason than Pro-Rep subjects, though this may simply be due to the non-representativeness, conditional on party, of the MTurk sample. For instance, only 76 percent of Republicans in this sample approved of President Trump's performance; in a Gallup poll conducted contemporaneously (from June 25-July 1), 87 percent of Republicans approved of his performance (\cite{gallup18b}).} 

These results suggest that the magnitude of bias of motivated reasoning is broadly similar across demographics and that while the direction of motivated beliefs is heterogeneous by party, it is not very different across non-political demographics.

\subsection{Motivated Reasoning, Initial Beliefs, and Overprecision}
\label{moreas-overprecision}

We now consider two other relationships between initial beliefs and motivated reasoning: how much motivated reasoning can explain people's differences in beliefs, and how the bias relates to overprecise confidence intervals.

First, the data show that variation in this experiment's measure of motivated beliefs can explain a sizable fraction of variation in actual beliefs about these questions. I look at the relationship between motives and beliefs by correlating answers to politicized questions with differences in assessments between Pro-Rep and Pro-Dem news. For each politicized question, subjects' initial guesses are winsorized (at the 5-percent level), normalized, and signed; positive numbers correspond to more Pro-Rep. Next, for each subject, these normalized guesses are averaged (and re-normalized) to give a measure of how Pro-Rep her beliefs are. I correlate this value with the normalized average difference between Pro-Rep news assessments and Pro-Dem news assessments.

Using $R^2$, variation in news assessments explains 13 percent of the variation in beliefs. By comparison, all of the non-political demographics collected in this experiment --- age, gender, race, education, logged income, whether one is religious, and whether one is from a state that voted for Trump or Clinton in 2016 --- explain 7 percent of the variance in beliefs.\footnote{These are unadjusted values. Adjusted $R^2$ is 13 percent for assessments and 6 percent for demographics.} This suggests that my measure of motivated reasoning performs at least as well as oft-discussed demographics at predicting beliefs. 

Second, motivated reasoning can help explain a particular form of overprecision due to miscalibrated confidence intervals. In particular, we posit that motivated reasoners formed directionally-biased belief distributions, thereby leading them to overestimate the probability that the answers are within their confidence interval. Such a story would imply that overprecision is stronger for politicized than for neutral topics, and stronger for partisans than for moderates:
\begin{hypothesis}[Overprecision and partisanship]\label{hypothesis-overprecision}
\begin{itemize}
\item On politicized and performance questions, subjects' 50-percent confidence intervals contain the correct answer less than 50 percent of the time.
\item On politicized questions, the likelihood that subjects' confidence intervals contain the correct answer decreases in their partisanship.
\end{itemize}
\end{hypothesis}

In support of \Cref{hypothesis-overprecision}, subjects are overprecise in their beliefs about questions that evoke motivated beliefs. On politicized topics, subjects' confidence intervals contain the correct answer 46.6 percent of the time (s.e. 0.6 percent); this is statistically significantly less than 50 percent ($p<0.001$). Overprecision on these topics is primarily driven by partisans, whose intervals contain the correct answer 44.2 percent of the time (s.e. 0.9 percent). Moderates' intervals contain the correct answer 48.8 percent of the time (s.e. 0.8 percent). Partisans' overprecision is statistically significantly larger than moderates' ($p<0.001$). On the performance question, subjects' confidence intervals contain the correct answer 42.0 percent of the time (s.e. 1.6 percent), which is statistically significantly less than 50 percent ($p<0.001$). 

Overprecision is stronger for motivated questions than neutral questions, suggesting that these results are not simply driven by a bias towards overly narrow confidence intervals. On the ``Random Number'' question, which asks subjects to guess what a random number drawn uniformly from 0 to 100 will be, confidence intervals contain the correct answer 54.6 percent of the time, which indicates mild \textit{underprecision}.\footnote{On the other neutral questions, subjects also exhibit moderate underprecision.}

Finally, overprecision is correlated with over-trusting error-reinforcing Fake News, consistent with the notion that current beliefs are reflective of motivated beliefs. Subjects who are overprecise on a question assess Fake News to be 3.3 pp more likely to be truthful (s.e. 0.8 pp; $p<0.001$) compared to underprecise subjects, and subjects who are overprecise on a question assess True News to be 2.4 pp less likely to be truthful (s.e. 0.7 pp; $p=0.001$) compared to underprecise subjects.

\subsection{Discussion}

The experimental results strongly favor the hypothesis of motivated reasoning with politically-motivated beliefs over the null hypothesis of Bayesian updating. Motivated reasoning also fits the data better than unmotivated patterns of over- or under-inference. Subjects significantly over-trust Pro-Party news and Fake News in an environment with uninformative signals, real monetary stakes, and little room for self-deception.

These results show that political motives are necessary to consider in addition to prior beliefs when predicting how people respond to new information. In fact, motivated reasoning can partly explain one why people update further in the direction of their prior than a Bayesian would. When priors reflect motivated beliefs, a method that detects prior-confirming biases in belief updating may be detecting motivated reasoning instead (e.g. \cite{ER11}).

The results in \Cref{changing-guesses} relate to the effect of motivated reasoning on political polarization. Not only do subjects polarize in beliefs about the veracity of news, they polarize in their beliefs about the questions themselves, despite receiving uninformative signals. \textcite{GS11} find only modest differences in the media that liberals and conservatives consume, and motivated reasoning can help explain why people polarize even if they consume similar media outlets. 
Results also suggest that the beliefs people are motivated by are not necessarily optimistic ones, but are rather ones that support their political identity.\footnote{In fact, \textcite{T-WPa} shows that absent politics, optimism about the world plays a negligible role in motivated reasoning.} Furthermore, these results indicate that beliefs people find attractive are \textit{even further apart} than their current beliefs. One reason that people do not already hold their most-preferred beliefs is that motivated reasoners are still affected by information; the amount of distortion in updating is constrained by actual informational content. Motivated reasoners who receive precise signals would in fact become \textit{less} polarized.

\section{Conclusion}\label{conclusion}

This paper has shown that distortion of new information in favorable directions --- motivated reasoning --- plays a substantial role in people's assessment of the veracity of news and helps explain why people form inaccurate and polarized beliefs about the world around them. It developed a novel experimental paradigm that is able to identify the channel of motivated reasoning from Bayesian updating and other channels across a variety of settings. Results demonstrated the role that \textit{politically}-motivated reasoning plays in how people form beliefs about applied economic issues like income mobility, crime, and immigration. They also showed how this bias can lead to further belief polarization, overprecision, and an excess trust in Fake News. 

There are several avenues for future work with this design. For instance, the design can be used by applied researchers who are interested in detecting motivated reasoning. Researchers, working in many different contexts, can plug in any factual question with a real-valued answer to determine how people motivatedly reason about that question. Motives can be compared across varied contexts and populations. 

The notion of a susceptibility parameter also suggests a lever for debiasing people. This design enables us to identify and estimate the magnitude of motivated reasoning; interventions whose objective is debiasing can then use this estimate to test the efficacy of the treatment. One approach is to estimate susceptibility for a treatment group and a control group. Having an approach to reducing motivated reasoning would be a valuable way for researchers to combat polarization and biased beliefs about important issues, especially in a highly politicized society.

\vspace{5mm}

\nocite{*}
\printbibliography

\newpage 

\appendix

\section{Study Materials: Exact Question Wordings}
\label{question-wordings}

\begin{small}

\subsubsection*{Crime Under Obama}
\label{obama-crime-question}
\vspace{-1mm}
Some people believe that the Obama administration was too soft on crime and that violent crime increased during his presidency, while others believe that President Obama's pushes towards criminal justice reform and reducing incarceration did not increase violent crime.

This question asks how murder and manslaughter rates changed during the Obama administration. In 2008 (before Obama became president), the murder and manslaughter rate was 54 per million Americans.

In 2016 (at the end of Obama's presidency), what was the per-million murder and manslaughter rate?

\vspace{1mm}

\textit{Correct answer: 53.}

\textit{Source linked on results page: \url{http://bit.ly/us-crime-rate}}

\vspace{-2mm}

\subsubsection*{Upward Mobility}
\label{mobility-question}
\vspace{-1mm}
In 2017, Donald Trump signed into law the largest tax reform bill since Ronald Reagan's 1981 and 1986 bills. Some people believe that Reagan's reforms accelerated economic growth and allowed lower-income Americans to reap the benefits of lower taxes, while other people believe that this decreased the government's spending to help lower-income Americans get ahead.

This question asks whether children who grew up in low-income families during Reagan's tenure were able to benefit from his tax reforms.

Of Americans who were born in the lowest-income (bottom 20\%) families from 1980-1985, what percent rose out of the lowest-income group as adults?

(Please guess between 0 and 100.)

\vspace{1mm}

\textit{Correct answer: 64.9.}

\textit{Source linked on results page: \url{http://bit.ly/us-upward-mobility} (page 47)}

\vspace{-2mm}

\subsubsection*{Racial Discrimination}
\label{race-question}
\vspace{-1mm}
In the United States, white Americans have higher salaries than black Americans on average. Some people attribute these differences in income to differences in education, training, and culture, while others attribute them more to racial discrimination.

In a study, researchers sent fictitious resumes to respond to thousands of help-wanted ads in newspapers. The resumes sent had identical skills and education, but the researchers gave half of the (fake) applicants stereotypically White names such as Emily Walsh and Greg Baker, and gave the other half of the applicants stereotypically Black names such as Lakisha Washington and Jamal Jones.

9.65 percent of the applicants with White-sounding names received a call back. What percent of the applicants with Black-sounding names received a call back?

(Please guess between 0 and 100.)

\vspace{1mm}

\textit{Correct answer: 6.45.}

\textit{Source linked on results page: \url{http://bit.ly/labor-market-discrimination}}

\vspace{-2mm}

\subsubsection*{Gender and Math GPA}
\label{gender-question}
\vspace{-1mm}
In the United States, men are more likely to enter into mathematics and math-related fields. Some people attribute this to gender differences in interest in or ability in math, while others attribute it to other factors like gender discrimination.

This question asks whether high school boys and girls differ substantially in how well they do in math classes. A major testing service analyzed data on high school seniors and compared the average GPA for male and female students in various subjects.

Male students averaged a 3.04 GPA (out of 4.00) in math classes. What GPA did female students average in math classes?

(Please guess between 0.00 and 4.00.)

\vspace{1mm}

\textit{Correct answer: 3.15.}

\textit{Source linked on results page: \url{http://bit.ly/gender-hs-gpa}}

\vspace{-2mm}

\subsubsection*{Refugees and Violent Crime}
\label{refugees-question}
\vspace{-1mm}
Some people believe that the U.S. has a responsibility to accept refugees into the country, while others believe that an open-doors refugee policy will be taken advantage of by criminals and put Americans at risk.

In 2015, German leader Angela Merkel announced an open-doors policy that allowed all Syrian refugees who had entered Europe to take up residence in Germany. From 2015-17, nearly one million Syrians moved to Germany. This question asks about the effect of Germany's open-doors refugee policy on violent crime rates.

In 2014 (before the influx of refugees), the violent crime rate in Germany was 224.0 per hundred-thousand people.

In 2017 (after the entrance of refugees), what was the violent crime rate in Germany per hundred-thousand people?

\vspace{1mm}

\textit{Correct answer: 228.2.}

\textit{Sources linked on results page: Main site: \url{http://bit.ly/germany-crime-main-site}. 2014 and 2015 data: \url{http://bit.ly/germany-crime-2014-2015}. 2016 and 2017 data: \url{http://bit.ly/germany-crime-2016-2017}.}

\vspace{-2mm}

\subsubsection*{Climate change}
\label{global-warming-question}
\vspace{-1mm}
Some people believe that there is a scientific consensus that human activity is causing global warming and that we should have stricter environmental regulations, while others believe that scientists are not in agreement about the existence or cause of global warming and think that stricter environmental regulations will sacrifice jobs without much environmental gain.

This question asks about whether most scientists think that global warming is caused by humans. A major nonpartisan polling company surveyed thousands of scientists about the existence and cause of global warming.

What percent of these scientists believed that ``Climate change is mostly due to human activity''?

(Please guess between 0 and 100.)

\vspace{1mm}

\textit{Correct answer: 87.}

\textit{Source linked on results page: \url{http://bit.ly/scientists-climate-change}}

\vspace{-2mm}

\subsubsection*{Gun Reform}
\label{guns-question}
\vspace{-1mm}
The United States has a homicide rate that is much higher than other wealthy countries. Some people attribute this to the prevalence of guns and favor stricter gun laws, while others believe that stricter gun laws will limit Americans' Second Amendment rights without reducing homicides very much.

After a mass shooting in 1996, Australia passed a massive gun control law called the National Firearms Agreement (NFA). The law illegalized, bought back, and destroyed almost one million firearms by 1997, mandated that all non-destroyed firearms be registered, and required a lengthy waiting period for firearm sales.

Democrats and Republicans have each pointed to the NFA as evidence for/against stricter gun laws. This question asks about the effect of the NFA on the homicide rate in Australia.

In the five years before the NFA (1991-1996), there were 319.8 homicides per year in Australia. In the five years after the NFA (1998-2003), how many homicides were there per year in Australia?

\vspace{1mm}

\textit{Correct answer: 318.6.}

\textit{Sources linked on results page: \url{http://bit.ly/australia-homicide-rate} (Suicides declined substantially, however. For details: \url{http://bit.ly/impact-australia-gun-laws}.)}

\vspace{-2mm}

\subsubsection*{Media Bias}
\label{media-question}
\vspace{-1mm}
Some people believe that the media is unfairly biased towards Democrats, while some believe it is balanced, and others believe it is biased towards Republicans.

This question asks whether journalists are more likely to be Democrats than Republicans.

A representative sample of journalists were asked about their party affiliation. Of those either affiliated with either the Democratic or Republican Party, what percent of journalists are Republicans?

(Please guess between 0 and 100.)

\vspace{1mm}

\textit{Correct answer: 19.8.}

\textit{Source linked on results page: \url{http://bit.ly/journalist-political-affiliation}}

\vspace{-2mm}



\subsubsection*{Democrats' Relative Performance}
\label{party-performance-question}
\vspace{-1mm}
This question asks whether you think Democrats or Republicans did better on this study about political and U.S. knowledge. I've compared the average points scored by Democrats and Republicans among 100 participants (not including yourself).

The Republicans scored 70.83 points on average.

How many points do you think the Democrats scored on average?

(Please guess between 0 and 100)

\vspace{1mm}

\textit{Correct answer: 72.44.}

\vspace{-2mm}

\subsubsection*{Republicans' Relative Performance}
\vspace{-1mm}
This question asks whether you think Democrats or Republicans did better on this study about political and U.S. knowledge. I've compared the average points scored by Democrats and Republicans among 100 participants (not including yourself).

The Democrats scored 72.44 points on average.

How many points do you think the Republicans scored on average?

(Please guess between 0 and 100)

\vspace{1mm}

\textit{Correct answer: 70.83.}

\vspace{-2mm}

\subsubsection*{Own Relative Performance}
\label{own-performance-question}
\vspace{-1mm}
How well do you think you performed on this study about political and U.S. knowledge? I've compared the average points you scored for all questions (prior to this one) to that of 100 other participants.

How many of the 100 do you think you scored higher than?

(Please guess between 0 and 100.)

\vspace{1mm}

\textit{Correct answer: Depends on participant's performance.}

\vspace{-2mm}

\subsubsection*{Random Number}
\label{random-number-question}
\vspace{-1mm}
A computer will randomly generate a number between 0 and 100. What number do you think the computer chose?

(As a reminder, it is in your best interest to guess an answer that is close to the computer's choice, even if you don't perfectly guess it.)

\vspace{1mm}

\textit{Correct answer: Randomly generated for each participant.}

\subsubsection*{Latitude of Center of the United States}
\label{latitude-question}
\vspace{-1mm}
The U.S. National Geodetic Survey approximated the geographic center of the continental United States. (This excludes Alaska and Hawaii, and U.S. territories.)

How many degrees North is this geographic center?

(Please guess between 0 and 90. The continental U.S. lies in the Northern Hemisphere, the Equator is 0 degrees North, and the North Pole is 90 degrees North.)

\vspace{1mm}

\textit{Correct answer: 39.833.}

\textit{Source linked on results page: \url{http://bit.ly/center-of-the-us}}

\vspace{-2mm}

\subsubsection*{Longitude of Center of the United States}
\label{longitude-question}
\vspace{-1mm}
The U.S. National Geodetic Survey approximated the geographic center of the continental United States. (This excludes Alaska and Hawaii, and U.S. territories.)

How many degrees West is this geographic center?

(Please guess between 0 and 180. The continental U.S. lies in the Western Hemisphere, which ranges from 0 degrees West to 180 degrees West.)

\vspace{1mm}

\textit{Correct answer: 98.583.}

\textit{Source linked on results page: \url{http://bit.ly/center-of-the-us}}

\vspace{-2mm}

\subsubsection*{Comprehension Check: Current Year}
\label{comprehension-question}
\vspace{-1mm}
In 1776 our fathers brought forth, upon this continent, a new nation, conceived in Liberty, and dedicated to the proposition that all men are created equal.

What is the year right now?

This is not a trick question and the first sentence is irrelevant; this is a comprehension check to make sure you are paying attention. For this question, your lower and upper bounds should be equal to your guess if you know what year it currently is.

\vspace{1mm}

\textit{Correct answer: 2018.}

\textit{Source linked on results page: \url{http://bit.ly/what-year-is-it}}

\end{small}

\newpage

\newpage

%

\newpage

\section{Appendix: Additional Experimental Results}
\label{appendix-experiment}

\begin{center}
\begin{threeparttable}
\begin{footnotesize}
\caption{Prior Beliefs by Party}
\label{appendix-prior}
{
\def\sym#1{\ifmmode^{#1}\else\(^{#1}\)\fi}
\begin{tabular}{l*{4}{c}}
\hline\hline
                    &\multicolumn{1}{c}{Pro-Rep}         &\multicolumn{1}{c}{Pro-Dem}         &\multicolumn{1}{c}{Difference}          &\multicolumn{1}{c}{Answer}         \\
\hline
Obama Crime Guess   &   55.907&   49.560&     6.348   &  53          \\
                    &  (0.765)         &  (0.391)         &      (0.858)            \\
Mobility Guess      &   30.185&   22.152&     8.034  & 64.9           \\
                    &  (1.048)         &  (0.611)         &     (1.211)             \\
Race Guess          &   12.349&    8.051&     4.298   & 6.45          \\
                    &  (0.874)         &  (0.436)         &      (0.975)            \\
Gender Guess        &    3.059&    3.086&      -0.027  & 3.15           \\
                    &  (0.015)         &  (0.008)         &       (0.017)           \\
Refugees Guess      &  287.640&  239.004&    48.637  & 228.2            \\
                    &  (5.894)         &  (2.353)         &       (6.335)           \\
Climate Guess       &   75.226&   85.366&      -10.140   & 87         \\
                    &  (1.056)         &  (0.572)         &        (1.200)          \\
Gun Laws Guess      &  230.013&  184.478&    45.535   & 318.6           \\
                    &  (5.950)         &  (3.914)         &        (7.113)           \\
Media Guess         &   36.656&   41.850&       -5.195   & 19.8        \\
                    &  (1.211)         &  (0.599)         &         (1.349)         \\
Rep Score Guess     &   71.563&   61.933&      9.630    & 70.83        \\
                    &  (0.787)         &  (0.614)         &         (0.997)         \\
Dem Score Guess     &   64.671&   73.277&     -8.606   & 72.44          \\
                    &  (0.771)         &  (0.415)         &           (0.875)       \\
\hline
Observations        &     2430         &     5643         &     8073   &       \\
\hline\hline
\multicolumn{4}{l}{\footnotesize Standard errors in parentheses}\\
\end{tabular}
}

\end{footnotesize}
\begin{tablenotes}
\vspace{3mm}
\begin{scriptsize}
\item \textbf{Notes:} OLS, robust standard errors. Guesses are winsorized at the 5-percent level. Third column represents mean Pro-Rep guess minus mean Pro-Dem guess. The sign of every coefficient points in the predicted motive direction from Table 1.
\end{scriptsize}
\end{tablenotes}
\end{threeparttable}
\end{center}

\newpage


\begin{center}
\begin{threeparttable}
\begin{footnotesize}
\caption{Balance Table}
\def\sym#1{\ifmmode^{#1}\else\(^{#1}\)\fi}
\begin{tabular}{p{2.5cm}*{4}{c}}
\hline\hline
&\multicolumn{1}{c}{Anti-Party News}&\multicolumn{1}{c}{Pro-Party News}&\multicolumn{1}{c}{Anti vs. Pro}&\multicolumn{1}{p{2.75cm}}{\centering p-value}\\
\hline
Partisanship  & 0.484 & 0.478 & 0.007 & 0.312 \\   & (0.005) & (0.005) & (0.007) & \\Rep vs. Dem & -0.237 & -0.236 & -0.001 & 0.937 \\   & (0.008) & (0.008) & (0.011) & \\Male  & 0.532 & 0.534 & -0.002 & 0.881 \\   & (0.008) & (0.008) & (0.011) & \\Age  & 35.261 & 35.400 & -0.139 & 0.573 \\   & (0.175) & (0.173) & (0.246) & \\Education  & 14.716 & 14.765 & -0.049 & 0.242 \\   & (0.029) & (0.030) & (0.042) & \\Log(income)  & 10.725 & 10.748 & -0.024 & 0.182 \\   & (0.012) & (0.013) & (0.018) & \\White  & 0.752 & 0.760 & -0.008 & 0.430 \\   & (0.007) & (0.007) & (0.010) & \\Black  & 0.075 & 0.081 & -0.006 & 0.303 \\   & (0.004) & (0.004) & (0.006) & \\Hispanic  & 0.066 & 0.062 & 0.004 & 0.499 \\   & (0.004) & (0.004) & (0.006) & \\Religious  & 0.443 & 0.457 & -0.014 & 0.214 \\   & (0.008) & (0.008) & (0.011) & \\Red State  & 0.567 & 0.558 & 0.009 & 0.431 \\   & (0.008) & (0.008) & (0.011) & \\
WTP elicited  & 0.490 & 0.476 & 0.014 & 0.213 \\   & (0.008) & (0.008) & (0.011) & \\Told 1/2 True & 0.333 & 0.344 & -0.011 & 0.309 \\ 
 & (0.007) & (0.008) & (0.011) & \\
\hline
\(N\)  & 3961 & 3941 & 7902 & \\
\hline
\hline
\end{tabular}
\label{balance-table}
\end{footnotesize}
\begin{tablenotes}
\item \begin{scriptsize} \textbf{Notes:} Standard errors in parentheses. Rep vs. Dem is the rating of the Republican Party minus the rating of the Democratic Party and is between -1 and 1. Partisanship is the absolute difference in these ratings. Education is in years. Religious is 1 if subject in any religious group. Red State is 1 if state voted for Trump in 2016 election. WTP elicited is 1 if subject in the WTP group and 0 if in the second-guess group. Told 1/2 True is 1 if subject is told that P(True News) is 1/2 and 0 if subject is not.
\end{scriptsize}
\end{tablenotes}
\end{threeparttable}
\end{center}

\clearpage
%
%

\begin{center}
\begin{threeparttable}
\begin{footnotesize}
\caption{The Effect of News Direction, Actual Veracity, and Skewed Prior Distributions on Perceived Veracity}
\label{unskewed-priors}
\begin{tabular}{l*{4}{c}}
\hline\hline
                                   &\multicolumn{1}{c}{(1)}&\multicolumn{1}{c}{(2)}&\multicolumn{1}{c}{(3)}&\multicolumn{1}{c}{(4)}\\
\hline
Unskewed                           &   -0.010&   -0.011&   -0.011&    0.007\\
                                   &  (0.009)&  (0.009)&  (0.010)&  (0.013)\\
Pro-Party News                     &    0.083&    0.037&    0.028&    0.075\\
                                   &  (0.007)&  (0.014)&  (0.008)&  (0.008)\\
Unskewed x Pro-Party               &    0.016&    0.015&    0.018&    0.007\\
                                   &  (0.012)&  (0.019)&  (0.013)&  (0.013)\\
Partisanship x Pro-Party           &         &    0.097&         &         \\
                                   &         &  (0.024)&         &         \\
Unskewed x Partisanship x Pro-Party&         &    0.006&         &         \\
                                   &         &  (0.032)&         &         \\
Anti-Party News                    &         &         &   -0.052&         \\
                                   &         &         &  (0.008)&         \\
Unskewed x Anti-Party              &         &         &    0.001&         \\
                                   &         &         &  (0.013)&         \\
True News                          &         &         &         &   -0.026\\
                                   &         &         &         &  (0.007)\\
Unskewed x True News               &         &         &         &   -0.026\\
                                   &         &         &         &  (0.013)\\
Question FE                        &      Yes&      Yes&       No&      Yes\\
Round FE                           &      Yes&      Yes&      Yes&      Yes\\
Subject FE                         &      Yes&      Yes&      Yes&      Yes\\
Neutral News                       &       No&       No&      Yes&       No\\
\hline
Observations                       &     7882&     7882&    10499&     7882\\
\(R^{2}\)                          &     0.25&     0.25&     0.21&     0.25\\
Mean                               &    0.574&    0.574&    0.575&    0.574\\
\hline\hline
\multicolumn{5}{l}{\footnotesize Standard errors in parentheses}\\
\end{tabular}

\end{footnotesize}
\begin{tablenotes}
\item \begin{scriptsize} \textbf{Notes:} OLS, errors clustered at subject level. Neutral News indicates that Pro-Party / Anti-Party compared to Neutral News, as defined in Table 1. Controls: race, gender, log(income), education (in years), religion, whether state voted for Trump or Clinton in 2016. Partisanship is absolute difference between Republican and Democratic ratings. Unskewed is 1 if initial guess is exactly halfway between lower / upper bounds and 0 otherwise. Observations dropped if lower bound equals upper bound.
\end{scriptsize}
\end{tablenotes}
\end{threeparttable}
\end{center}

\clearpage 

\begin{center}
\begin{threeparttable}
\begin{footnotesize}
\caption{The Effect of News Direction, Actual Veracity, and Previous News Directions on Perceived Veracity}
\label{news-independence}
\begin{tabular}{l*{4}{c}}
\hline\hline
                                   &\multicolumn{1}{c}{(1)}&\multicolumn{1}{c}{(2)}&\multicolumn{1}{c}{(3)}&\multicolumn{1}{c}{(4)}\\
\hline
Previous Pro-Party                 &   -0.001&   -0.001&   -0.002&   -0.002\\
                                   &  (0.002)&  (0.002)&  (0.002)&  (0.002)\\
Pro-Party News                     &    0.087&    0.040&    0.036&    0.076\\
                                   &  (0.006)&  (0.012)&  (0.006)&  (0.007)\\
Partisanship x Pro-Party           &         &    0.099&         &         \\
                                   &         &  (0.022)&         &         \\
Anti-Party News                    &         &         &   -0.048&         \\
                                   &         &         &  (0.007)&         \\
True News                          &         &         &         &   -0.034\\
                                   &         &         &         &  (0.006)\\
Question FE                        &      Yes&      Yes&       No&      Yes\\
Round FE                           &      Yes&      Yes&      Yes&      Yes\\
Subject FE                         &      Yes&      Yes&      Yes&      Yes\\
Neutral News                       &       No&       No&      Yes&       No\\
\hline
Observations                       &     7902&     7902&    10552&     7902\\
\(R^{2}\)                          &     0.25&     0.25&     0.21&     0.25\\
Mean                               &    0.574&    0.574&    0.575&    0.574\\
\hline\hline
\multicolumn{5}{l}{\footnotesize Standard errors in parentheses}\\
\end{tabular}

\end{footnotesize}
\begin{tablenotes}
\begin{scriptsize}
\item \textbf{Notes:} OLS, errors clustered at subject level. Neutral News indicates that Pro-Party / Anti-Party compared to Neutral News, as defined in Table 1. Controls: race, gender, log(income), education (in years), religion, whether state voted for Trump or Clinton in 2016. Partisanship is the absolute difference between Republican and Democratic ratings. Previous Pro-Party is the number of all previous pieces of news that are Pro-Party minus the number that are Anti-Party.
\end{scriptsize}
\end{tablenotes}
\end{threeparttable}
\end{center}

\clearpage

\begin{figure}[htb!]
\caption{CDF of Perceived Veracity for Pro-Performance and Anti-Performance News}
\label{cdf-good-bad-performance}
\vspace{-2mm}
\begin{center}
\includegraphics[width = .85\textwidth]{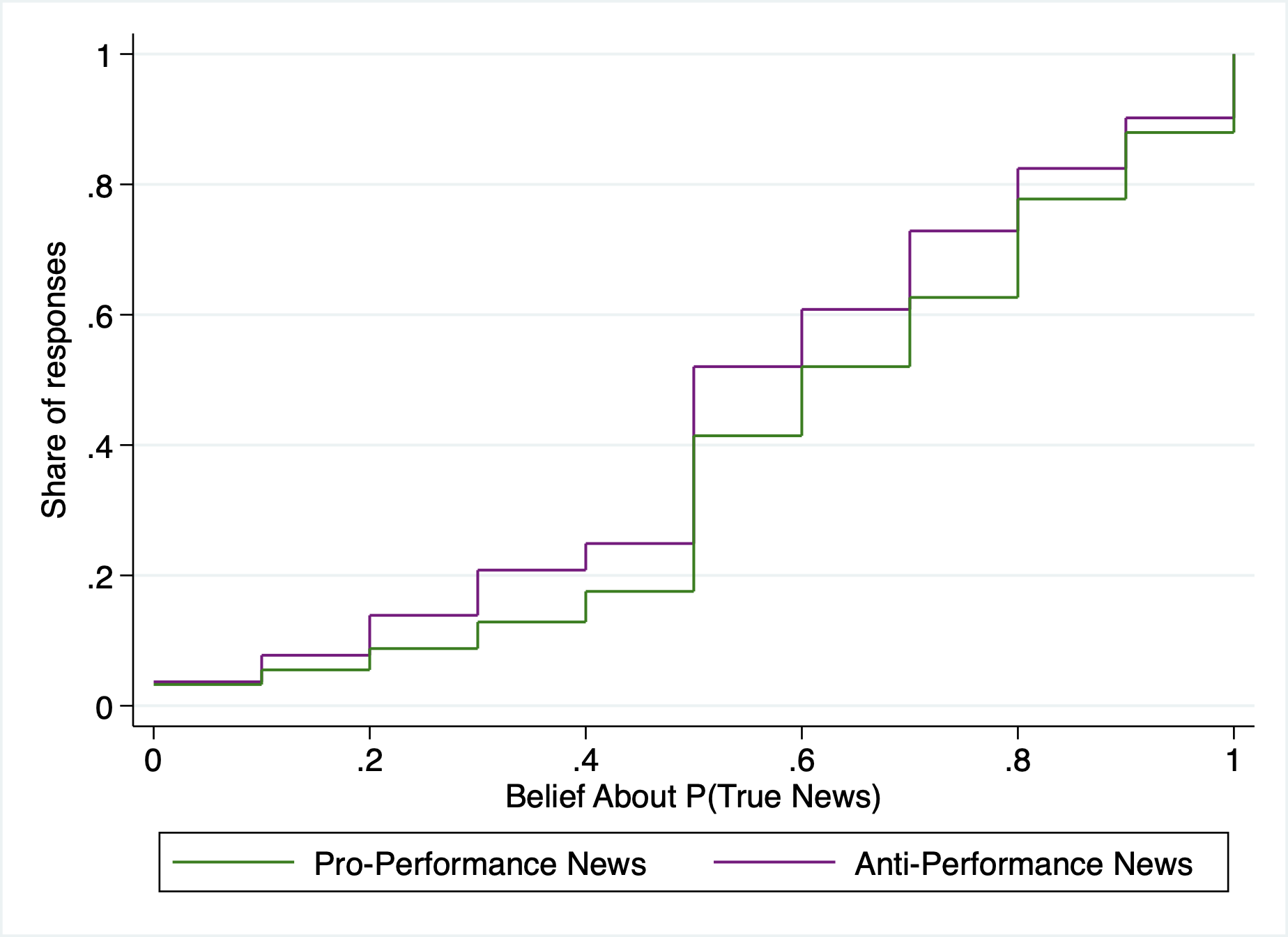}
\end{center}
\begin{threeparttable}
\begin{tablenotes}
\vspace{-6mm}
\begin{scriptsize}
\item \textbf{Notes:} Pro-Performance and Anti-Performance news are defined in Table 1. This figure shows that subjects trust Pro-Performance news more than Anti-Performance news. The x-axis measures subjects' beliefs about P(True | Pro-/Anti-Performance News). The y-axis measures the share of respondents that give at most that high of an assessment. Bayesians would have the same trust in news for Pro-Performance and Anti-Performance news, and the residual is motivated reasoning. 
\end{scriptsize}
\end{tablenotes}
\end{threeparttable}
\end{figure}

\clearpage

\begin{figure}[htb!]
\caption{CDF of Perceived Veracity for True News and Fake News}
\label{cdf-true-fake}
\vspace{-2mm}
\begin{center}
\includegraphics[width = .85\textwidth]{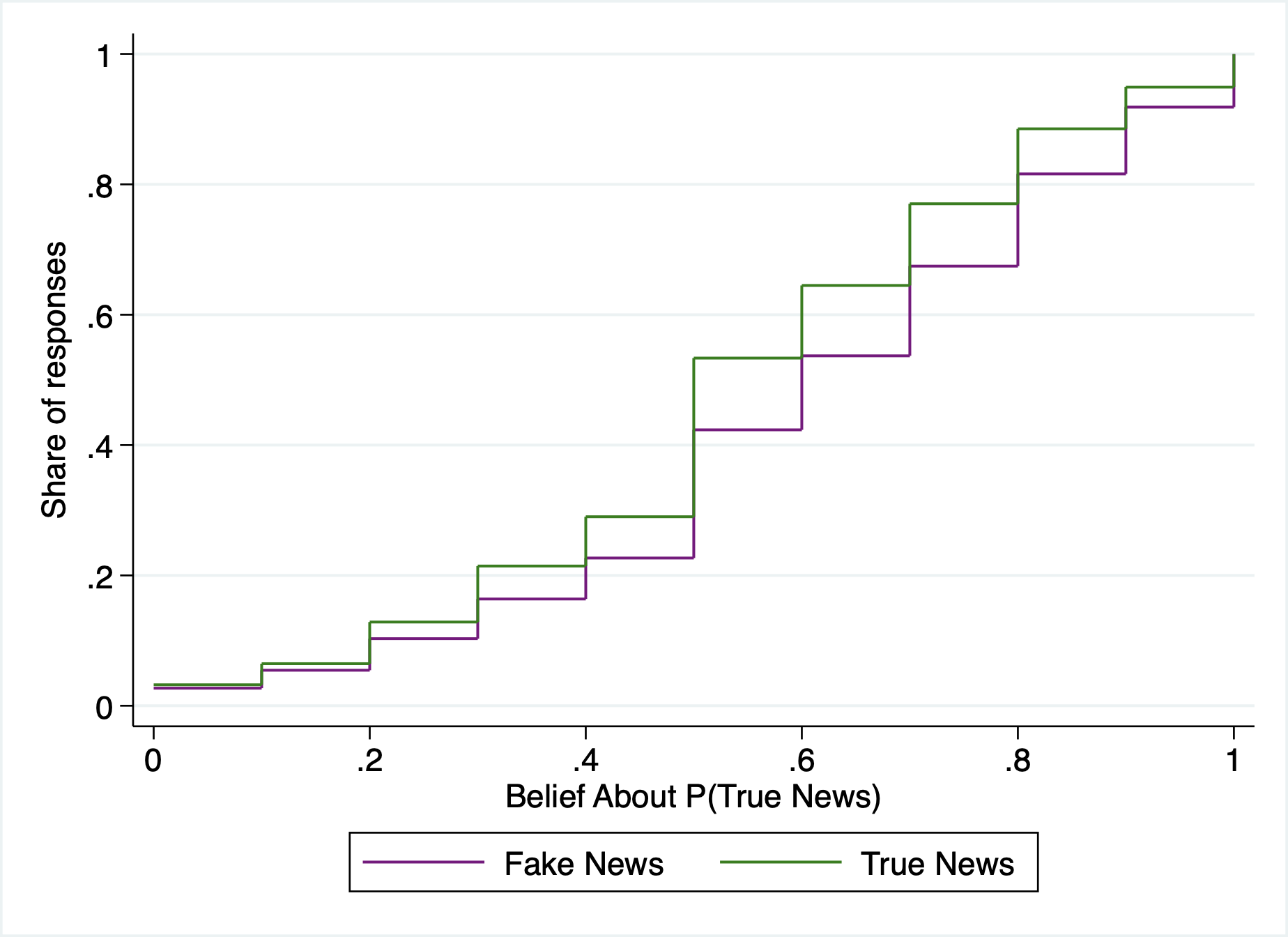}
\end{center}
\begin{threeparttable}
\begin{tablenotes}
\vspace{-6mm}
\begin{scriptsize}
\item \textbf{Notes:} Only political topics included. This figure shows that subjects trust Fake News more than True News. The x-axis measures subjects' beliefs about P(True | actual True/Fake News). The y-axis measures the share of respondents that give at most that high of an assessment. Bayesians would have the same trust in news for True and Fake News, and the residual is motivated reasoning. 
\end{scriptsize}
\end{tablenotes}
\end{threeparttable}
\end{figure}

\clearpage

\section{Online Appendix: Additional Theoretical Results}\label{appendix-theory}

\subsection{Inferring Motives from Beliefs}
\label{appendix-theory-priors}
Agents may have different beliefs because these beliefs reflect past instances of motivated reasoning. In this subsection, we consider motives that depend on the state $\theta_q$ and an econometrician who observes motivated-reasoning agents' updating process, inferring whether their motive function is increasing or decreasing in $\theta_q$. When motives are unobservable, an experimenter can learn about agents' motives by looking at their initial beliefs $\mu_{iq}$. Conceptually, an agent's error in beliefs can be partly explained by motivated reasoning, and therefore the direction of the error predicts the direction of the motive function. Two agents who motivatedly reason in different directions, and each who receive a signal drawn from the same distribution, will hold different median beliefs: A motivated reasoner with an increasing motive function will be more likely to hold a belief that $\mu_{iq} > \theta_q$, and a motivated reasoner with a decreasing motive function will be more likely to hold a belief that $\mu_{iq} < \theta_q$. This implies that an agent who believes $\mu_{iq} > \theta_q$ is more likely to have an increasing motive function than is an agent who believes $\mu_{iq} < \theta_q$. 

When the two agents then make news assessments using the structure above, agents will trust news that \textit{reinforces} the error in their beliefs more than news that \textit{mitigates} the error. This occurs even though signals are designed exactly so that their interpretation is distinct from $\mu_{iq}$. 

More formally, there is a state $\theta_q \in \mathbb{R}$. Consider a Bayesian (she) and a motivated reasoner (he) with the same prior $\theta_q \sim F_{\theta_q}$ and who receive a public signal $z_q \sim F_{zq}(\cdot)$. We now assume that motivated reasoners have heterogeneous motive functions $m_{iq}(\theta'_q)$ that are strictly monotonic in $\theta'_q$; that is, $m_{iq}(\theta'_q)$ is drawn from a distribution $F_{mq}(m_q(\cdot))$ that includes strictly increasing and strictly decreasing functions of $\theta'_q$. The econometrician observes the distributions $F_{zq}(\cdot)$ and $F_{mq}(m_q(\cdot))$, as well as the states $\theta_q$ and reported medians $\mu_{iq}$, but not the realizations $z_q$ or $m_{iq}(\theta'_q)$. We also assume that $z_q$ leads the Bayesian's posterior median $\mu_{Bq}$ to take values close to $\theta_q$ with positive probability, but that there is no point mass at exactly $\theta_q$. That is, for all $\delta > 0$, there exists some $\delta' > 0$ such that $\mathbb{P}(|\mu_{Bq} - \theta_q| < \delta) > \delta'$, while $\mathbb{P}_{Bq}(\mu_{Bq} = \theta_q) = 0$. 

Without loss of generality, consider what the econometrician infers about a motivated reasoner who has a strictly increasing $m_{iq}(\theta'_q)$. Since the log-likelihood of the motive is strictly increasing, his posterior distribution first-order stochastically dominates the Bayesian's posterior distribution. In addition, for every such motive function, there exists a $\delta>0$ such that for all signals leading to the Bayesian having a posterior median $\mu_{Bq} \in (\theta_q - \delta, \theta_q)$, the motivated reasoner has posterior median $\mu_{iq} > \theta_q$. Since there is a probability of at least $\delta'>0$ of such a signal, this high-$\theta_q$-motivated reasoner is \textit{strictly} more likely than the Bayesian to state a median belief that is greater than $\theta_q$. By the same argument, a low-$\theta_q$ motivated reasoner is strictly less likely than the Bayesian to state a median that is less than $\theta_q$. 

If some people have monotonically-increasing motives and others have monotonically-decreasing motives, then the econometrician will believe that: 
\[ \mathbb{P}_{\text{econometrician}}(m_{iq}(\theta_q) \text{ increasing } | \text{ }\mu_{iq} > \theta_q) > \mathbb{P}_{\text{econometrician}}(m_{iq}(\theta_q) \text{ increasing } | \text{ }\mu_{iq} < \theta_q). \]

Recall that message $G_{iq}$ says that $\theta_q > \mu_{iq}$ and that message $L_{iq}$ says $\theta_q < \mu_{iq}$. Then, for the econometrician, $\mathbb{E}[a(G_{iq}), \mu_{iq} > \theta_q] > \mathbb{E}[a(G_{iq}, \mu_{iq} < \theta_q]$ and $\mathbb{E}[a(L_{iq}), \mu_{iq} > \theta_q] < \mathbb{E}[a(L_{iq}), \mu_{iq} < \theta_q]$ when motives are heterogeneous. Since $G_{iq}$ and $L_{iq}$ are equally likely by construction, this means that agents trust the source of \textit{error-reinforcing} messages more than the source of \textit{error-mitigating} messages on questions when motivated reasoning plays a role. 

Using this signal structure, error-mitigating messages come from \textit{True News} and error-reinforcing messages come from \textit{Fake News}. Therefore, agents give higher assessments to Fake News than True News, with and without controls for observable party preference:
\begin{fact}[Motivated reasoning leads to over-trusting Fake News, under-trusting True News]
\label{fact-Fake-News}
Suppose that agents motivatedly reason with a strictly monotonic motive. Then:
\begin{itemize}
\item $a(\text{Fake News}_{iq}) > a(\text{True News}_{iq})$. 
\item $a(\text{Fake News}_{iq}; \text{Pro-Party news}_{iq}) \geq a(\text{True News}_{iq}; \text{Pro-Party news}_{iq})$. 
\item $a(\text{Fake News}_{iq}; \text{Anti-Party news}_{iq}) \geq a(\text{True News}_{iq}; \text{Anti-Party news}_{iq})$. 
\end{itemize}
Suppose also that the sign of the slope of the motive function is heterogeneous within party. That is, the probability of an agent having $\frac{\partial m_{iq}(\theta_q)}{\partial \theta_q} > 0$ is strictly between 0 and 1, conditional on the agent's party. Then:
\begin{itemize}
\item $a(\text{Fake News}_{iq}; \text{Pro-Party news}_{iq}) > a(\text{True News}_{iq}; \text{Pro-Party news}_{iq})$. 
\item $a(\text{Fake News}_{iq}; \text{Anti-Party news}_{iq}) > a(\text{True News}_{iq}; \text{Anti-Party news}_{iq})$. 
\end{itemize}
\end{fact}
The stark result that motivated reasoners will trust Fake News more than True News is particular to the uninformativeness of the messages. However, the prediction that agents will trust Fake News more than a \textit{Bayesian} will is quite general, only relying on unobservable inputs into current beliefs. This prediction only holds for motivated states, psychologically differentiating this theory from unmotivated explanations of over-trusting Fake News (such as confirmation bias). Practically, it suggests that excessive trust in disinformation is more prominent when people hold motivated beliefs.

\subsection{Motivated Reasoning and Overprecision}
\label{moreas-consequences}

Under further functional form assumptions, the model of motivated reasoning also predicts that agents will overestimate the probability that the answers to the questions are within a subjective confidence interval: \textit{overprecision}. Unlike other determinants of overprecision, this will be because of miscalibrated beliefs due to motivated reasoning, and we do not assume a general bias towards overly-narrow confidence intervals. Unlike the previous subsection, we now assume that Nature draws $\theta_q$ from $\mathcal{N}(\mu_{q0}, 1 / \tau^2_{q0})$ independently across $q$, and that Nature gives agents a noisy public signal $z_q = \theta_q + \epsilon_{qz}$, where $\epsilon_{qz} \sim \mathcal{N}(0, 1 / \tau^2_{qz})$. 

Slightly abusing notation, we assume also that motivated reasoners have motives linear in $\theta_q$: $m_{iq}(\theta_q) = m_{iq} \cdot \theta_q$.\footnote{We also keep the assumption of constant $\varphi$ but note that there are many reasons to believe that $\varphi$ ought to depend on $z_q$. With normally-distributed signals, one way that $\varphi$ could depend on $z_q$ is through its precision. For instance, $\varphi(z_q, \tau_{qz}) = \min\{\varphi_c \tau_{qz}, \bar{\varphi}\}$. Using this form, the susceptibility of two weak signals is equal to the sum of their precisions, but there is a maximum level of susceptibility.} In the political context, $|m_{iq}|$ can be thought of as political partisanship. Finally, we additionally assume that the prior probability of True News is $p$ and is constant for all agents. 

After observing $z_q$ (but before playing the game in \Cref{moreas-subsection}), a Bayesian forms the posterior:
\[f_{Bq}(\theta'_q | z_q) = \mathcal{N}\left(\frac{\tau_{q0} \mu_{q0} + \tau_{qz} z_q}{\tau_{q0} + \tau_{qz}}, \tau_{q0} + \tau_{qz} \right), \]
and a motivated reasoner forms the posterior:
\[f_{iq}(\theta'_q | z_q) = \mathcal{N}\left(\frac{\tau_{q0} \mu_{q0} + \tau_{qz} z_q + \varphi \cdot m_{iq}}{\tau_{q0} + \tau_{qz}}, \tau_{q0} + \tau_{qz} \right). \]

Notably, the two agents have the same posterior variance, but the motivated reasoner's distribution is miscalibrated. Consider their $(1-Q)/2$-quantile and $(1+Q)/2$-quantile beliefs, and call this the \textit{$Q$-confidence interval}. Then:
\begin{fact}[Motivated reasoning and overprecision]\label{fact-overprecision}
Suppose that a motivated reasoner has normally-distributed priors and receives a signal normally distributed with mean equal to $\theta_q$, as above. When $\varphi>0$, the probability that his $Q$-confidence interval contains $\theta_q$ is equal to $Q$ for $m_{iq}=0$ and strictly decreases in $|m_{iq}|$.
\end{fact}
Since Bayesian updating is equivalent to motivated reasoning with $m_{iq}=0$, Bayesians will appear to be appropriately precise and motivated reasoners will appear overprecise due to their miscalibrated confidence intervals. They will be both be appropriately precise on neutral topics, where $m_{iq} = 0$. Note that the direction of the bias relies both on linear motives and the normal-normal functional form.\footnote{For instance, suppose that the state space contains two values and a Bayesian infers from a signal that either one value has a $(1+Q)/2$ likelihood of occurring or the other value has a $(1+Q)/2$ likelihood of occurring. Then, her confidence interval would contain one point, and a motivated reasoner may have a confidence interval that contains both points.}

In the context of politics, I posit that partisanship is a proxy for $|m_{iq}|$ on politicized questions. On such questions, the probability that agents' $Q$-confidence interval contains $\theta_q$ will then be decreasing in their partisanship.

\clearpage

\section{Online Appendix: Replication}
\label{replication}

I preregistered a replication for the findings from this paper. I ran this in conjunction with a debiasing treatment; the replication tests whether the control group from that sample satisfies the hypotheses from this experiment. This section reports all replication results that were specified in the pre-analysis plan in \textcite{T19R}. 

There are a few differences between the replication sample and the original sample. The replication was conducted approximately one year later, on July 8-9, 2019. The replication questions included additional topics and variants of the original questions.\footnote{In particular, two new politicized topics were added: Wage Growth and Healthcare. On six of the politicized topics, subjects received slightly different versions of the original question.} There were no neutral questions. 

The sample includes 1,050 subjects recruited from Amazon's Mechanical Turk platform that passed pre-specified comprehension checks that are akin to those in the original experiment. There are 982 subjects who are either Pro-Rep or Pro-Dem in the replication sample, and these subjects give 5,314 news veracity assessments on politicized topics. 

\subsection{Primary Outcomes}
The most important primary outcome results are all strongly replicated. As seen in the first column of \Cref{political-replication}, subjects give statistically significantly higher assessments to Pro-Party news than to Anti-Party news $(p<0.001)$.\footnote{The coefficient is smaller in the replication, due in large part to the new added questions. On the questions with the exact same wording as the original study, the treatment effect is 7.1 percentage points (s.e. 1.2 percentage points). On other politicized questions, the treatment effect is 3.5 percentage points (s.e. 1.0 percentage points). Of the original questions, the effects on the following topics were significant at $p<0.05$ in the predicted direction: Climate Change, Race, Refugees, Gun Laws, Party Performance, Own Performance. The effects on the following topics were not significant at $p<0.05$: Obama and Crime, Gender, Media.} The second column shows that this gap is increasing in partisanship ($p=0.006$).

The next-most important primary outcome results are strongly replicated. \Cref{political-replication} shows that subjects give statistically significantly higher assessments to Fake News than to True News. This holds both when Pro-Party / Anti-Party news is not controlled for (column 3) and when Pro-Party / Anti-Party news is controlled for (column 4); both results are statistically significant at $p<0.001$.

The main alternative measure of motivated reasoning is suggestively replicated. As seen in the first column of \Cref{change-message-replication}, results suggest that subjects are more likely to update in the direction of the Pro-Party message compared to the Anti-Party message ($p=0.055$).\footnote{As with the main effect, the coefficient is smaller in the replication, due in large part to the new questions. On the questions with the exact same wording as the original study, the treatment effect is 5.7 percentage points (s.e. 2.6 percentage points). On other politicized questions, the treatment effect is 2.0 percentage points (s.e. 2.6 percentage points).} The third column shows that, as in \Cref{robustness-checks}, this difference vanishes once the news veracity assessment measure is controlled for.

\subsection{Secondary Outcomes}
The underperformance result (not discussed in the main text) is strongly replicated. Subjects score 66.3 points (s.e. 0.4 points) on politicized news assessments and 65.5 points (s.e. 1.6 points) on performance news assessments on average. Both of these are statistically significantly lower than 75 points, the score that subjects would receive if they had answered ``5/10 chance the message came from True News'' ($p<0.001$).

The result that subjects' confidence intervals are overprecise is strongly replicated. On politicized topics, subjects' 50 percent confidence intervals contain the correct answer 44.1 percent of the time (s.e. 0.8 percent); this is statistically significantly different from 50 percent ($p<0.001$). As seen in \Cref{overprecision-replication}, the result that this measure of overprecision is increasing in partisanship is suggestive ($p=0.066$). 

The two polarization results are replicated. On politicized topics, \Cref{change-message-replication} shows that subjects are statistically significantly more likely to follow Polarizing news than anti-Polarizing news ($p=0.031$).\footnote{As in \Cref{robustness-checks}, this difference vanishes once the news assessment measure is controlled for.} Subjects also state initial medians that are more likely to be in the Pro-Party direction ($p<0.001$). 

\subsection{Untested Replications}
I did not pre-register replication tests for performance-driven motivated reasoning (or anything involving neutral topics) given the limited sample size. Results, however, are broadly similar to those in the main experiment. Subjects assess Pro-Performance news to be 8.0 percentage points higher than Anti-Performance news (s.e. 2.6 percentage points; $p=0.003$). Demographic heterogeneity, robustness exercises, and minor results were also not tested. Further work can test whether these results also replicate with a larger sample.

\subsection*{Replication Tables}

\begin{center}
\begin{threeparttable}[htb!]
\caption{The Effect of News Direction and Actual Veracity on Perceived Veracity: Replication}
\begin{footnotesize}
\begin{tabular}{l*{4}{c}}
\hline\hline
                    &\multicolumn{1}{c}{(1)}&\multicolumn{1}{c}{(2)}&\multicolumn{1}{c}{(3)}&\multicolumn{1}{c}{(4)}\\
\hline
Pro-Party News \hspace{30mm}    &    0.088&    0.041&         &    0.077\\
                    &  (0.006)&  (0.012)&         &  (0.006)\\
Partisanship x Pro-Party      &         &    0.099&         &         \\
           &         &  (0.022)&         &         \\
True News           &         &         &   -0.059&   -0.034\\
                    &         &         &  (0.006)&  (0.006)\\
Question FE         &      Yes&      Yes&      Yes&      Yes\\
Round FE            &      Yes&      Yes&      Yes&      Yes\\
Subject FE          &      Yes&      Yes&      Yes&      Yes\\
\hline
Observations        &     7902&     7902&     7902&     7902\\
\(R^{2}\)           &     0.25&     0.25&     0.23&     0.25\\
Mean                &    0.574&    0.574&    0.574&    0.574\\
\hline\hline
\multicolumn{5}{l}{\footnotesize Standard errors in parentheses}\\
\end{tabular}

\end{footnotesize}
\label{political-replication}
\begin{tablenotes}
\begin{scriptsize}
\item \textbf{Notes:} OLS, errors clustered at subject level. Only Pro-Party / Anti-Party news observations. Partisanship is the absolute difference between ratings of the Republican and Democratic parties.
\end{scriptsize}
\end{tablenotes}
\end{threeparttable}
\end{center}

\vspace{5mm}

\begin{center}
\begin{threeparttable}[htb!]
\caption{Changing Guess to Follow Message: Replication}
\begin{footnotesize}
{
\def\sym#1{\ifmmode^{#1}\else\(^{#1}\)\fi}
\begin{tabular}{l*{4}{c}}
\hline\hline
                    &\multicolumn{1}{c}{(1)}         &\multicolumn{1}{c}{(2)}         &\multicolumn{1}{c}{(3)}         &\multicolumn{1}{c}{(4)}         \\
\hline
Pro-Party News  \hspace{30mm}     &    0.038  &                  &   -0.020         &                  \\
                    &  (0.020)         &                  &  (0.018)         &                  \\
Polarizing News     &                  &    0.040 &                  &   -0.018         \\
                    &                  &  (0.019)         &                  &  (0.017)         \\
P(True)             &                  &                  &    1.108&    1.107\\
                    &                  &                  &  (0.055)         &  (0.055)         \\
Question FE         &      Yes         &      Yes         &      Yes         &      Yes         \\
Round FE            &      Yes         &      Yes         &      Yes         &      Yes         \\
Subject FE          &      Yes         &      Yes         &      Yes         &      Yes         \\
\hline
Observations        &     5314         &     5314         &     5314         &     5314         \\
\(R^{2}\)           &     0.34         &     0.34         &     0.48         &     0.48         \\
Mean                &    0.654         &    0.654         &    0.654         &    0.654         \\
\hline\hline
\multicolumn{5}{l}{\footnotesize Standard errors in parentheses}\\
\end{tabular}
}

\end{footnotesize}
\label{change-message-replication}
\begin{tablenotes}
\begin{scriptsize}
\item \textbf{Notes:} OLS, errors clustered at subject level. Only Pro-Party / Anti-Party news observations. Polarizing News: tells subjects that, compared to their initial guess, the answer is in the opposite direction from the population mean. Dependent variable is 1 if subjects change their guess upwards when the message says ``Greater Than'' or downwards when the message says ``Less Than,'' -1 if they change it in the opposite direction, and 0 if they do not change it.
\end{scriptsize}
\end{tablenotes}
\end{threeparttable}
\end{center}

\vspace{5mm}

\begin{center}
\begin{threeparttable}[htb!]
\caption{Overprecision and Partisanship: Replication}
\begin{footnotesize}
\begin{tabular}{l*{2}{c}}
\hline\hline
                                   &\multicolumn{1}{c}{\hspace{10mm}(1)} \hspace{10mm}&\multicolumn{1}{c}{(2)}\\
\hline
Partisanship \hspace{40mm}         &   0.055&    0.056\\
                                   &  (0.030)&  (0.030)\\
Subject controls                   &       No&      Yes\\
\hline
Observations                       &     5314&     5314\\
\(R^{2}\)                          &     0.00&     0.01\\
Mean                               &    0.061&    0.061\\
\hline\hline
\multicolumn{3}{l}{\footnotesize Standard errors in parentheses}\\
\end{tabular}

\end{footnotesize}
\label{overprecision-replication}
\begin{tablenotes}
\begin{scriptsize}
\item \textbf{Notes:} OLS, errors clustered at subject level. Only politicized topics included. Partisanship is the absolute difference between ratings of the Republican and Democratic parties. Subject controls are race, gender, age, log(income), education, religion, and whether home state voted for Trump or Clinton in 2016.
\end{scriptsize}
\end{tablenotes}
\end{threeparttable}
\end{center}

\newpage

\section{Online Appendix: Further Heterogeneity and Robustness Analyses}

This section presents results on heterogeneity in magnitude of motivated reasoning and additional robustness checks for the main results in Table 2. Results are similar for each randomization arm, if I include subjects who fail comprehension checks, and if the dependent variable is the logit probability of news veracity assessments. 

\subsection*{Heterogeneity in Magnitude}

\Cref{heterogeneity-magnitude} plots the coefficients from the regression of news assessments on the interaction of whether the news was ``good'' or ``bad'' and partisanship, as well as on binarized observable demographics. We see that the effect of non-political demographics are small, and most are statistically insignificantly different from zero.

\begin{figure}[htb!]
\caption{Heterogeneity in the Magnitude of Motivated Reasoning}
\begin{center}
\vspace{-5mm}
\includegraphics[width = .76\textwidth]{fne_heterogeneity_magnitude.png}
\end{center}
\label{heterogeneity-magnitude}
\begin{threeparttable}
\begin{tablenotes}
\begin{scriptsize}
\vspace{-10mm}
\item \textbf{Notes:} This figure plots the relative treatment effect of seeing Pro-Party / Performance news versus Anti-Party / Performance news on subjects' news assessments by partisanship and demographics. These are OLS regression coefficients, errors clustered at subject level. FE included for subject, round number, and topic. Partisanship is from 0 to 1: abs(Republican Party rating - Democratic Party rating). Older: above the median age in the experiment. High income: above median income in the experiment. Red State: state voted for Trump in 2016. Religious: subject affiliates with any religion. 
\end{scriptsize}
\end{tablenotes}
\end{threeparttable}
\end{figure}

\newpage

\subsection*{Main Results by Randomization Group}

We consider Table 2 for each randomization group. Recall that subjects either give a second guess or see a WTP page, and subjects are either given a prior P(True) = 0.5 or are not. Neither arm affects the main results or the average news veracity assessment substantially.

\begin{center}
\begin{threeparttable}[htb!]
\caption{Motivated Reasoning and Perceived Truthfulness of News: Second-Guess Group}
\begin{footnotesize}
\begin{tabular}{l*{6}{c}}
\hline\hline
                    &\multicolumn{1}{c}{(1)}&\multicolumn{1}{c}{(2)}&\multicolumn{1}{c}{(3)}&\multicolumn{1}{c}{(4)}&\multicolumn{1}{c}{(5)}&\multicolumn{1}{c}{(6)}\\
\hline
Pro-Party News      &    0.090&    0.092&    0.041&    0.031&         &    0.081\\
                    &  (0.009)&  (0.009)&  (0.018)&  (0.009)&         &  (0.009)\\
Partisanship x      &         &         &    0.107&         &         &         \\
Pro-Party           &         &         &  (0.034)&         &         &         \\
Anti-Party News     &         &         &         &   -0.057&         &         \\
                    &         &         &         &  (0.010)&         &         \\
True News           &         &         &         &         &   -0.061&   -0.035\\
                    &         &         &         &         &  (0.009)&  (0.009)\\
Question FE         &      Yes&      Yes&      Yes&       No&      Yes&      Yes\\
Round FE            &      Yes&      Yes&      Yes&      Yes&      Yes&      Yes\\
Subject FE          &       No&      Yes&      Yes&      Yes&      Yes&      Yes\\
Subject controls    &      Yes&       No&       No&       No&       No&       No\\
Neutral News        &       No&       No&       No&      Yes&       No&       No\\
\hline
Observations        &     4085&     4085&     4085&     5455&     4085&     4085\\
\(R^{2}\)           &     0.05&     0.24&     0.25&     0.20&     0.23&     0.25\\
Mean                &    0.578&    0.578&    0.578&    0.581&    0.578&    0.578\\
\hline\hline
\multicolumn{7}{l}{\footnotesize Standard errors in parentheses}\\
\end{tabular}

\end{footnotesize}
\label{regression-assessment-secondguess}
\begin{tablenotes}
\begin{scriptsize}
\item \textbf{Notes:} OLS, errors clustered at subject level. Neutral News indicates that Pro-Party / Anti-Party news assessments are compared to assessments on Neutral topics. These classifications are defined in Table 1. Controls: race, gender, log(income), years of education, religion, and state. Partisanship is the absolute difference between ratings of the Republican and Democratic parties. Observations only for Second-Guess group.
\end{scriptsize}
\end{tablenotes}
\end{threeparttable}
\end{center}

\newpage

\begin{center}
\begin{threeparttable}[htb!]
\caption{Motivated Reasoning and Perceived Truthfulness of News: Willingness-to-Pay Group}
\begin{footnotesize}
\begin{tabular}{l*{6}{c}}
\hline\hline
                    &\multicolumn{1}{c}{(1)}&\multicolumn{1}{c}{(2)}&\multicolumn{1}{c}{(3)}&\multicolumn{1}{c}{(4)}&\multicolumn{1}{c}{(5)}&\multicolumn{1}{c}{(6)}\\
\hline
Pro-Party News      &    0.094&    0.085&    0.042&    0.043&         &    0.074\\
                    &  (0.009)&  (0.009)&  (0.017)&  (0.009)&         &  (0.009)\\
Partisanship x      &         &         &    0.087&         &         &         \\
Pro-Party           &         &         &  (0.029)&         &         &         \\
Anti-Party News     &         &         &         &   -0.039&         &         \\
                    &         &         &         &  (0.009)&         &         \\
True News           &         &         &         &         &   -0.056&   -0.032\\
                    &         &         &         &         &  (0.009)&  (0.009)\\
Question FE         &      Yes&      Yes&      Yes&       No&      Yes&      Yes\\
Round FE            &      Yes&      Yes&      Yes&      Yes&      Yes&      Yes\\
Subject FE          &       No&      Yes&      Yes&      Yes&      Yes&      Yes\\
Subject controls    &      Yes&       No&       No&       No&       No&       No\\
Neutral News        &       No&       No&       No&      Yes&       No&       No\\
\hline
Observations        &     3817&     3817&     3817&     5097&     3817&     3817\\
\(R^{2}\)           &     0.04&     0.25&     0.26&     0.22&     0.24&     0.26\\
Mean                &    0.570&    0.570&    0.570&    0.569&    0.570&    0.570\\
\hline\hline
\multicolumn{7}{l}{\footnotesize Standard errors in parentheses}\\
\end{tabular}

\end{footnotesize}
\label{regression-assessment-wtp}
\begin{tablenotes}
\begin{scriptsize}
\item \textbf{Notes:} OLS, errors clustered at subject level. Neutral News indicates that Pro-Party / Anti-Party news assessments are compared to assessments on Neutral topics. These classifications are defined in Table 1. Controls: race, gender, log(income), years of education, religion, and state. Partisanship is the absolute difference between ratings of the Republican and Democratic parties. Observations only for Willingness-to-Pay group.
\end{scriptsize}
\end{tablenotes}
\end{threeparttable}
\end{center}

\newpage

\begin{center}
\begin{threeparttable}[htb!]
\caption{Motivated Reasoning and Perceived Truthfulness of News: Given 50-50 Prior}
\begin{footnotesize}
\begin{tabular}{l*{6}{c}}
\hline\hline
                    &\multicolumn{1}{c}{(1)}&\multicolumn{1}{c}{(2)}&\multicolumn{1}{c}{(3)}&\multicolumn{1}{c}{(4)}&\multicolumn{1}{c}{(5)}&\multicolumn{1}{c}{(6)}\\
\hline
Pro-Party News      &    0.091&    0.088&    0.067&    0.046&         &    0.078\\
                    &  (0.011)&  (0.010)&  (0.020)&  (0.011)&         &  (0.011)\\
Partisanship x      &         &         &    0.049&         &         &         \\
Pro-Party           &         &         &  (0.035)&         &         &         \\
Anti-Party News     &         &         &         &   -0.040&         &         \\
                    &         &         &         &  (0.012)&         &         \\
True News           &         &         &         &         &   -0.056&   -0.029\\
                    &         &         &         &         &  (0.011)&  (0.011)\\
Question FE         &      Yes&      Yes&      Yes&       No&      Yes&      Yes\\
Round FE            &      Yes&      Yes&      Yes&      Yes&      Yes&      Yes\\
Subject FE          &       No&      Yes&      Yes&      Yes&      Yes&      Yes\\
Subject controls    &      Yes&       No&       No&       No&       No&       No\\
Neutral News        &       No&       No&       No&      Yes&       No&       No\\
\hline
Observations        &     2674&     2674&     2674&     3568&     2674&     2674\\
\(R^{2}\)           &     0.05&     0.27&     0.28&     0.22&     0.25&     0.27\\
Mean                &    0.573&    0.573&    0.573&    0.572&    0.573&    0.573\\
\hline\hline
\multicolumn{7}{l}{\footnotesize Standard errors in parentheses}\\
\end{tabular}

\end{footnotesize}
\label{regression-assessment-prior}
\begin{tablenotes}
\begin{scriptsize}
\item \textbf{Notes:} OLS, errors clustered at subject level. Neutral News indicates that Pro-Party / Anti-Party news assessments are compared to assessments on Neutral topics. These classifications are defined in Table 1. Controls: race, gender, log(income), years of education, religion, and state. Partisanship is the absolute difference between ratings of the Republican and Democratic parties. Observations only if Given 50-50 Prior.
\end{scriptsize}
\end{tablenotes}
\end{threeparttable}
\end{center}

\newpage

\begin{center}
\begin{threeparttable}[htb!]
\caption{Motivated Reasoning and Perceived Truthfulness of News: Not Given 50-50 Prior}
\begin{footnotesize}
\begin{tabular}{l*{6}{c}}
\hline\hline
                    &\multicolumn{1}{c}{(1)}&\multicolumn{1}{c}{(2)}&\multicolumn{1}{c}{(3)}&\multicolumn{1}{c}{(4)}&\multicolumn{1}{c}{(5)}&\multicolumn{1}{c}{(6)}\\
\hline
Pro-Party News      &    0.093&    0.088&    0.025&    0.033&         &    0.077\\
                    &  (0.008)&  (0.008)&  (0.015)&  (0.008)&         &  (0.008)\\
Partisanship x      &         &         &    0.131&         &         &         \\
Pro-Party           &         &         &  (0.027)&         &         &         \\
Anti-Party News     &         &         &         &   -0.052&         &         \\
                    &         &         &         &  (0.008)&         &         \\
True News           &         &         &         &         &   -0.061&   -0.037\\
                    &         &         &         &         &  (0.007)&  (0.007)\\
Question FE         &      Yes&      Yes&      Yes&       No&      Yes&      Yes\\
Round FE            &      Yes&      Yes&      Yes&      Yes&      Yes&      Yes\\
Subject FE          &       No&      Yes&      Yes&      Yes&      Yes&      Yes\\
Subject controls    &      Yes&       No&       No&       No&       No&       No\\
Neutral News        &       No&       No&       No&      Yes&       No&       No\\
\hline
Observations        &     5228&     5228&     5228&     6984&     5228&     5228\\
\(R^{2}\)           &     0.05&     0.24&     0.24&     0.20&     0.22&     0.24\\
Mean                &    0.575&    0.575&    0.575&    0.577&    0.575&    0.575\\
\hline\hline
\multicolumn{7}{l}{\footnotesize Standard errors in parentheses}\\
\end{tabular}

\end{footnotesize}
\label{regression-assessment-noprior}
\begin{tablenotes}
\begin{scriptsize}
\item \textbf{Notes:} OLS, errors clustered at subject level. Neutral News indicates that Pro-Party / Anti-Party news assessments are compared to assessments on Neutral topics. These classifications are defined in Table 1. Controls: race, gender, log(income), years of education, religion, and state. Partisanship is the absolute difference between ratings of the Republican and Democratic parties. Observations only for Not Given 50-50 Prior.
\end{scriptsize}
\end{tablenotes}
\end{threeparttable}
\end{center}

\newpage

\subsection*{Results Without Comprehension Checks}

The main results do not include subjects who fail attention and comprehension checks. As such, 313 of 1300 subjects are removed from the analysis. This table repeats the analysis without removing subjects; results do not significantly change.

\begin{center}
\begin{threeparttable}[htb!]
\caption{Motivated Reasoning and Perceived Truthfulness of News: Including Subjects Who Fail Comprehension}
\begin{footnotesize}
\begin{tabular}{l*{6}{c}}
\hline\hline
                    &\multicolumn{1}{c}{(1)}&\multicolumn{1}{c}{(2)}&\multicolumn{1}{c}{(3)}&\multicolumn{1}{c}{(4)}&\multicolumn{1}{c}{(5)}&\multicolumn{1}{c}{(6)}\\
\hline
Pro-Party News      &    0.076&    0.071&    0.027&    0.031&         &    0.064\\
                    &  (0.005)&  (0.005)&  (0.010)&  (0.005)&         &  (0.005)\\
Partisanship x      &         &         &    0.097&         &         &         \\
Pro-Party           &         &         &  (0.018)&         &         &         \\
Anti-Party News     &         &         &         &   -0.038&         &         \\
                    &         &         &         &  (0.006)&         &         \\
True News           &         &         &         &         &   -0.043&   -0.026\\
                    &         &         &         &         &  (0.005)&  (0.005)\\
Question FE         &      Yes&      Yes&      Yes&       No&      Yes&      Yes\\
Round FE            &      Yes&      Yes&      Yes&      Yes&      Yes&      Yes\\
Subject FE          &       No&      Yes&      Yes&      Yes&      Yes&      Yes\\
Subject controls    &      Yes&       No&       No&       No&       No&       No\\
\hline
Observations        &    10478&    10478&    10478&    13991&    10478&    10478\\
\(R^{2}\)           &     0.03&     0.30&     0.31&     0.27&     0.29&     0.30\\
Mean                &    0.561&    0.561&    0.561&    0.562&    0.561&    0.561\\
\hline\hline
\multicolumn{7}{l}{\footnotesize Standard errors in parentheses}\\
\end{tabular}

\label{regression-assessment-failedcheck}
\end{footnotesize}
\begin{tablenotes}
\begin{scriptsize}
\item \textbf{Notes:} OLS, errors clustered at subject level. Neutral News indicates that Pro-Party / Anti-Party news assessments are compared to assessments on Neutral topics. These classifications are defined in Table 1. Controls: race, gender, log(income), years of education, religion, and state. Partisanship is the absolute difference between ratings of the Republican and Democratic parties. Observations \textit{include} subjects who failed comprehension checks.
\end{scriptsize}
\end{tablenotes}
\end{threeparttable}
\end{center}

\newpage

\subsection*{Results Using Logit Veracity Assessments}

The model suggests that the relevant dependent variable is logit(P(True)) instead of P(True). \Cref{regression-assessment-logit} is the same as Table 2 but with this new dependent variable. Technically, since logit(0) and logit(1) are undefined, they are replaced here with logit(0.025) and logit(0.975).\footnote{Subjects choose P(True) = 0 to maximize expected earnings if and only if they believe P(True) $\in [0, 0.05]$. 0.025 is the midpoint of this range. Results are similar if 0.05 is chosen or if these observations are removed.}

\begin{center}
\begin{threeparttable}[htb!]
\caption{Motivated Reasoning and Perceived Truthfulness of News: Logit Veracity Assessments}
\begin{footnotesize}
\begin{tabular}{l*{6}{c}}
\hline\hline
                    &\multicolumn{1}{c}{(1)}&\multicolumn{1}{c}{(2)}&\multicolumn{1}{c}{(3)}&\multicolumn{1}{c}{(4)}&\multicolumn{1}{c}{(5)}&\multicolumn{1}{c}{(6)}\\
\hline
Pro-Party News      &    0.473&    0.453&    0.206&    0.173&         &    0.396\\
                    &  (0.033)&  (0.033)&  (0.065)&  (0.034)&         &  (0.034)\\
Partisanship x      &         &         &    0.515&         &         &         \\
Pro-Party           &         &         &  (0.117)&         &         &         \\
Anti-Party News     &         &         &         &   -0.263&         &         \\
                    &         &         &         &  (0.037)&         &         \\
True News           &         &         &         &         &   -0.306&   -0.178\\
                    &         &         &         &         &  (0.032)&  (0.032)\\
Question FE         &      Yes&      Yes&      Yes&       No&      Yes&      Yes\\
Round FE            &      Yes&      Yes&      Yes&      Yes&      Yes&      Yes\\
Subject FE          &       No&      Yes&      Yes&      Yes&      Yes&      Yes\\
Subject controls    &      Yes&       No&       No&       No&       No&       No\\
Neutral News        &       No&       No&       No&      Yes&       No&       No\\
\hline
Observations        &     7902&     7902&     7902&    10552&     7902&     7902\\
\(R^{2}\)           &     0.04&     0.25&     0.25&     0.21&     0.23&     0.25\\
Mean                &    0.374&    0.374&    0.374&    0.383&    0.374&    0.374\\
\hline\hline
\multicolumn{7}{l}{\footnotesize Standard errors in parentheses}\\
\end{tabular}

\end{footnotesize}
\label{regression-assessment-logit}
\begin{tablenotes}
\begin{scriptsize}
\item \textbf{Notes:} Dependent variable is logit(P(True)). OLS, errors clustered at subject level. Neutral News indicates that Pro-Party / Anti-Party news assessments are compared to assessments on Neutral topics. These classifications are defined in Table 1. Controls: race, gender, log(income), years of education, religion, and state. Partisanship is the absolute difference between ratings of the Republican and Democratic parties. 
\end{scriptsize}
\end{tablenotes}
\end{threeparttable}
\end{center}

\newpage

\begin{figure}[h]
\centering
\caption{Round-by-Round Effects of News Direction on Perceived Veracity}
\includegraphics[width = .85\textwidth]{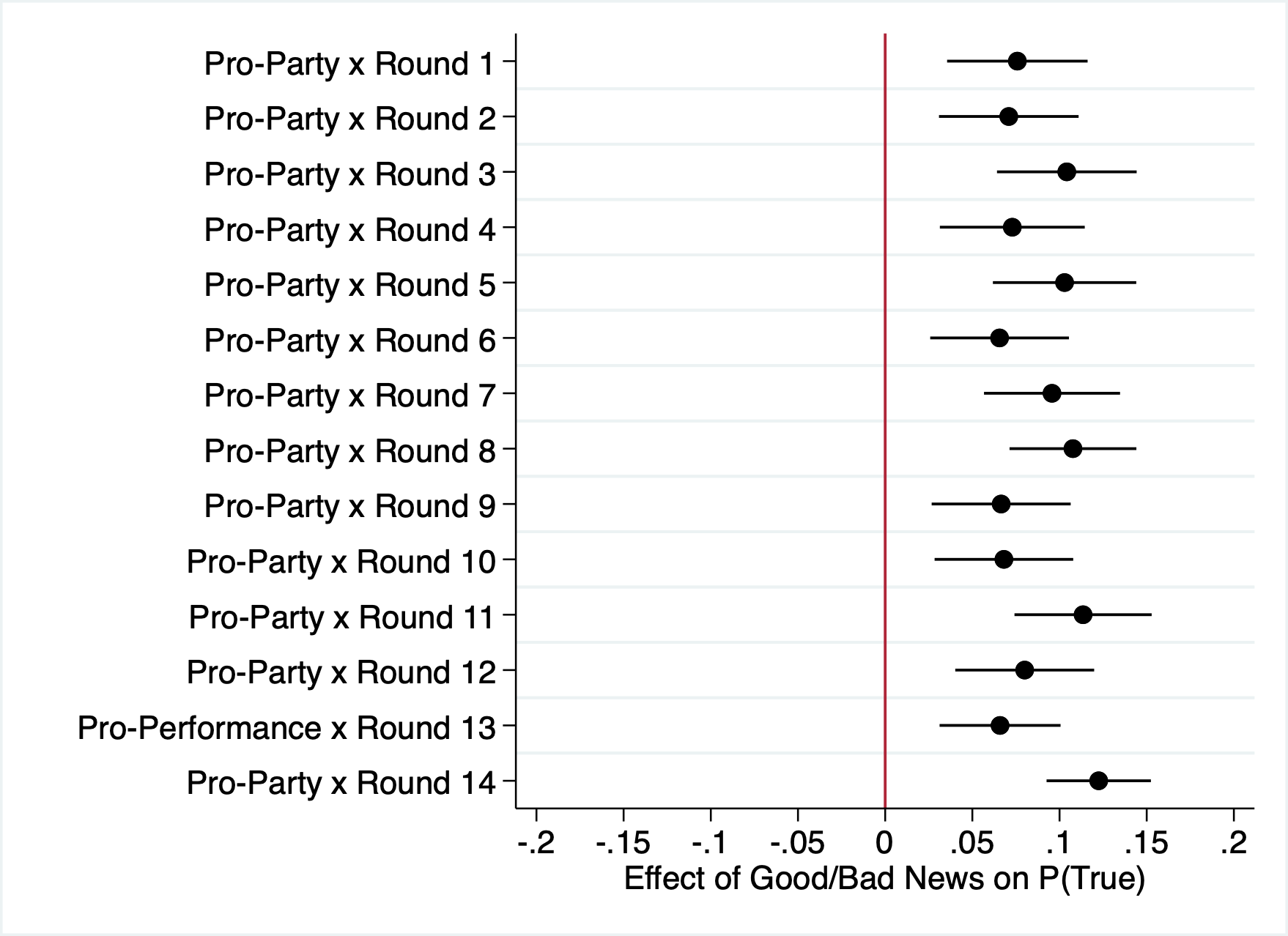}
\label{moreas-by-round}
\begin{threeparttable}
\begin{tablenotes}
\begin{scriptsize}
\vspace{-7mm}
\item \textbf{Notes:} OLS regression coefficients, errors clustered at subject level. FE included for subject, round number, and topic. Pro-Party (vs. Anti-Party) and Pro-Performance (vs. Anti-Performance news is defined in Table 1. Performance news is only seen in Round 13. Error bars correspond to 95 percent confidence intervals. 
\end{scriptsize}
\end{tablenotes}
\end{threeparttable}
\end{figure}

\newpage

\section{Study Materials: Experiment Flow and Screenshots (Not For Publication)}

\subsection{Flow of Experiment}

Subjects see a series of pages in the following order:
\begin{itemize}
\item Introduction and Consent
\item Demographics and Current Events Quiz
\item Opinions
\item Instructions for Question Pages
\item Question 1
\item Instructions for News Assessment Pages
\item News Assessment 1
\item Question 2, News Assessment 2, \dots, Question 14, News Assessment 14
\item Feedback
\item Results and Payment
\end{itemize}
Screenshots for each of the pages are in the following subsection. Red boxes are not shown to subjects and are included for illustration purposes only. Results pages here are cut off after three questions, but all results are shown to subjects. Choices on the Demographics page and statements on the Opinions page are randomly ordered.

Subjects in the Willingness-To-Pay group see the News Valuation page between Question 12 and News Assessment 12. They see the black bar page if their elicited valuation is lower than the random number.

Subjects in the Second Guess group see the version of the News Assessment page with the message ``After seeing this message and assessing its truthfulness, what is your guess of the answer to the original question?''

\newpage

\subsection{Screenshots of Study Materials}

\begin{center}
\includegraphics[width = \textwidth]{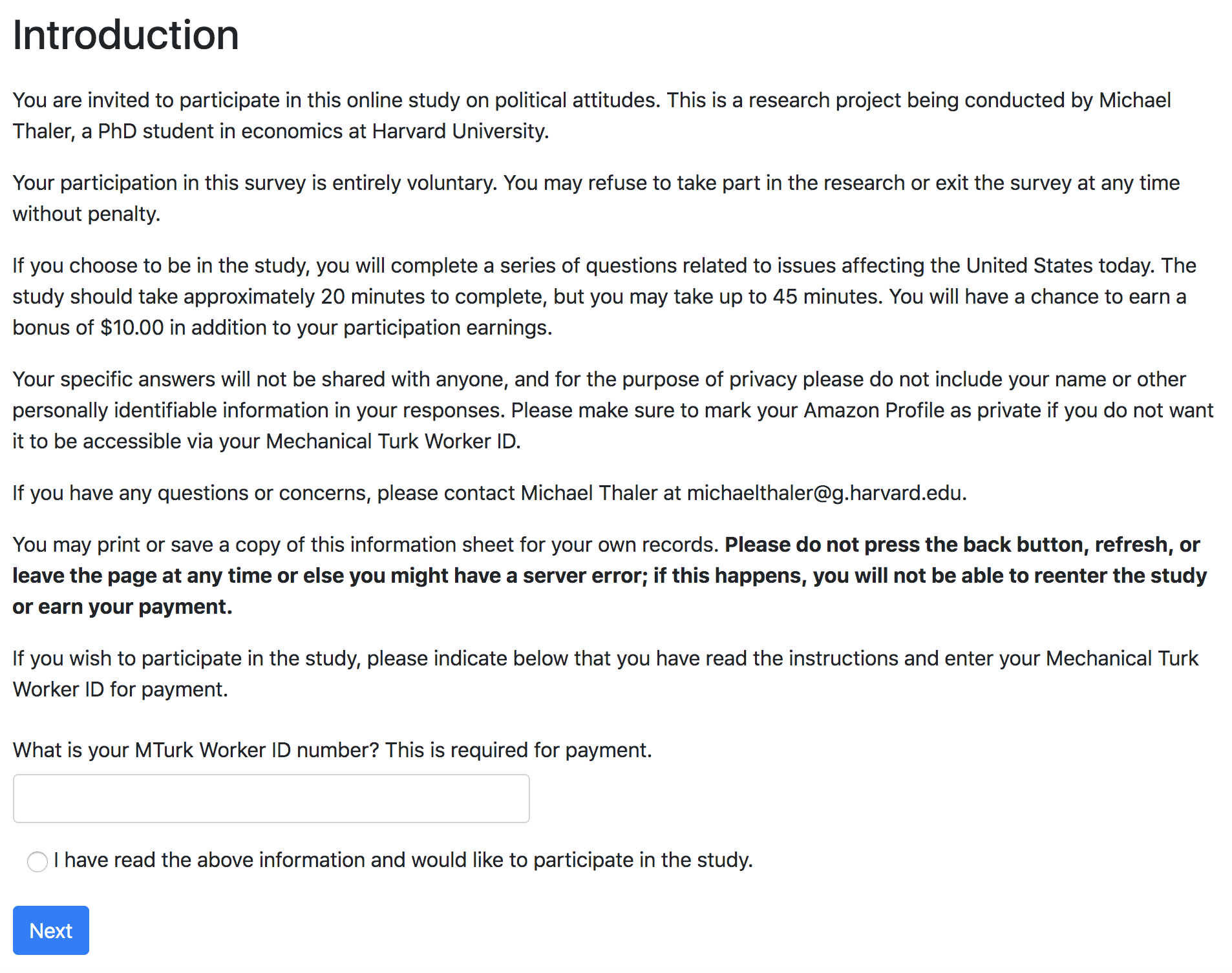}
\end{center}

\newpage

\begin{center}
\includegraphics[height = \textheight]{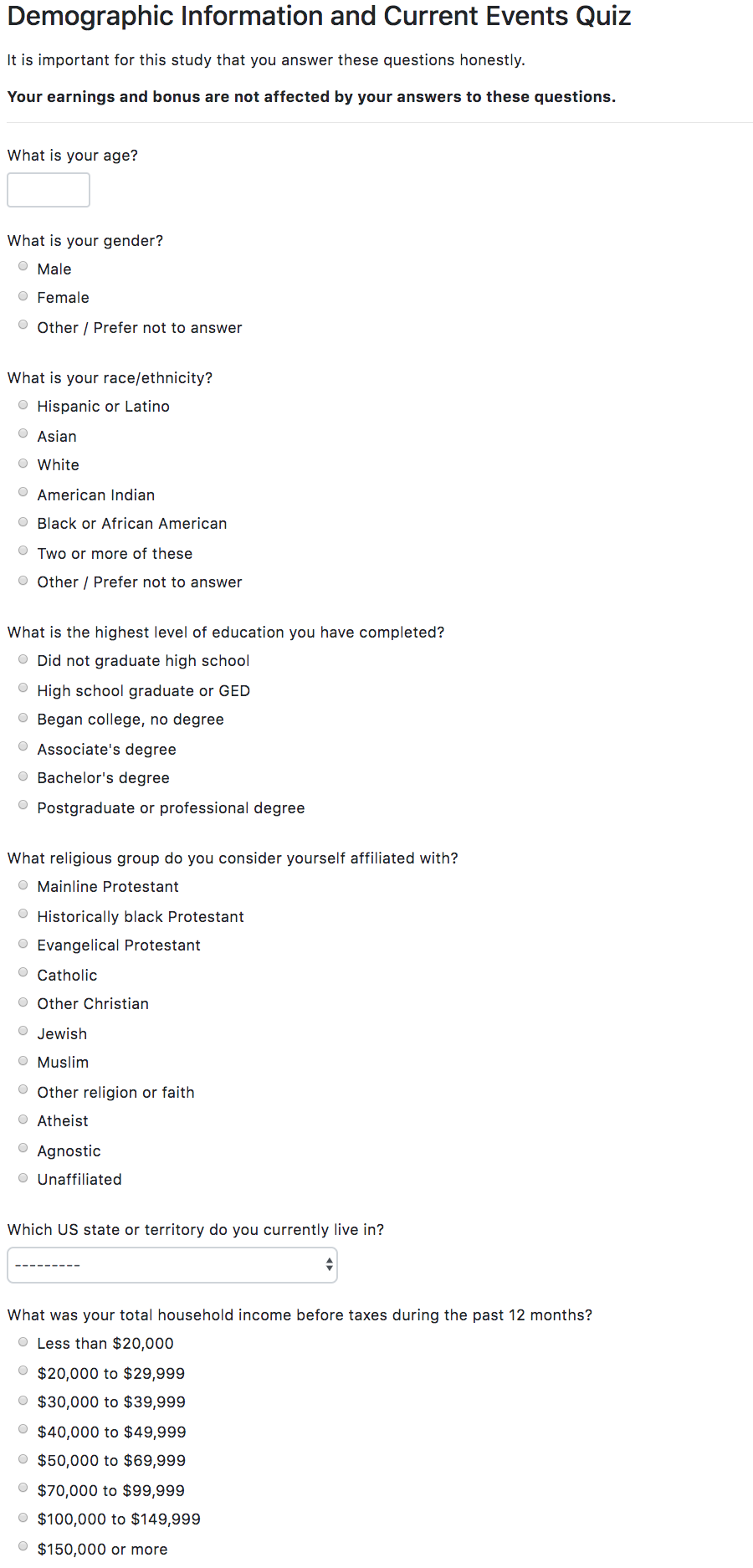}
\end{center}

\newpage

\begin{center}
\includegraphics[height = \textheight]{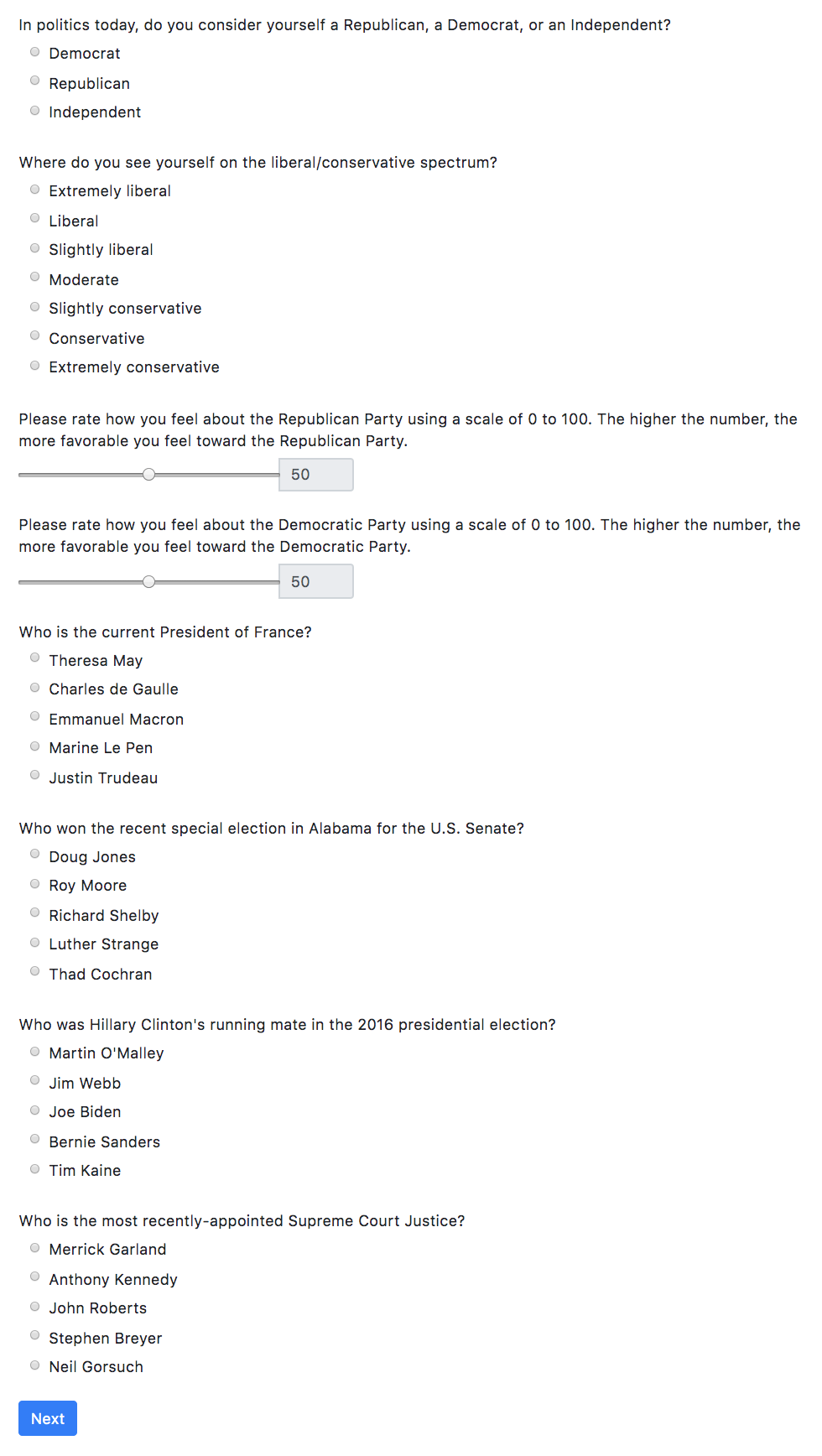}
\end{center}

\newpage


\label{instructions_question_text}
\begin{center}
\includegraphics[width = \textwidth]{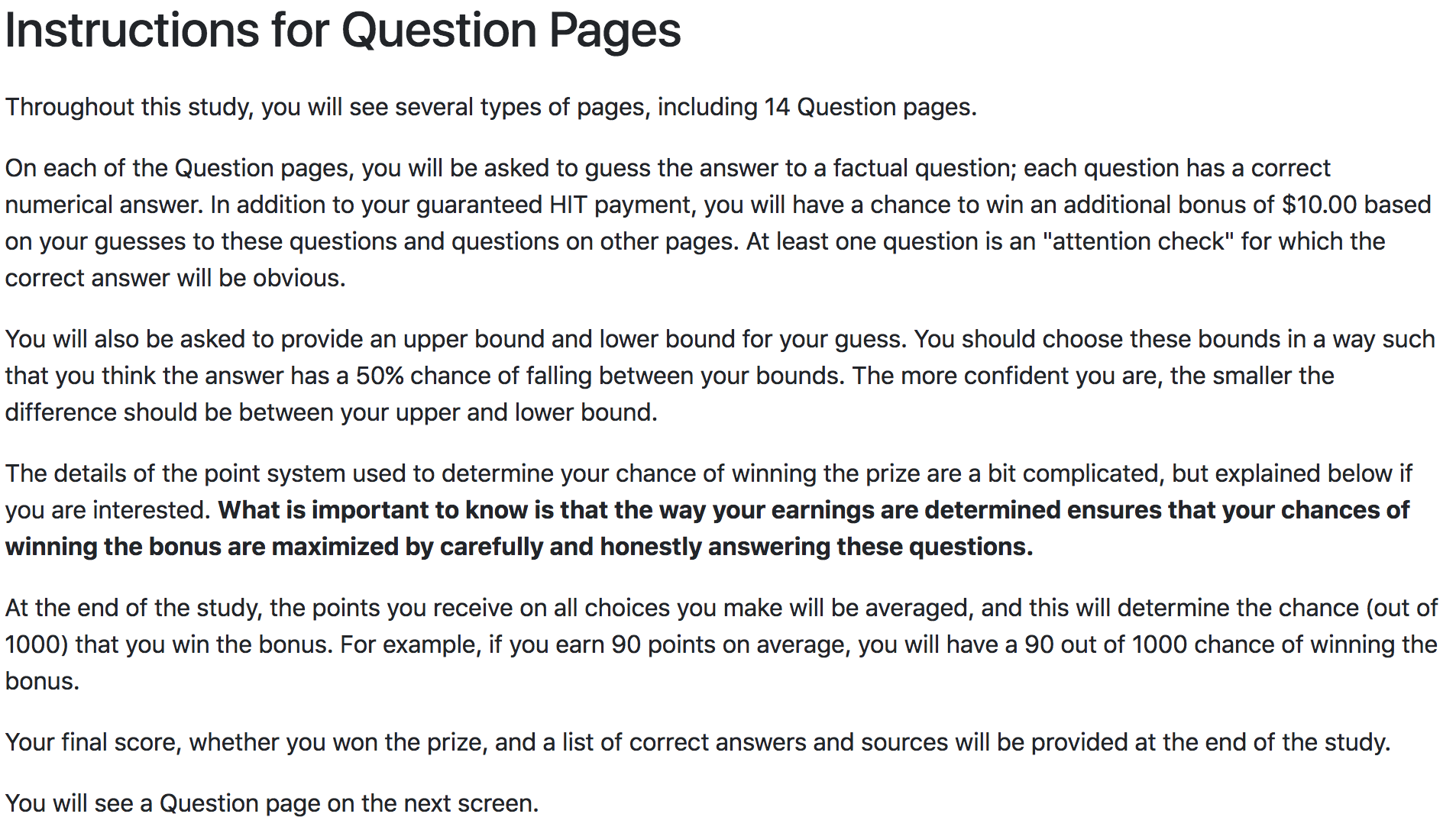}
\end{center}

\label{instructions_question_points}
\begin{center}
\includegraphics[width = \textwidth]{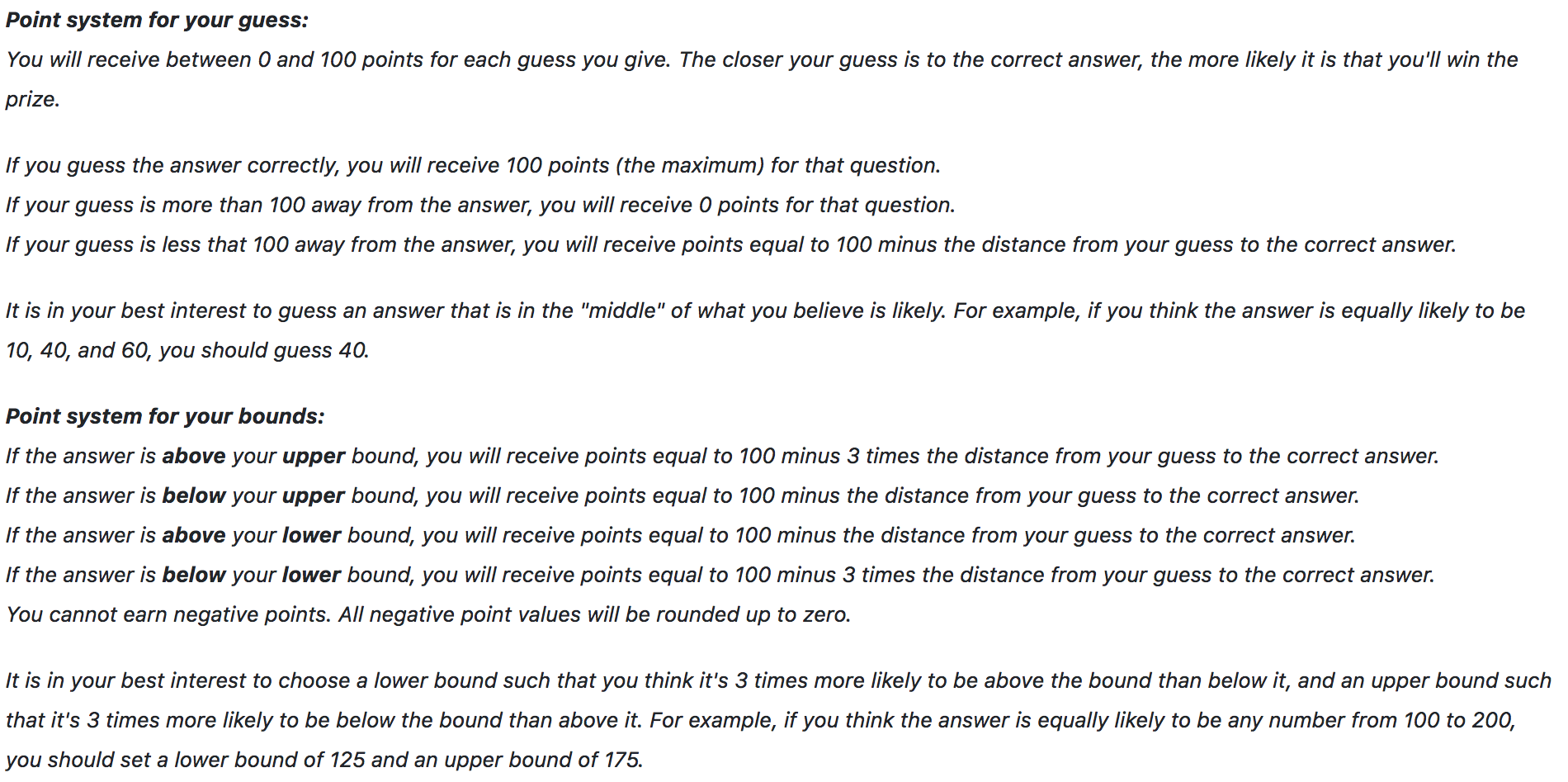}
\end{center}

\newpage


\begin{figure}[h]
\begin{center}
\includegraphics[width = \textwidth]{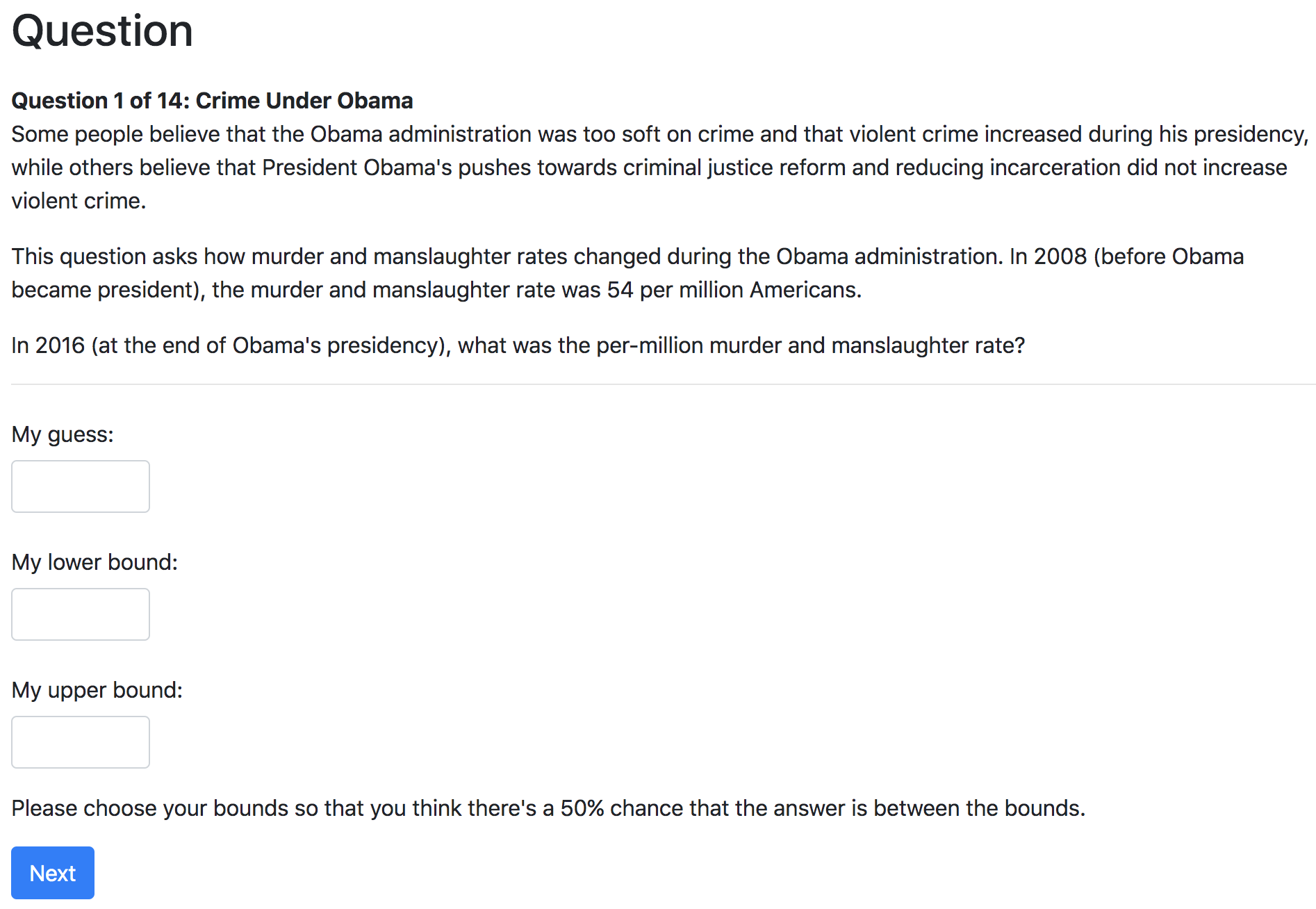}
\end{center}
\caption{Crime Under Obama question page.}
\label{question3}
\end{figure}

\newpage


\label{instructions_news_text}
\begin{center}
\includegraphics[width = \textwidth]{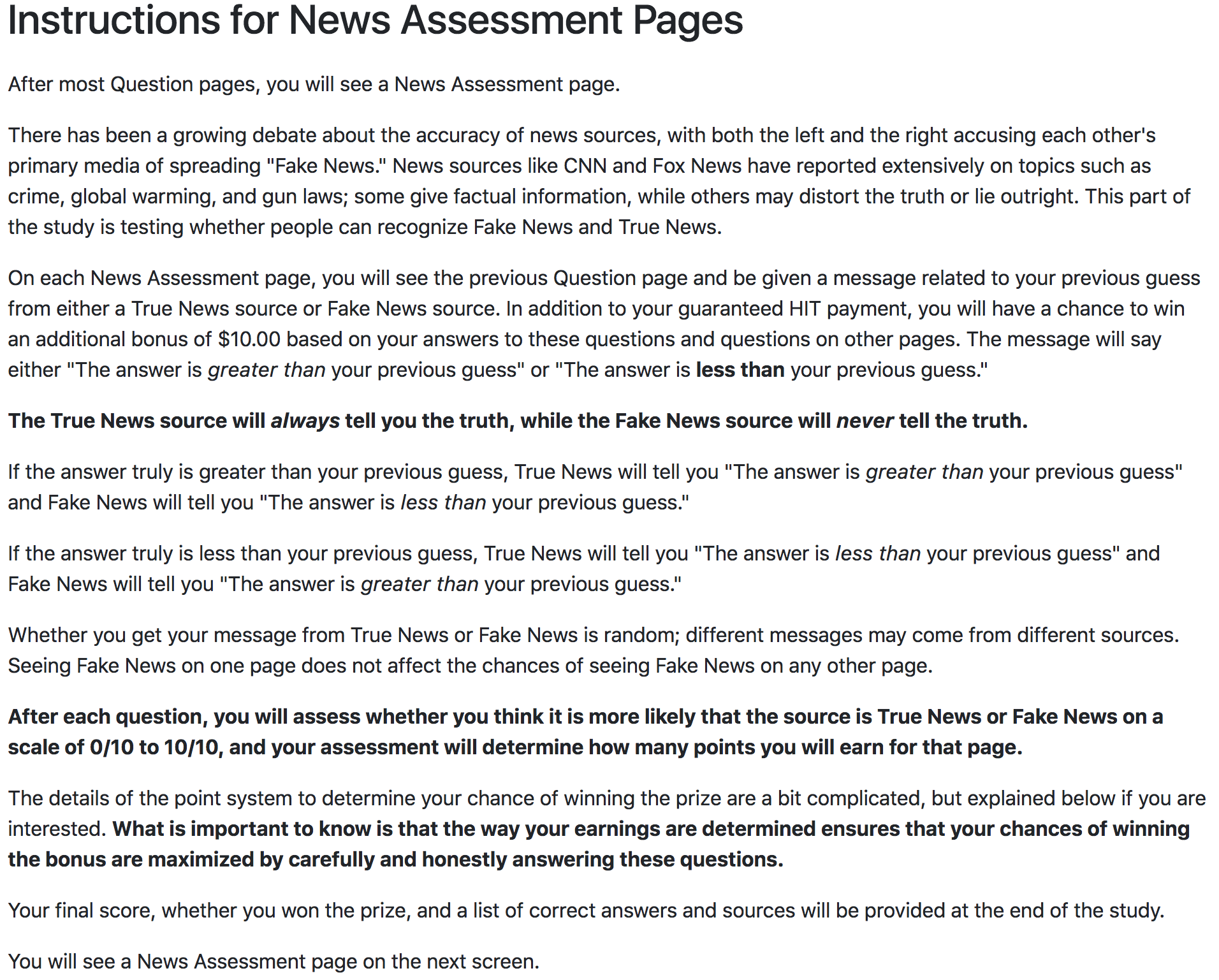}
\end{center}

\label{instructions_news_points}
\begin{center}
\includegraphics[width = \textwidth]{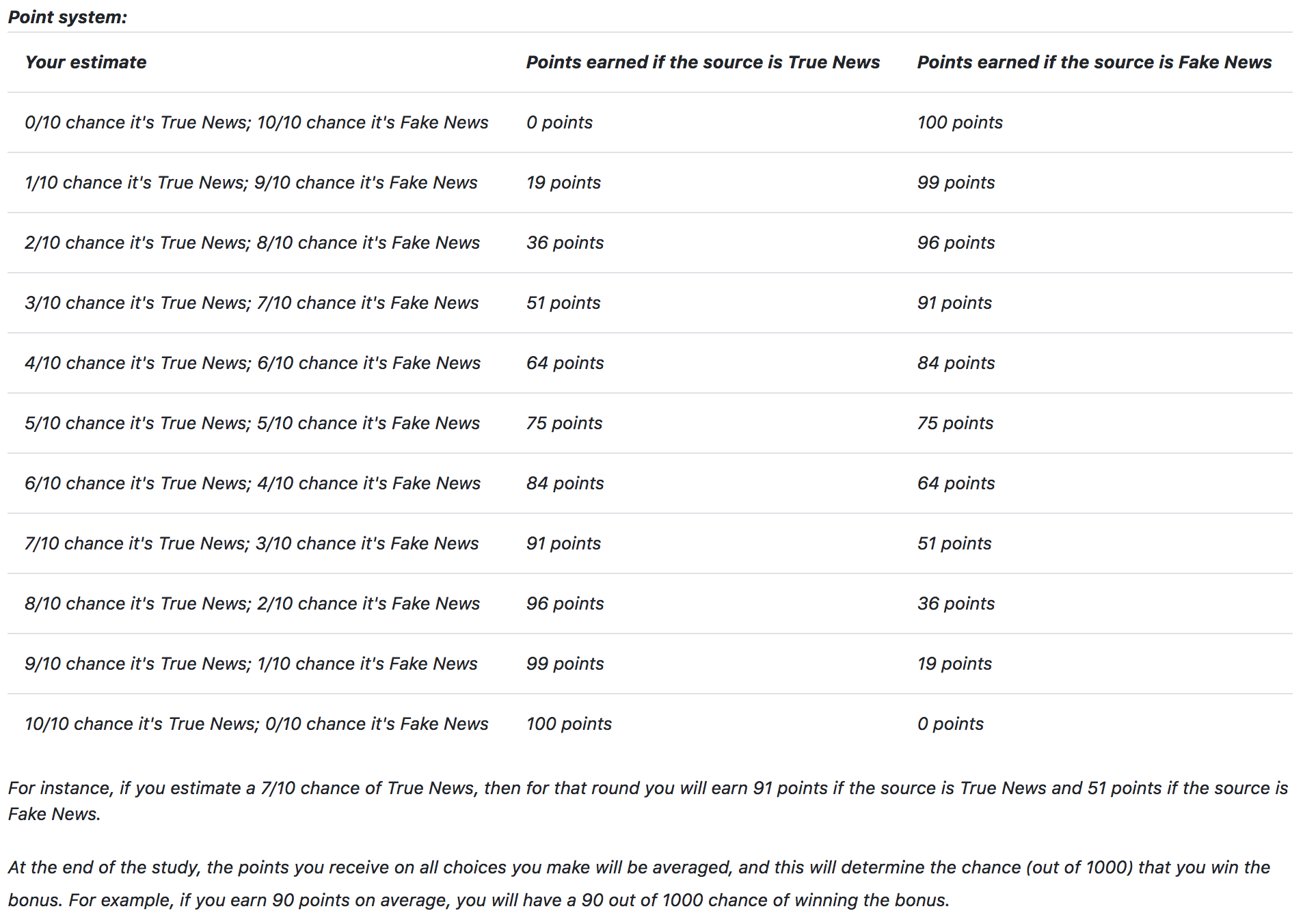}
\end{center}

\newpage

\begin{figure}
\begin{center}
\includegraphics[width = \textwidth]{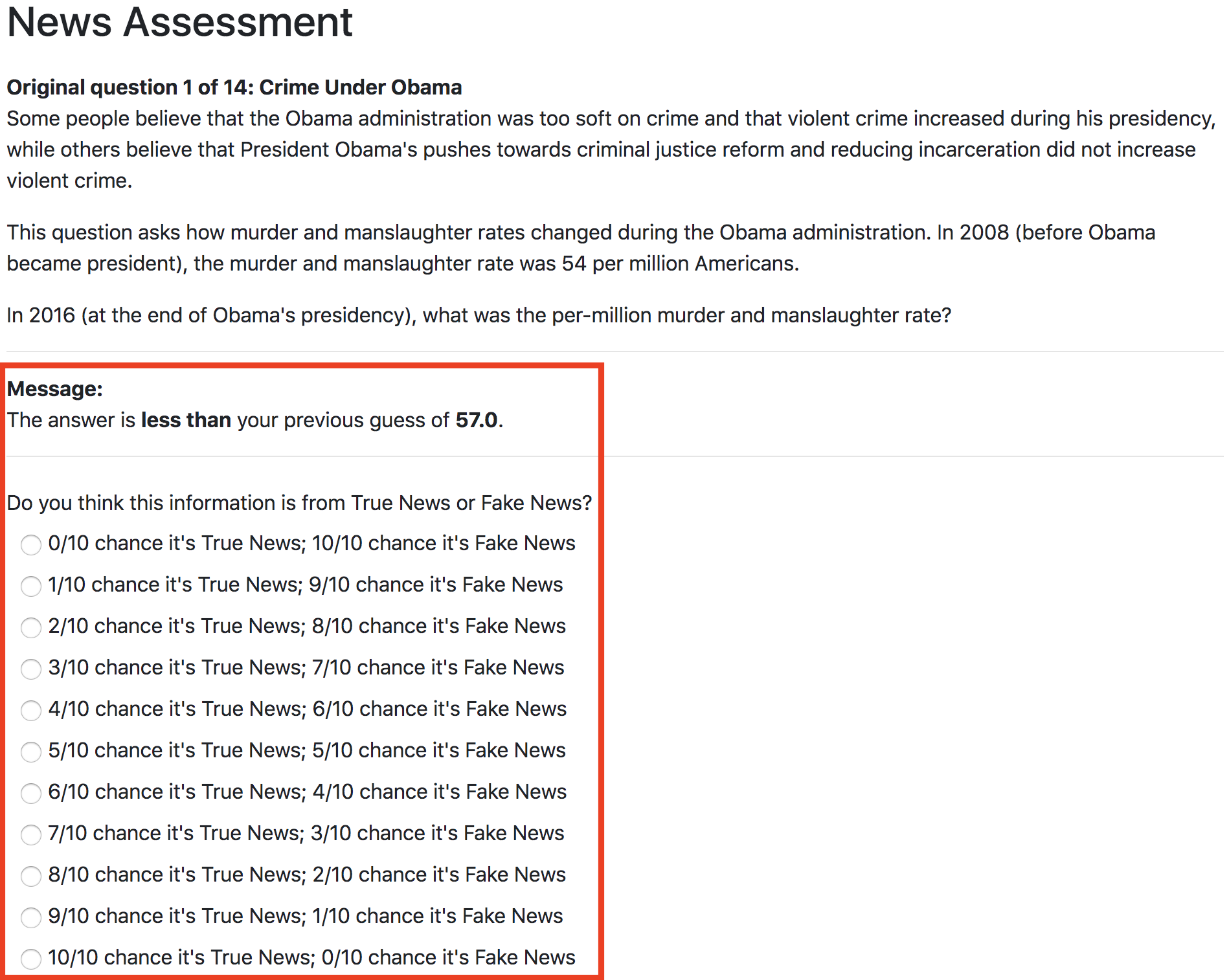}
\end{center}
\caption{Crime Under Obama news assessment page.}
\label{news3}
\end{figure}

\clearpage

\begin{figure}
\begin{center}
\includegraphics[width = \textwidth]{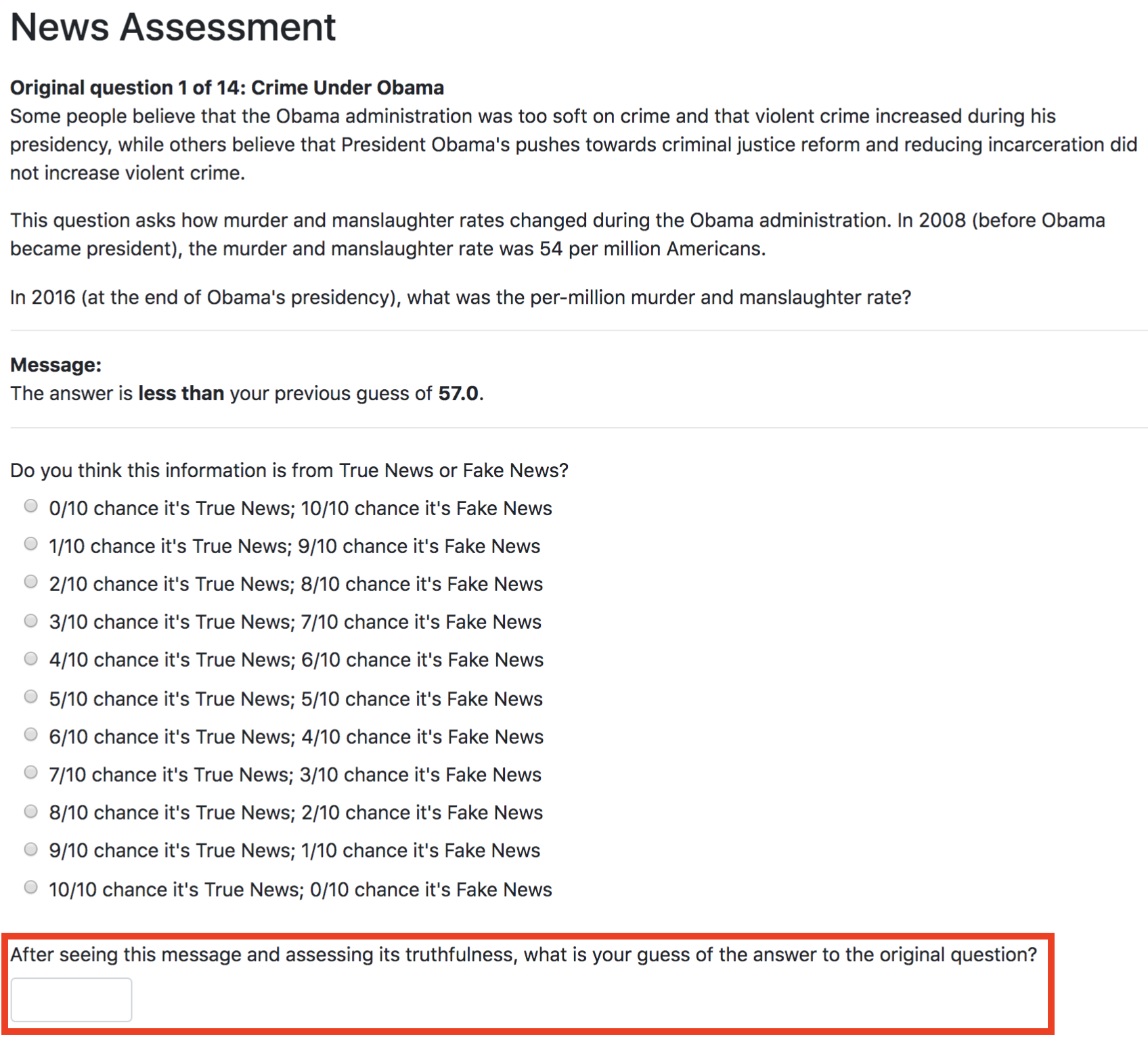}
\end{center}
\caption{Crime Under Obama news assessment page: Second Guess question.}
\label{secondguess3}
\end{figure}

\clearpage

\begin{center}
\includegraphics[width = \textwidth]{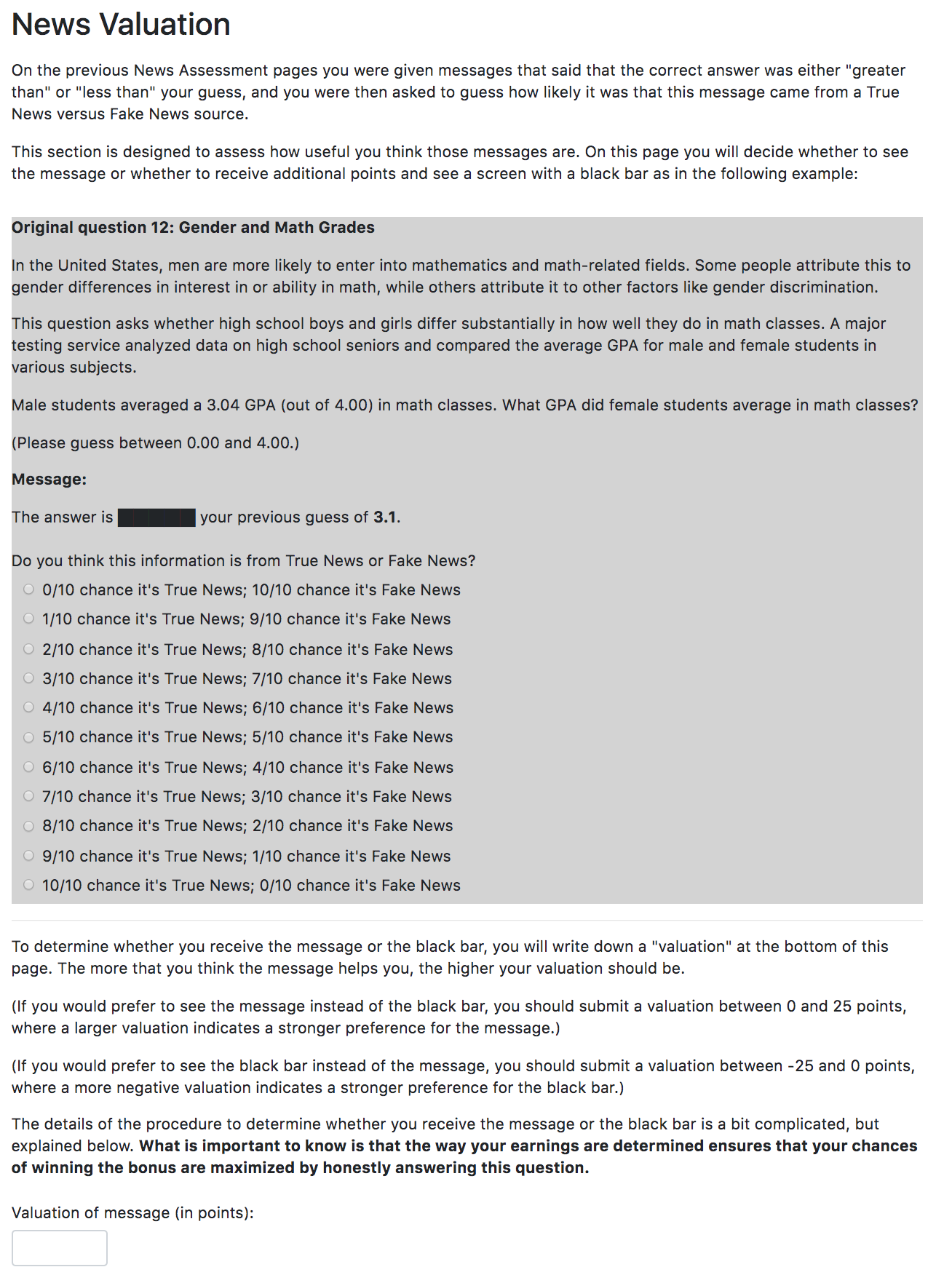}
\end{center}

\clearpage

\begin{center}
\includegraphics[width = \textwidth]{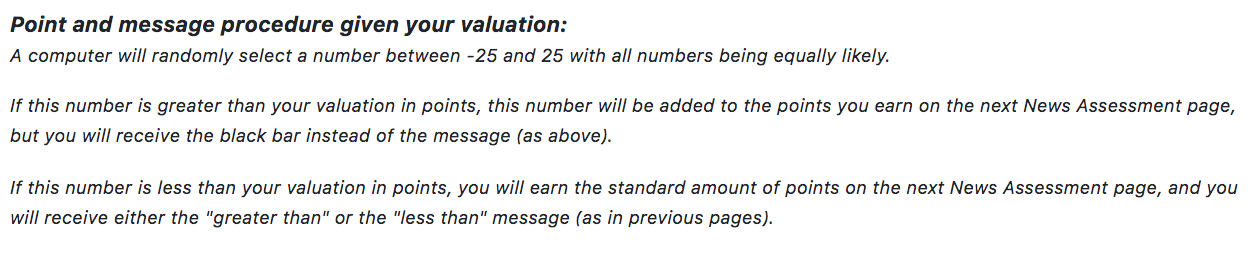}
\end{center}

\newpage

\begin{center}
\includegraphics[width = \textwidth]{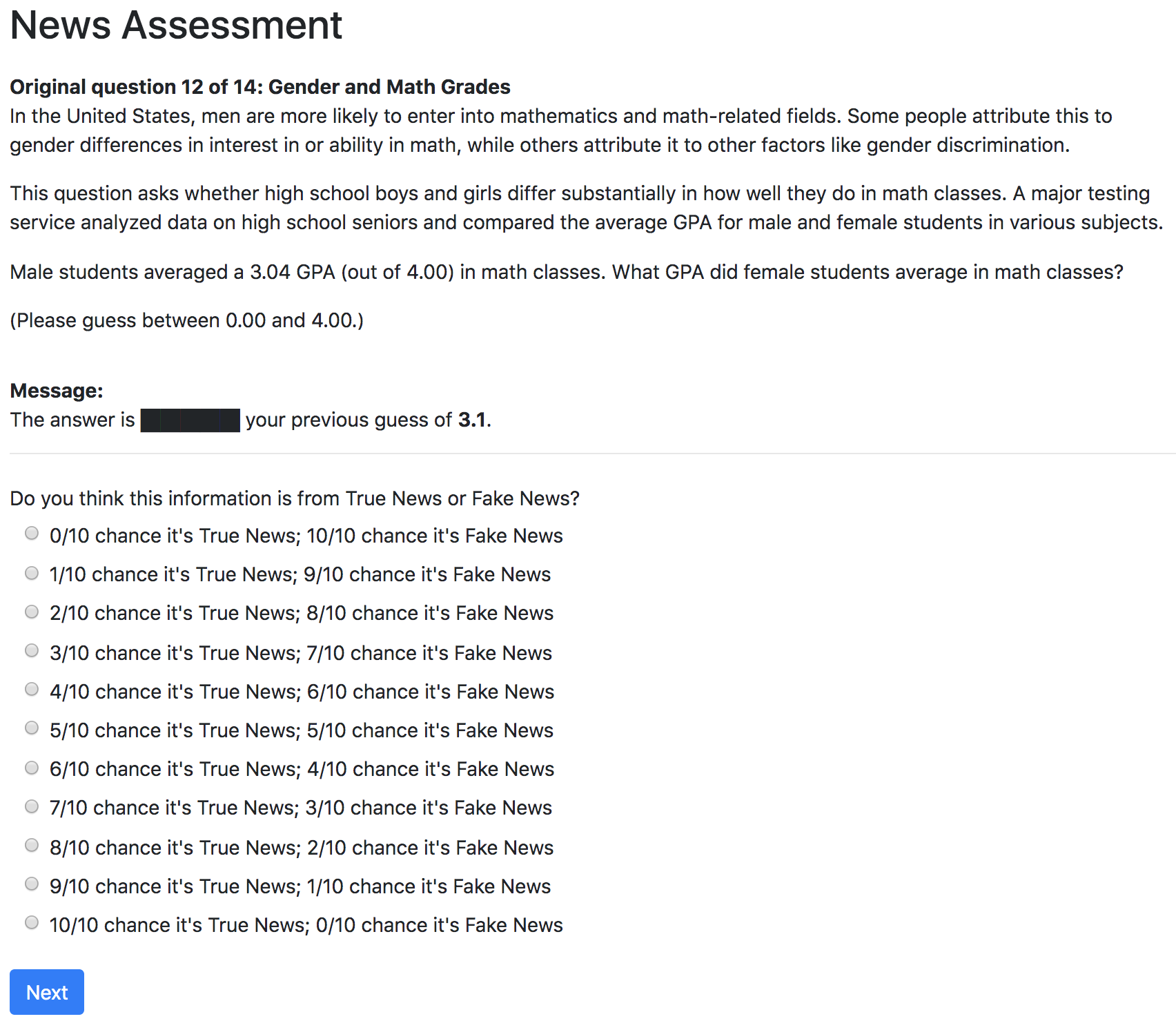}
\end{center}

\newpage

\begin{center}
\includegraphics[width = \textwidth]{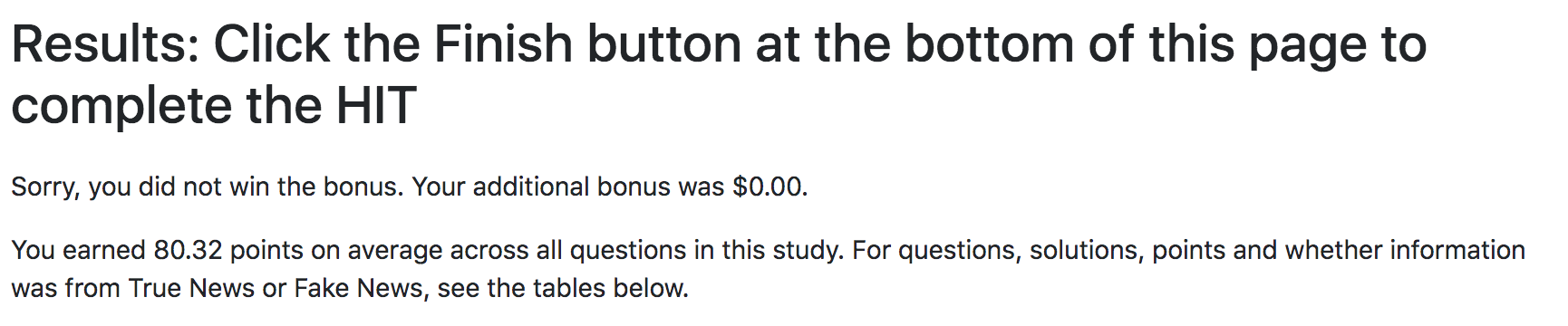}
\end{center}

\begin{center}
\includegraphics[width = \textwidth]{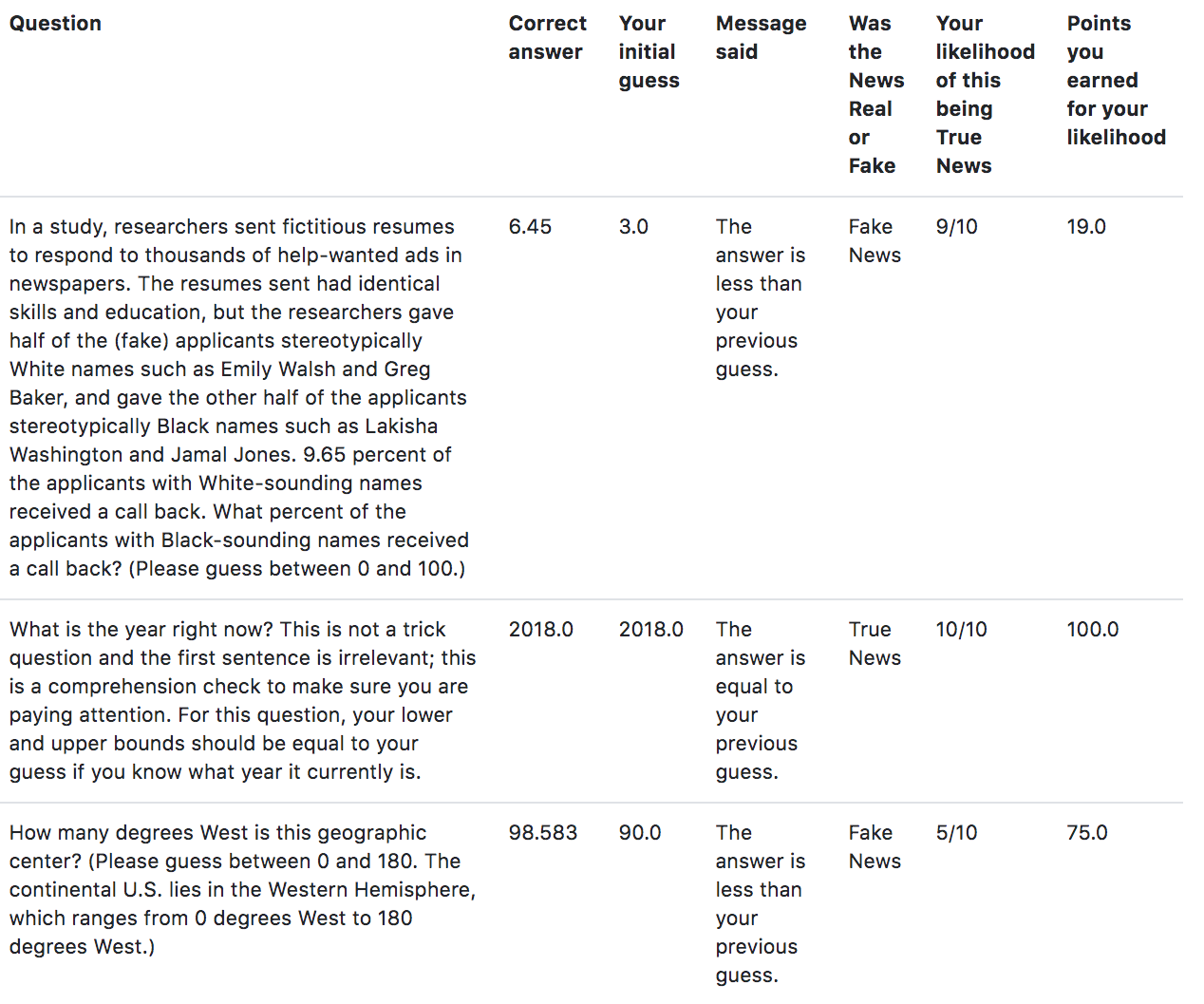}
\end{center}

\newpage

\begin{center}
\includegraphics[width = \textwidth]{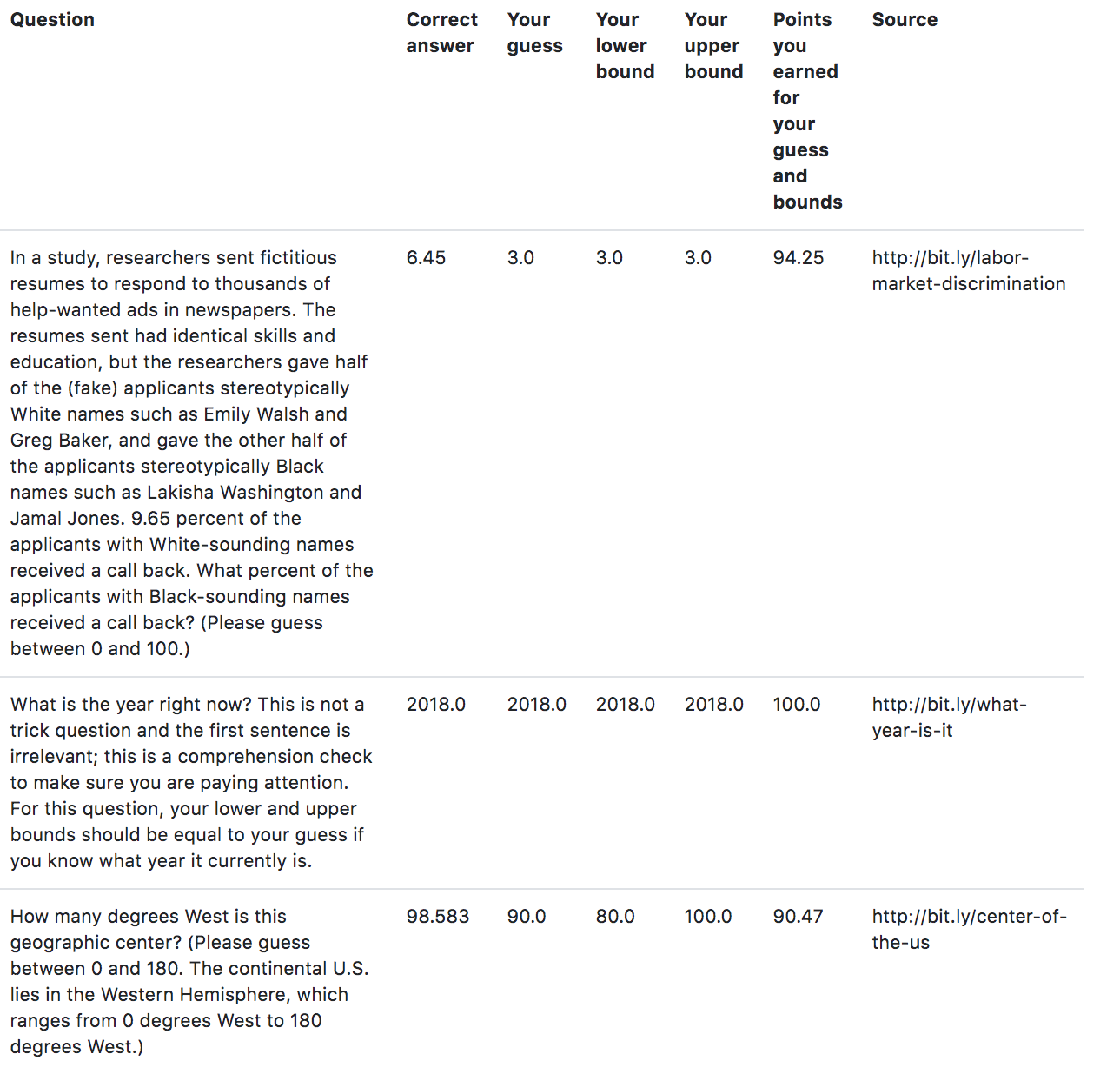}
\end{center}

\end{document}